\documentclass[twocolumn, trackchanges]{aastex7}

\begin{document}

\title{SMA Observations Reveal Abundant HNC Chemistry in Transition Disks}

\correspondingauthor{Kyle Gresko}

\author[0009-0008-7113-4125]{Kyle Gresko}
\affiliation{Department of Astronomy, University of Virginia, Charlottesville, VA 22904, USA}
\affiliation{Department of Physics and Astronomy, University of Texas San Antonio, San Antonio, TX 78249, USA}
\affiliation{Space Science Division, Southwest Research Institute, San Antonio, TX 78238, USA}
\email[show]{kyle.gresko@utsa.edu}  

\author[0000-0003-1413-1776]{Charles J.\ Law}
\altaffiliation{NASA Hubble Fellowship Program Sagan Fellow}
\affiliation{Department of Astronomy, University of Virginia, Charlottesville, VA 22904, USA}
\email{cjl8rd@virginia.edu}  

\author[0000-0003-3840-7490]{Deryl E. Long}
\affiliation{Department of Astronomy, University of Virginia, Charlottesville, VA 22904, USA}
\email{del6h@virginia.edu}

\author[0000-0003-2076-8001]{L. Ilsedore Cleeves}
\affiliation{Department of Astronomy, University of Virginia, Charlottesville, VA 22904, USA}
\email{lic3f@virginia.edu}  

\author[0000-0002-7607-719X]{Feng Long}
\altaffiliation{NASA Hubble Fellowship Program Sagan Fellow}
\affiliation{Lunar and Planetary Laboratory, University of Arizona, Tucson, AZ 85721, USA}
\email{fenglong@arizona.edu}

\author[0000-0003-3283-6884]{Yuri Aikawa}
\affiliation{Department of Astronomy, The University of Tokyo, 7-3-1 Hongo, Bunkyo-ku, Tokyo 113-0033, Japan}
\email{aikawa@astron.s.u-tokyo.ac.jp}

\author[0000-0002-8716-0482]{Jennifer B. Bergner}
\affiliation{Department of Chemistry, University of California, Berkeley, University Avenue and, Oxford St, Berkeley, CA 94720, USA}
\email{jbergner@berkeley.edu}

\author[0000-0003-1837-3772]{Romane Le Gal}
\affiliation{Université Grenoble Alpes, CNRS, IPAG, F-38000 Grenoble, France}
\affiliation{Institut de Radioastronomie Millimetrique (IRAM), 300 rue de la piscine, F-38406 Saint-Martin d’Hères, France}
\email{romane.le-gal@univ-grenoble-alpes.fr}

\author[0000-0003-4784-3040]{Viviana V. Guzm\'{a}n}
\affiliation{Instituto de Astrof\'isica, Pontificia Universidad Cat\'olica de Chile, Av. Vicu\~na Mackenna 4860, 7820436 Macul, Santiago, Chile}
\affiliation{Millennium Nucleus on Young Exoplanets and their Moons (YEMS), Chile}
\email{viviana.guzman@uc.cl}

\author[0000-0001-8798-1347]{Karin I. \"Oberg}
\affiliation{Center for Astrophysics \textbar\, Harvard \& Smithsonian, 60 Garden St., Cambridge, MA 02138, USA}
\email{koberg@cfa.harvard.edu}

\author[0000-0003-1526-7587]{David J. Wilner}
\affiliation{Center for Astrophysics \textbar\, Harvard \& Smithsonian, 60 Garden St., Cambridge, MA 02138, USA}
\email{dwilner@cfa.harvard.edu}

\begin{abstract}
The physical and chemical conditions of protoplanetary disks shape the properties of nascent planetary systems. The line ratios and relative abundances of the HCN-HNC isomer pair are well-suited for tracing these gas conditions, since isomer chemistry is linked to the underlying temperature, elemental abundances, and irradiation environment of the emitting gas. While HCN emits bright lines regularly observed in disks, the fainter HNC lines are targeted significantly less often, precluding our ability to calibrate the HNC-to-HCN ratio as a tracer of disk properties. Here, we present new Submillimeter Array observations of five transition disks around the T~Tauri stars GM~Aur, J1604, LkCa~15, GG~Tau, and V4046~Sgr, covering the J=3--2 and J=4--3 lines of HCN and HNC. We detected at least one line of both HCN and HNC in each source and measured disk-integrated HNC-to-HCN flux and column density ratios of ${\approx}$0.1-0.7 and 0.1-0.4, respectively. For all sources, measured HNC fluxes exceed predictions from models of full~(non-transition)~disks by ${\approx}$3-10$\times$, while HCN fluxes appear typical. The relative brightness of HNC vs. HCN in transition disks strongly suggests a link between the presence of a cavity and efficient HNC production. We corroborate this trend using chemical models of the DM~Tau transition disk, showing HNC production and destruction are connected to the radiation environment. In our sample, disk-integrated HNC-to-HCN~column density ratio shows no trend with disk gas temperature but positively correlates with disk mass due to increased HNC abundance in the larger reservoirs of cooler gas in more massive disks.
\end{abstract}



\section{Introduction} 

The gas composition, structure, and physical conditions of protoplanetary disks are largely responsible for shaping the properties of the resultant planetary systems \citep[e.g.,][]{Oberg11, Mordasini16}. Molecular line emission at (sub)-millimeter wavelengths has been used as a powerful probe of these properties, including gas surface density, elemental ratios, ionization levels, thermal structures, and kinematics \citep[e.g.,][]{Cleeves15, Bergin16, Schwarz16, Aikawa21, Law21_thermal, Yoshida22, Teague25}, all of which influence disk evolution and ongoing planet formation. In particular, many disks show bright emission from abundant carbon- and nitrogen-rich species (e.g., CN, HCN, HC$_3$N, CH$_3$CN), whose chemistry is sensitive to the underlying stellar and disk properties \citep[e.g.,][]{Thi04, Chapillon12, Guzman15, Oberg15, Bergner19, Terwisga19, Ilee21, Calahan23, Paneque24}.

The hydrogen cyanide (HCN) molecule and its isomer, hydrogen isocyanide (HNC), are particularly well-suited for tracing gas conditions, since isomer chemistry is often tightly-linked to the temperature and elemental abundances of the emitting gas \citep{Loison14, Long21}. Both HCN and HNC are produced by the dissociative recombination of HCNH$^+$, which is generated via pathways starting from atomic N, N$_2$, and NH$_3$ \citep{Loison14, Visser18}. Given that this production route has nearly equal branching ratios and both species share similar excitation properties \citep[e.g.,][]{Semaniak01, Mendes12}, their line ratios and relative abundances are expected to be set by several temperature-sensitive HNC destruction reactions. Some of these reactions can directly convert HNC into HCN \citep{Graninger14, Loison14}. With increasing gas temperature, an increasing number of these destruction routes are efficient and thus, to first order, a higher HNC-to-HCN line ratio indicates that the emission is originating from colder gas. However, these routes also include a dependence on elemental abundances \citep{Loison14}, and both models and observations imply relationships with the local UV luminosity \citep{Bublitz19, Long21, Harada24} and cosmic-ray (CR) ionization rate \citep{Behrens22, Behrens24}. Overall, this means that the HNC-to-HCN ratio encodes a variety of valuable information about the gas from which it emits.

While the HNC-to-HCN line ratio has been observed in many astrophysical settings, including the interstellar medium, Solar System objects, star-forming regions, starless cores, proto-brown dwarfs, planetary nebula, and external galaxies \citep[e.g.,][]{Schilke92, Hebrard12, Graninger14, Jin15, Riaz18, Bublitz19, Hacar20, Behrens22, SantaMaria23, Harada24, Lee24, Tasa25, Stuber25}, it has been little explored in disks. HCN is among the most abundant molecules in disks and has bright lines that are regularly observed \citep[e.g.,][]{Oberg10, Guilloteau16, Bergner19, Bergner21, Guzman21}, but the fainter HNC lines are rarely targeted \citep[e.g.,][]{Dutrey97, Thi04, Graninger15}. To date, only five Class~II protoplanetary disks have HNC detections--namely, the DM~Tau \citep{Dutrey97}, HD~163296 \citep{Graninger15}, GG~Tau \citep{Phuong21}, TW~Hya \citep{Long21}, and J160830.7-382827 \citep[hereafter, `J1608';][]{Long21} disks. Thus, this limited sample means that we do not understand the typical HNC-to-HCN line and abundance ratios in disks, or how these ratios are influenced by stellar and disk properties. Moreover, the relatively few constraints on HNC chemistry inhibit further improvements to models of disk isomer chemistry.

\setlength{\tabcolsep}{1.5pt}
\begin{deluxetable*}{lccccccccccc}
\tablecaption{Stellar and Disk Properties \label{tab:source_prop}}
\tablewidth{0pt}
\tablehead{
\colhead{Source} & \colhead{Dist.} & \colhead{SpT} & \colhead{M$_*$\tablenotemark{a}} & \colhead{L$_*$} & \colhead{Age} & \colhead{v$_{\rm{sys}}$\tablenotemark{a}} & \colhead{incl.} & \colhead{PA} &  \colhead{r$_{\rm{mm, cavity}}$} & \colhead{r$_{\rm{mm, edge}}$} & \colhead{$\dot{M}$} \\ 
\colhead{} & \colhead{} & \colhead{(pc)}  & \colhead{(M$_{\odot}$)} & \colhead{(L$_{\odot}$)} & \colhead{(Myr)} & \colhead{(km~s$^{-1}$)} &  \colhead{($^{\circ}$)} & \colhead{($^{\circ}$)} & \colhead{(au)} &\colhead{(au)} & \colhead{(10$^{-8}$~M$_{\odot}$~yr~$^{-1}$)}
}
\startdata
GM~Aur & 159$^{[1]}$ & K6$^{[2]}$ & 1.1$^{[3]}$ & 1.2$^{[4]}$ & ${\sim}$2.5$^{[5]}$ & 5.6$^{[3]}$ & 53.2$^{[6]}$ & 57.2$^{[6]}$ & 40$^{[6]}$ & 170$^{[6]}$ & 0.3--2.2$^{[24,25]}$ \\
J1604\tablenotemark{b} & 145$^{[1]}$ & K2$^{[7]}$ & 1.2$^{[8]}$ & 0.6$^{[9]}$ & ${\sim}$5-10$^{[10]}$ & 4.6$^{[8]}$ & 6.0$^{[10]}$ & 258.8$^{[8]}$ & 83$^{[8]}$ & 265$^{[8]}$ & 0.003$^{[26]}$ \\
LkCa~15 & 157$^{[1]}$ & K5$^{[11]}$ & 1.2$^{[12]}$ & 1.1$^{[11]}$ & ${\sim}$5$^{[11]}$ & 6.3$^{[12]}$ & 50.2$^{[13]}$ & 61.9$^{[13]}$ & 76$^{[13]}$ & 153$^{[13]}$ & 0.063$^{[11]}$ \\
GG~Tau\tablenotemark{c} & 150$^{[1]}$ & M0+M2,~M3$^{[14]}$ & 1.2$^{[14]}$ & 0.67$^{[15]}$ & ${\sim}$3$^{[14]}$ & 6.5$^{[16]}$ & 35.0$^{[17]}$ & 7.0$^{[17]}$ & 224$^{[18]}$ & 260$^{[18]}$ & 6.4$^{[27]}$ \\
V4046~Sgr\tablenotemark{c} & 73$^{[1]}$ & K5+K7$^{[19]}$ & 1.8$^{[20]}$ & 0.86$^{[21]}$ & ${\sim}$12-23$^{[22]}$ & 2.9$^{[21]}$ & 34.7$^{[21]}$ & 75.7$^{[21]}$ & 31$^{[23]}$ & 100$^{[23]}$ & 0.16$^{[28]}$
\enddata
\tablecomments{References are: 1.~\citet{Gaia23}; 2.~\citet{Espaillat10}; 3.~\citet{Teague21_MAPS}; 4.~\citet{Macias18}; 5.~\citet{Kraus09}; 6.~\citet{Huang20_GMAur}; 7.~\citet{Preibisch05}; 8.~\citet{Stadler23}; 9.~\citet{Carpenter14}; 10.~\citet{Dong17}; 11.~\citet{Donati19}; 12.~\citet{Law_23_surfaces}; 13.~\citet{Facchini20}; 14.~\citet{Keppler20}; 15.~\citet{Hartigan03}; 16.~\citet{Guilloteau99}; 17.~\citet{Phuong21}; 18.~\citet{Phuong20}; 19.~\citet{Quast00}; 20.~\citet{Flaherty20}; 21.~\citet{Rosenfeld12}; 22.~\citet{Mamajek14}; 23.~\citet{Martinez22}; 24.~\citet{Espaillat19}; 25.~\citet{Wendeborn24}; 26.~\citet{Manara20}; 27.~\citet{Phuong20_Macc}; 28.~\citet{Donati11}. \\ }
\tablenotetext{a}{Dynamical stellar masses and systemic velocities (in the LSR frame) were derived via resolved observations of CO isotopologue lines.}
\tablenotetext{b}{We hereafter refer to 2MASS J16042165-2130284 (also known as RXJ1604.3-2130~A) as ``J1604".}
\tablenotetext{c}{For the multiple star systems GG~Tau and V4046~Sgr, the stellar mass and bolometric luminosity are the sum of all stars in each system.}
\end{deluxetable*}
\setlength{\tabcolsep}{4pt}

\begin{deluxetable*}{lccccccc}
\tablecaption{Observational Details\label{tab:calb_details}}
\tablewidth{0pt}
\tablehead{
\colhead{Source}  & \colhead{UT Date} & \colhead{Num.} & \colhead{SMA} & \colhead{$\tau$} & \multicolumn3c{Calibrators} \\ \cline{6-8}
\colhead{} & \colhead{} & \colhead{Ant.\tablenotemark{a}} & \colhead{config.} &\colhead{(225~GHz)} & \colhead{Flux} & \colhead{Passband} & \colhead{Gain}
}
\startdata
GM~Aur& 2023 Nov 05 & 6 & COM & 0.04 & Titan & 3C~84 & 0510+180, 3C~111 \\
       &   2023 Oct 22 & 7 & EXT & 0.07  & Callisto & 3C~84  & 0510+180, 3C~111 \\
J1604  & 2023 Jun 26 & 7 & COM & 0.05 & Ceres & 3C~454.3 & 1507-168, 1517-243 \\
LkCa~15 & 2023 Nov 06 & 6 & COM & 0.03 & Titan  & 3C84  & 0510+180, 3C~111 \\
       &   2023 Oct 20 & 7 & EXT & 0.12 & Titan & 3C~84 & 0510+180, 3C~111  \\
GG~Tau& 2023 Nov 12 & 7 & COM & 0.11 & Uranus & 3C~84  & 0510+180, 3C~120 \\
       &   2023 Oct 30 & 7  & EXT & 0.09  & Titan  &  3C~84 & 0510+180, 3C~120 \\
V4046~Sgr & 2023 Jul 14 & 7 & COM & 0.06 & Callisto & 3C~279 & 1700-261, NRAO~530  \\
\enddata
\tablenotetext{a}{Number of antennas remaining after flagging.}
\end{deluxetable*}

To expand the sample of HNC detections, transition disks featuring large central dust cavities are of particular interest. All but one of the existing detections are found in disks with inner cavities. The detection of HNC~J=1--0 in the transition disk around J1608 by \citet{Long21} is especially intriguing. Not only was the HNC line flux one order-of-magnitude higher than predicted in full disk models, but the spatial distribution of the HNC emission was centrally-peaked within the central dust cavity \citep{Long21} -- a notable mismatch from model predictions and observational inferences of a ring-like reservoir of HNC located in cold gas at larger radii in full disks \citep{Graninger15, Long21}. Although \citet{Long21} acknowledge that full disk models are likely not appropriate for the elevated UV fluxes within dust cavities, this nonetheless hints at a rich and potentially distinct HNC chemistry occurring in transition disks. Taken together, these findings point to transition disks as a powerful opportunity to expand the number of HNC detections and investigate the influence of disk structure on the resulting HNC-to-HCN ratios.

In this paper, we present new SMA observations of the J=3--2 and J=4--3 lines of HCN and HNC in a sample of five transition disks. Using these data, we determine rotational temperatures and column densities for the disks in our sample, which we use to assess the nature of HNC chemistry in disk environments. In Section \ref{sec:disk_sample}, we briefly describe our disk sample and summarize the observations used in this work in Section \ref{sec:observations_overview}. We present our results in Section \ref{sec:results}. In Section \ref{sec:discussion}, we explore how HNC chemistry varies across our sample and discuss the chemical origins of HNC in planet-forming disks. We summarize our conclusions in Section \ref{sec:conclusions}.

\section{Disk Sample} \label{sec:disk_sample}
Our sample includes five disks around the T~Tauri stars GM~Aur, J1604, LkCa~15, GG~Tau, and V4046~Sgr. All sources host large (100s of au), gas-rich transition (GM~Aur, J1604, LkCa~15) or circumbinary disks (GG~Tau, V4046~Sgr) with resolved, central dust-depleted cavities ranging from 31~au to as large as 224~au \citep[e.g.,][]{Phuong20, Martinez22, Huang20_GMAur, Facchini20, Law21_MAPSIII, Long22, Stadler23, Curone25}. Stellar hosts are M- or K-type with total system masses of 1.1-1.8~M$_{\odot}$ and bolometric luminosities of 0.6-1.2~L$_{\odot}$. V4046~Sgr and GG~Tau are known binary and multiple systems, respectively \citep{White99, Quast00}. Stellar accretion rates span more than three orders of magnitude from 0.003$\times 10^{-8}$ to 6.4$\times 10^{-8}$~M$_{\odot}$~yr~$^{-1}$, while system ages range from a few Myr to over 10~Myr. All disks are nearby with distances ranging from 73~pc to 159~pc. Table \ref{tab:source_prop} summarizes the stellar and disk properties of each system.

All disks in our sample have been the subject of extensive prior chemical observations, including as part of the Molecules with ALMA at Planet-forming Scales (MAPS) \citep[][]{Oberg21} and exoALMA \citep[][]{Teague25} ALMA Large Programs, and thus, have well-constrained physical and chemical structures. Additionally, each disk has existing detections of bright and spatially-resolved HCN molecular line emission \citep[e.g.,][]{Kastner18, Bergner19, Bergner21, Guzman21, Phuong21}. This sample of well-studied disks with known HCN-rich gas and comparable host star spectral types provides an ideal testbed to explore how HNC chemistry varies with disk structural properties, such as cavity size and disk mass, as well as stellar properties, including mass accretion rate.

\section{Observations}
\label{sec:observations_overview}

\subsection{Observational Details} \label{sec:observations_details}

We observed the GM~Aur, J1604, LkCa~15, GG~Tau, and V4046~Sgr disks between 26 June 2023 and 12 Nov 2023 with the Submillimeter Array (SMA)\footnote{The Submillimeter Array is a joint project between the Smithsonian Astrophysical Observatory and the Academia Sinica Institute of Astronomy and Astrophysics and is funded by the Smithsonian Institution and the Academia Sinica.} as part of project 2023A-S019 (PI: C. Law). All sources were observed in the compact (COM) configuration, while GM~Aur, GG~Tau, and LkCa~15 had additional data taken in the extended (EXT) configuration. The COM and EXT configurations had maximum baselines of ${\approx}$70~m and ${\approx}$220~m, respectively. Each observation used either 6 or 7 antennas with a $\tau_{225\rm{GHz}}$ between 0.03-0.12. Table~\ref{tab:calb_details} provides a summary of the observations.

Each observation used the RX240 and RX345 receivers and the upgraded SWARM correlator, which provided 48~GHz of bandwidth. The lower frequency receiver covered 244.3-256.6~GHz and 264.3-276.6~GHz, and the higher frequency receiver covered 333.8-346.1~GHz and 353.8-366.1~GHz. The tuning selection was motivated by the desire to simultaneously cover the J=3--2 and J=4--3 lines of both HCN and HNC (see Table \ref{tab:line_prop}).

We converted the raw data to CASA measurement sets using the \texttt{pyuvdata} \citep{Hazelton17} SMA reduction pipeline\footnote{\url{ https://github.com/Smithsonian/sma-data-reduction}} in CASA version \texttt{v6.3} \citep{McMullin_etal_2007, CASATeam20}. To reduce data volume, we channel-binned by a factor of four for initial calibration from the native SMA spectral resolution (140~kHz). Depending on the source and configuration, we used 3C~84, 3C~454.3, or 3C~279 for our passband calibrators, while Titan, Callisto, Ceres, or Uranus served as flux calibrators. Two of the following quasars (0510+180, 3C~111, 3C~120, 1507-168, 1517-243, 1700-261, NRAO~530) were used as gain calibrators for each source. Table~\ref{tab:calb_details} lists all calibrators used.

\begin{deluxetable}{lccccc}
\tablecaption{Molecular Line Properties\label{tab:line_prop}}
\tablewidth{0pt}
\tablehead{
\colhead{Species} & \colhead{Line}  & \colhead{Frequency } & \colhead{E$_{\rm{u}}$ (K)} & \colhead{A$_{\rm{ul}}$ (log$_{10}$ s$^{-1}$)} & \colhead{g$_{\rm{u}}$}
}
\startdata
HCN & J=3--2 & 265.886434 & 25.5 & $-$3.0766 & 21 \\
HCN & J=4--3 & 354.505478 & 42.5 & $-$2.6860 & 27 \\
HNC & J=3--2 & 271.981142 & 26.1 & $-$3.0298 & 7 \\
HNC & J=4--3 & 362.630303 & 43.5 & $-$2.6392 & 9 \\
\enddata
\tablecomments{The spectroscopic constants for all lines are taken from the CDMS database \citep{Muller01, Muller05, Endres16} with measurements for HCN and HNC by \citet{Ahrens02} and \citet{Saykally76, Creswell76, Okabayashi93, Thorwirth00}, respectively.}
\end{deluxetable}

\subsection{Imaging} \label{sec:imaging}

We first subtracted the continuum by manually identifying line-free channels using the \texttt{uvcontsub} task in CASA with a first-order polynomial. We then imaged all lines using the \texttt{tclean} task with Briggs weighting (\texttt{robust} = 2.0) and Keplerian masks generated with the \texttt{keplerian\_mask} \citep{rich_teague_2020_4321137} code. Keplerian mask parameters for each disk were based on the known stellar properties and disk geometry from the literature, including stellar masses, systematic velocities, inclinations, and position angles (Table \ref{tab:source_prop}). Images were binned in velocity to 1~km~s$^{-1}$, and in a few cases 2~km~s$^{-1}$, to maximize the signal-to-noise ratio (SNR). All images were \texttt{CLEAN}ed down to a 3$\sigma$ level, where $\sigma$ was the RMS noise, measured across five line-free channels of the dirty image. Table~\ref{tab:image_cube_prop} summarizes all image properties. The typical beam size was ${\approx}$1$^{\prime \prime}$-3$^{\prime \prime}$, while the RMS noise ranged from ${\approx}$40-360~mJy~beam$^{-1}$ depending on the source, line, and channel spacing.

\subsection{Line Detection Criteria} \label{sec:detection_stat}

We consider an individual line to be detected if it: (1) has a disk-integrated flux density that exceeds $3\sigma$ \underline{and} (2) shows a peak intensity of 3$\times$RMS or higher in at least two consecutive channels. Lines that met only one of these criteria were classified as tentative detections, and all others were classified as non-detections. 

Here, we restricted our focus to the disk-integrated fluxes, as we do not spatially-resolve either the HCN or HNC line emission in our observations of J1604 or V4046~Sgr, which were taken in the COM configuration. For GM~Aur, GG~Tau, and LkCa~15, where we had both COM and EXT observations, we marginally resolved the bright HCN lines but lack sufficient SNR in the HNC lines to robustly analyze the emission morphology.

\begin{deluxetable*}{lcccccc}
\tablecaption{Image Cube Properties and Integrated Flux Measurements \label{tab:image_cube_prop}}
\tablehead{
\colhead{Source} & \colhead{Line} & \colhead{Beam} & Mask Radius & \colhead{$\delta$v}  & \colhead{Channel RMS} & \colhead{Int. Flux}   \\
\colhead{} & \colhead{} & \colhead{($^{\prime \prime} \times ^{\prime \prime}$, $\deg$)} & \colhead{($\prime \prime$)}  & \colhead{(km~s$^{-1}$)} & \colhead{(Jy~beam$^{-1}$)} & \colhead{(Jy~km~s$^{-1}$)} 
 }
\startdata
GM~Aur & HCN J=3--2 & 2.25 $\times$ 1.30, 88.4 & 4.0 & 1.0 & 0.043 & 1.688 $\pm$ 0.244\\
       & HCN J=4--3 & 1.81 $\times$ 1.04, 45.6 & 4.0 & 1.0 & 0.087 & 2.325 $\pm$ 0.421 \\
       & HNC J=3--2 & 2.21 $\times$ 1.27, 46.0 & 4.0 & 2.0 & 0.046 & 1.203 $\pm$ 0.375 \\
       & HNC J=4--3 & 1.71 $\times$ 0.98, 47.6 & 4.0 & 2.0 & 0.135 &  $<$ 1.162\\
 \\
J1604 & HCN J=3--2 & 3.45 $\times$ 2.38,  7.9 & 4.0 & 1.0 & 0.113 & 6.576 $\pm$ 0.696 \\
           & HCN J=4--3 & 2.69 $\times$ 1.85, 11.7 & 4.0 & 1.0 & 0.268 & 10.619 $\pm$ 1.365 \\
           & HNC J=3--2 & 3.37 $\times$ 2.33,  8.0 & 4.0 & 1.0 & 0.131 & 2.612 $\pm$ 0.494 \\
           & HNC J=4--3 & 2.68 $\times$ 1.78, 13.3 & 4.0 & 2.0 & 0.360 & \textit{2.242 $\pm$ 1.950} \\
 \\
LkCa~15 & HCN J=3--2 & 2.29 $\times$ 1.31, 47.7 & 4.0 & 1.0 & 0.048 & 3.934 $\pm$ 0.427 \\
        & HCN J=4--3 & 1.87 $\times$ 1.09, 47.4 & 4.0 & 2.0 & 0.093 & 4.096 $\pm$ 0.734 \\
        & HNC J=3--2 & 2.23 $\times$ 1.28, 48.1 & 4.0 & 1.0 & 0.054 & 1.452 $\pm$ 0.343 \\
        & HNC J=4--3 & 1.75 $\times$ 0.99, 48.4 & 4.0 & 2.0 & 0.124 & $<$ 1.365 \\
 \\
GG~Tau & HCN J=3--2 & 1.69 $\times$ 1.25, 4.6 & 5.0 & 1.0 & 0.050 & 3.077 $\pm$ 0.426 \\
       & HCN J=4--3 & 1.21 $\times$ 0.94, 64.9 & 5.0 & 1.0 & 0.120 &  $<$ 1.139 \\
       & HNC J=3--2 & 1.66 $\times$ 1.19, 74.9 & 5.0 & 2.0 & 0.058 &  \textit{0.658 $\pm$ 0.615} \\
       & HNC J=4--3 & 1.12 $\times$ 0.79, 67.6 & 5.0 & 2.0 & 0.185 &  $<$ 4.008 \\
 \\
V4046~Sgr & HCN J=3--2 & 3.95 $\times$ 2.34, 9.2 & 4.0 & 1.0 & 0.098 & 13.926 $\pm$ 1.453 \\
          & HCN J=4--3 & 2.95 $\times$ 1.86, 6.5 & 4.0 & 1.0 & 0.254 & 19.532 $\pm$ 2.383 \\
          & HNC J=3--2 & 3.86 $\times$ 2.31, 9.5 & 4.0 & 1.0 & 0.112 & 2.096 $\pm$ 0.680 \\
          %& HNC J=4--3 & 2.91 $\times$ 1.81, 4.7 & 4.0 & 2.0 & 0.297 & \textbf{4.932} $\pm$ 3.067 \\
          & HNC J=4--3 & 2.91 $\times$ 1.81, 4.7 & 4.0 & 1.0 & 0.451 & 4.932 $\pm$ 1.056 \\
\enddata
\tablecomments{Fluxes in italics indicate tentative detections while we report 3$\sigma$ upper limits for non-detections. Flux errors include an additional 10\% systemic flux calibration uncertainty.}
\end{deluxetable*}

\section{Results} \label{sec:results}

\begin{figure*}
    \includegraphics[width=1.0\linewidth]{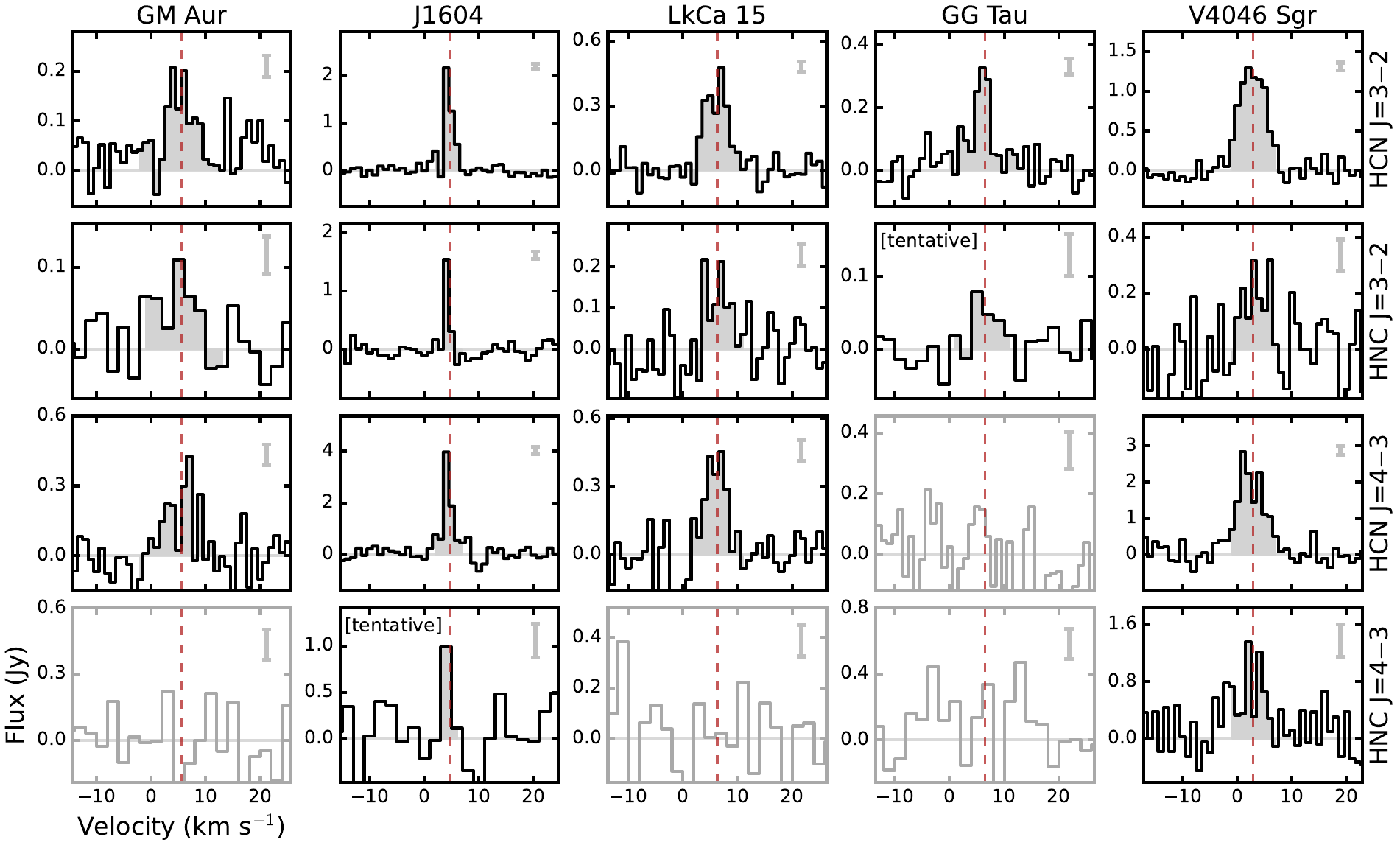}
    \caption{Gallery of spectra of the HCN and HNC J=3--2 and J=4--3 lines (\textit{rows}) for each disk (\textit{columns}) in our sample. All spectra were extracted in a circular aperture centered on the disk position. Shaded regions show the velocity ranges over which integrated fluxes were extracted. The dashed red lines show the systemic velocity of each disk and vertical errorbars show the 1$\sigma$ RMS. Non-detections are indicated as grayed-out boxes.}
    \label{fig:spectra_gallery}
\end{figure*}

\begin{figure*}
    \centering
    \includegraphics[width=\linewidth]{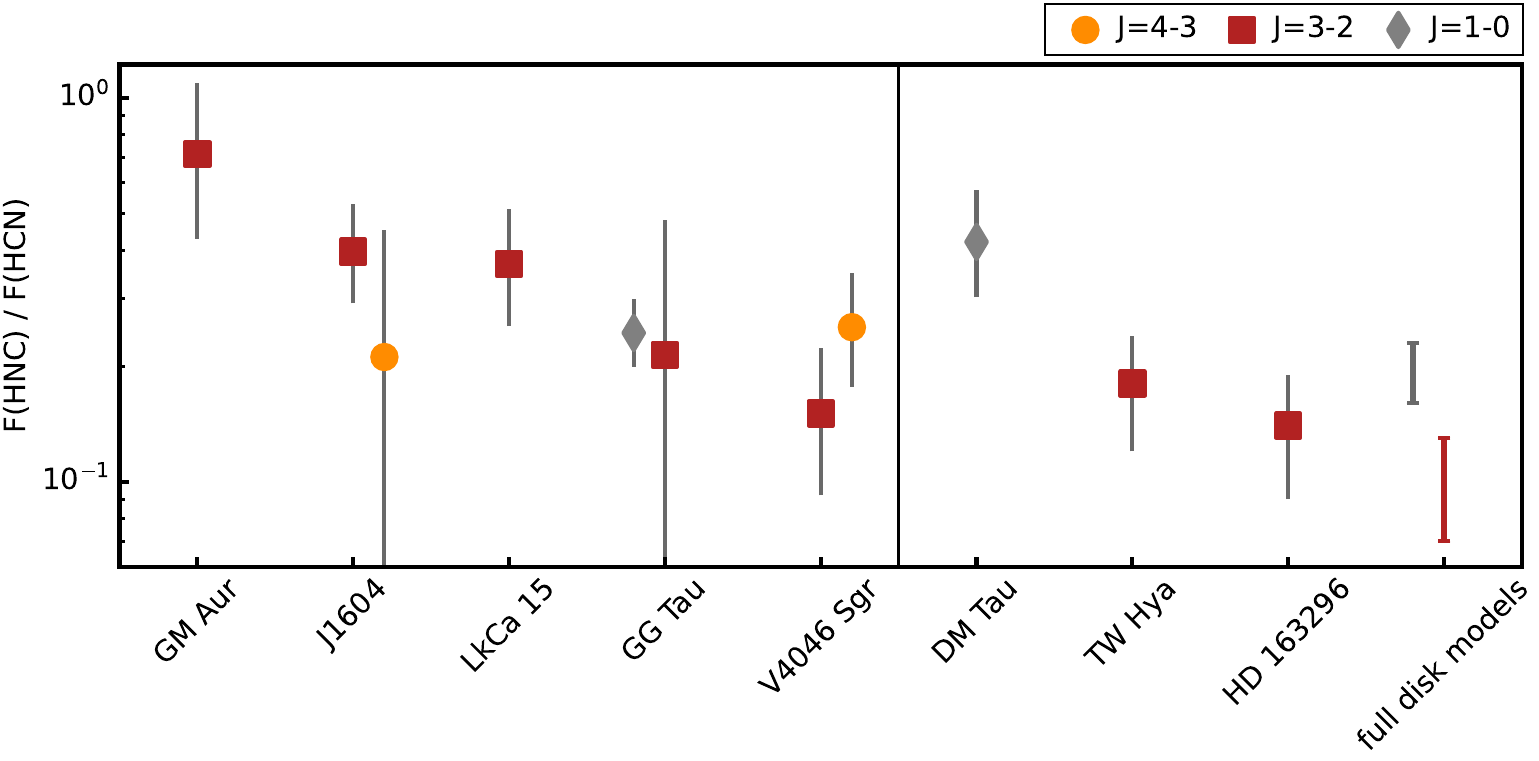}
    \caption{Disk-integrated HNC-to-HCN flux ratios in our SMA sample for J=4--3 \textit{(orange circles)}, J=3--2 \textit{(red squares)}, and J=1--0 \textit{(grey triangles)}. Literature measurements of the J=1--0 line in GG~Tau are from \citet{Phuong21}; J=3--2 in TW~Hya and HD~163296 are from \citet{Graninger15}; and J=1--0 in DM~Tau are from \citet{Dutrey97}. The DM~Tau values are from IRAM~30m-single dish observations and are not sensitive to the inner disk within ${\approx}$100~au from the star. Model predictions for full disks around T~Tauri systems with comparable disk masses to our sample are shown as solid lines \citep{Long21}.}
    \label{fig:flux_ratios}
\end{figure*}

\subsection{Line Detections and Integrated Fluxes} \label{sec:integrated_fluxes}

We measured the disk-integrated J=3--2 and J=4--3 fluxes of HCN and HNC using spectra extracted with the \texttt{gofish} \citep{Teague19JOSS} code. We adopt a conservative approach to spectral extraction. Each spectrum in Figure \ref{fig:spectra_gallery} is extracted using a circular mask, covering the full linewidth in velocity and with radii based on the known gas disk sizes \citep[e.g.,][]{Bergner19, Guzman21, Phuong21}. Given the angular resolution of our observations, we adopted intentionally inflated- mask sizes to confidently capture all flux. We adopted consistent mask sizes for each disk, which are listed in Table \ref{tab:image_cube_prop}. All HCN fluxes included potential contributions from hyperfine lines \citep[e.g.,][]{Bergner19, Cataldi21}, but at the sensitivity of our SMA observations, these were not clearly resolved. We estimated uncertainties with a bootstrapping approach, where we randomly-generated 500 circular masks centered on the disk position, but spanning only line-free channels, following \citet{Bergner19}. We took the uncertainty as: the standard deviation of the integrated fluxes within these bootstrapped masks, added in quadrature to 10\% of the measured flux (to account for additional systemic flux calibration uncertainties). Table \ref{tab:image_cube_prop} lists the measured fluxes and uncertainties.

\begin{figure*}
    \centering
    \includegraphics[width=0.9\linewidth]{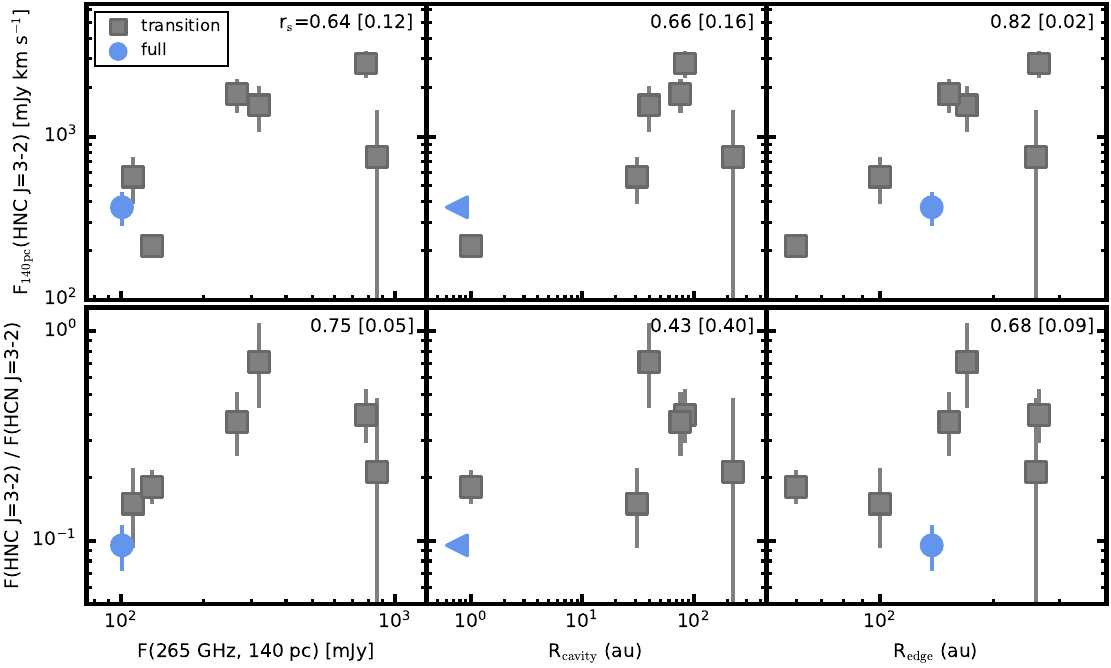}
    \caption{Disk-integrated HNC J=3--2 flux normalized to a distance of 140~pc (\textit{top row}) and HNC-to-HCN J=3--2 line ratios (\textit{bottom row}) versus a variety of disk properties (\textit{columns}). Continuum fluxes are normalized to 140~pc and scaled to 265~GHz via $\nu^{2.2}$ \citep{Andrews20}. Properties of the TW~Hya and HD~163296 disks are taken from \citet{Graninger15, Kastner15, Andrews16, Herczeg23} and \citet{Fairlamb15, Andrews18, Wichittanakom20, Oberg21}, respectively. Spearman correlation coefficients and associated p-values (computed excluding upper limits) are shown in the upper right corner of each panel.}
    \label{fig:flux_plots}
\end{figure*}

Figure \ref{fig:spectra_gallery} shows a gallery of the HCN and HNC spectra extracted from our SMA observations. Based on these spectra and the measured fluxes, we detected HCN J=3--2, HCN J=4--3, and HNC J=3--2 in all disks in our sample except for the GG~Tau disk, where HNC J=3--2 was tentatively detected and HCN J=4--3 was not detected. We only firmly detected HNC J=4--3 in the V4046~Sgr disk -- the first detection of this line in a planet-forming disk -- along with an additional tentative detection in the J1604 disk. 

For all disks with HCN detections, there are existing measurements of the J=3--2 and J=4--3 fluxes, except for J1604 and GM~Aur, where we detect J=4--3 for the first time. Our measurements are consistent with literature values in all sources \citep{Dutrey97, Thi04, Kastner14, Kastner18, Bergner19, Guzman21}, with the exception of J1604, where our inferred HCN J=3--2 flux is ${\approx}$30\% lower than reported in \citet{Bergner19}. 

Our high detection rates of HNC lines suggest an abundant reservoir of HNC gas in our transition disk sample. Both HNC J=3--2 and J=4--3, when detected, show consistently bright emission (${\gtrsim}$1~Jy~km~s$^{-1}$) and in a few sources, have fluxes within a factor of a few of the primary HCN molecule. In particular, the V4046~Sgr disk has a bright HNC J=4--3 integrated line flux (${\approx}$5~Jy~km~s$^{-1}$) that is more than twice that of the J=3--2 line.

\subsection{HNC Fluxes and HNC-to-HCN Flux Ratios} \label{sec:line_flux_ratios}

Here, we first searched for trends in the measured disk-integrated line fluxes and HNC-to-HCN flux ratios. 

\begin{figure*}
    \centering
    \includegraphics[width=\linewidth]{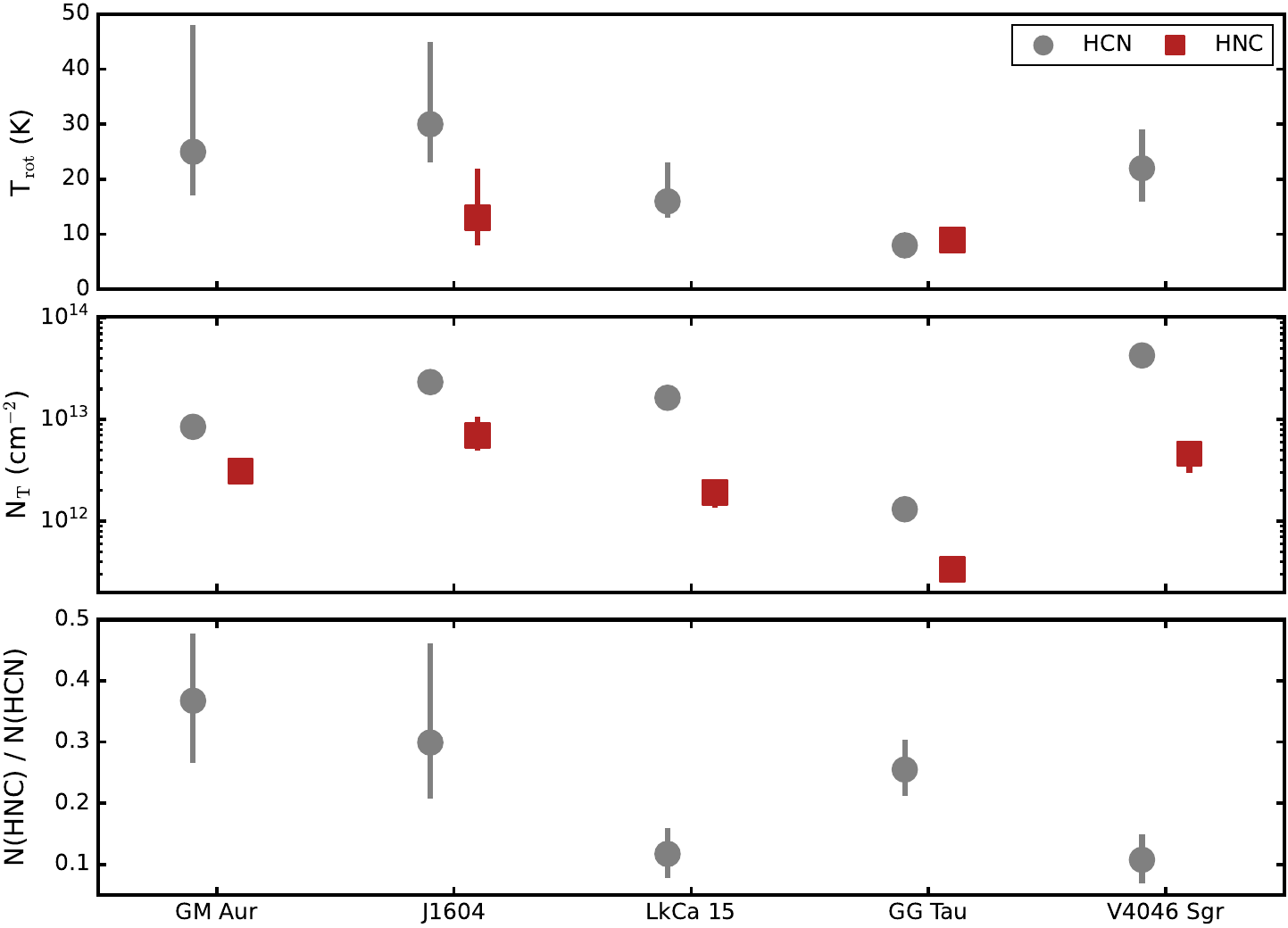}
    \caption{Rotational temperatures (\textit{top}) and column densities (\textit{middle}) for HCN (\textit{gray}) and HNC (\textit{red}), when available, and the HNC-to-HCN column density ratios (\textit{bottom}) for our sample. In a few cases, the uncertainties are smaller than the markers.}
    \label{fig:hnc_hcn_ncol_by_disk}
\end{figure*}

Figure \ref{fig:flux_ratios} shows the measured HNC-to-HCN line flux ratios. To augment our sample, we also included literature measurements for the TW~Hya and HD~163296 disks \citep{Graninger15} as well as the HNC-to-HCN J=1--0 line ratios in GG~Tau and DM~Tau from \citet{Phuong21} and \citet{Dutrey97}, respectively. In our sample, the measured flux ratio ranges from ${\approx}$0.1 to 0.7 with a median HNC-to-HCN J=3--2 line ratio  of 0.37 in our transition disk sample. GG~Tau and V4046~Sgr have low flux ratios similar to existing measurements in TW~Hya and HD~163296, as well as predicted flux ratios from the full disk models of \citet{Long21}, while GM~Aur, J1604, LkCa~15, and DM~Tau show higher ratios by up to a factor of a few. For those systems (J1604, GG~Tau, V4046~Sgr) with measurements of multiple lines, the ratios are consistent within uncertainties.

In Figure \ref{fig:flux_plots}, we consider a variety of disk properties, including mm dust flux, disk size and cavity size, all of which may influence the observed HNC chemistry. Here, we focus on the J=3--2 line as it is detected or tentatively-detected in both HCN and HNC in our entire sample. For all properties, we also computed the Spearman’s correlation coefficient using the \texttt{spearmanr} function in \texttt{scipy.stats} \citep{Virtanen20} to determine the significance of potential correlations. We use a Spearman correlation, which measures the strength of monotonic relationships, as we do not necessarily expect correlations in our sample to be linear.

For the HNC fluxes, we identify positive trends with mm dust flux, disk size, and cavity size, suggesting that more massive and extended disks have more abundant HNC, which is in agreement with model predictions \citep[e.g.,][]{Aikawa99, Long21}. The only significant correlation (p=0.02) observed is between disk size and HNC flux, while the cavity size and mm flux trends are only tentative. This is consistent with increased HNC production due to large cold gas reservoirs in larger disks. The same general trends are also reflected in the line flux ratios, with both cavity and disk size trends being positive, but tentative. However, we do find a modestly stronger correlation between mm dust flux and HNC-to-HCN flux ratio (p=0.05), compared to mm dust flux and absolute HNC fluxes alone (p=0.12), which may be explained by dust more effectively shielding HNC (versus HCN) from destruction via UV photons. We discuss the observed trends as they relate to the chemical origins of HNC in Section \ref{sec:discussion}. 

\subsection{Rotational Diagram Analysis} \label{sec:rot_diags}

\subsubsection{Fitting Rotational Diagrams} 

To infer the rotational temperatures and column densities of HCN and HNC, we use a rotational diagram approach \citep{goldsmith_population_1999}. Specifically, we closely followed the approach outlined in \citet{loomis_distribution_2018}. We assumed all lines are in local thermal equilibrium (LTE), as both HCN and HNC are expected to originate in the warm molecular layers of the disk \citep[e.g.,][]{Aikawa99, Agundez18}. For a molecule transitioning from rotational state $u$ to $l$, we can relate the line emission surface brightness ($I_{\nu}$) multiplied by the line width ($\Delta v$) to the column density of molecules in their respective upper energy state $N_{u}$:

\begin{equation}
    I_{\nu} \Delta v = \frac{A_{ul}N^{thin}_{u}hc}{4\pi},
\end{equation}

\noindent where $A_{ul}$ is the Einstein coefficient, $h$ is the Planck constant, and $c$ is the speed of light. The disk-averaged emission intensity is given by $I_{\nu} = S_{\nu}/\Omega$, or flux density divided by solid angle $\Omega$. Here, we use the maximum radius of the known HCN emission to define $\Omega$, as listed in Table \ref{tab:rotate_data}. For all analysis, we assume that HCN and HNC are co-spatial.

To calculate the rotational temperature T$_{\rm{rot}}$ and total column density N$^{\rm{thin}}_{\rm{T}}$ for each species, we use the Boltzmann equation:

\begin{equation}
    N_{T} = \frac{N^{thin}_{T} Q(T_{rot})}{g_u} e^{E_u / k_B T_{rot}},
\end{equation}

\noindent where $k_B$ is the Boltzmann constant, $E_u$ is the upper state energy, $g_u$ is the upper state degeneracy, and $Q$ is the partition function. All spectroscopic line data were obtained from the CDMS catalog, as listed in Table \ref{tab:line_prop}. Partition functions were linearly interpolated from catalog values.

As we do not know the optical depth of the observed HCN and HNC transitions, we must include the relevant correction factors to the inferred column densities, i.e., N$_u$ = C$_{\tau}$ N$_u^{\rm{thin}}$, where C$_{\tau}$ = $\tau$/(1 $-$ $e^{-\tau}$). This allows us to relate the optical depths of individual lines to the upper state level populations:

\begin{equation}
    \tau_{ul} = \frac{A_{ul} c^3 N_u}{8\pi \nu^3}(e^{(h\nu / k_B T_{rot})} - 1).
\end{equation}

We can then apply the Markov Chain Monte Carlo code \texttt{emcee} \citep{Foreman13} to calculate posterior probability distributions for $T_{\rm{rot}}$ and $N_u$, which we can then use to estimate $\tau$. We adopt the following uniform priors: $10^{10}$~cm$^{-2} < \rm{N}_{\rm{T}} < 10^{20}$~cm$^{-2}$ and 5~K $ < \rm{T}_{\rm{rot}} < 100$~K. All fits used thermal line widths. We used 256 walkers with 2000 burn-in steps and an additional 1000 steps to sample the posterior distribution. The best-fit values and uncertainties are taken as the median, and 16th-84th percentiles, respectively, of the posterior samples.

\subsubsection{Rotational Temperatures and Column Densities} 

Table \ref{tab:rotate_data} lists the derived rotational temperatures and column densities for all disks and species, along with the inferred line optical depths. The detection of two HCN lines in all sources allowed for the determination of T$_{\rm{rot}}$ and N$_{\rm{T}}$. In the case of GG Tau, we adopted the fluxes of HCN J=1--0 and HNC J=1--0 from \citet{Phuong21}. HCN typically showed optically thick emission ($\tau\approx~$2-9), with inferred $\tau$-corrected N$_{\rm{T}}$ ranging from $10^{12}$~cm$^{-2}$ to a few times $10^{13}$~cm$^{-2}$ and T$_{\rm{rot}}$ values from 8~K to 30~K. We list both the inferred N$_{\rm{T}}^{\rm{thin}}$ and optical-depth corrected column densities in Table \ref{tab:rotate_data}. Overall, N$_{\rm{T}}$, T$_{\rm{rot}}$, and line optical depths are consistent with prior inferences in the same sources based on previous analysis using the HCN hyperfine structure and isotopologue line emission \citep[e.g.,][]{Bergner19, Bergner21, Cataldi21, Phuong21, Guzman21}.

\begin{deluxetable*}{lcccccccccc}
\tablecaption{Disk-Averaged HCN and HNC Rotational Temperatures and Column Densities \label{tab:rotate_data}}
\tablewidth{0pt}
\tablehead{
\colhead{Source} & \colhead{Molecule}  & \colhead{R$_{\rm{max}}$} & \colhead{T$_{\rm{rot}}$} &\colhead{N$_{\rm{T}}^{\rm{thin}}$} & \colhead{N$_{\rm{T}}$} & \multicolumn3c{$\tau$} \\ \cline{7-9} 
\colhead{} & \colhead{} & \colhead{(au)}  & \colhead{(K)} & \colhead{(cm$^{-2}$)} & \colhead{(cm$^{-2}$)}  & \colhead{J=1--0} & \colhead{J=3--2} &  \colhead{J=4--3} 
}
\startdata
GM~Aur & HCN & 300 & 25$^{+23}_{-8}$ & 1.25$^{+0.30}_{-0.19}\times$10$^{12}$ & 8.49$^{+0.39}_{-0.37}\times$10$^{12}$ & \ldots & 6.0 & 5.1 \\
& HNC & 300 & [25] & 3.12$^{+0.76}_{-0.76}\times$10$^{12}$ & \ldots & \ldots & \ldots & \ldots  \\
J1604 & HCN & 230 & 30$^{+15}_{-7}$ & 4.21$^{+0.49}_{-0.40}\times$10$^{12}$ & 2.34$^{+0.04}_{-0.04}\times$10$^{13}$ & \ldots& 5.6 & 6.0 \\
& HNC & 230 & 13$^{+9}_{-5}$ & 6.19$^{+6.14}_{-2.41}\times$10$^{12}$ & 7.00$^{+3.60}_{-2.06}\times$10$^{12}$ & \ldots& 0.4 & 0.6 \\
LkCa~15 & HCN & 350 & 16$^{+7}_{-3}$ & 2.18$^{+0.87}_{-0.61}\times$10$^{12}$ & 1.64$^{+0.11}_{-0.10}\times$10$^{13}$ & \ldots& 5.6 & 4.0  \\
& HNC & 350 & [16] & 1.92$^{+0.54}_{-0.55}\times$10$^{12}$ & \ldots & \ldots & \ldots & \ldots \\
GG~Tau & HCN & 800 & 8$^{+1}_{-1}$ & 6.15$^{+0.54}_{-0.56}\times$10$^{11}$ & 1.31$^{+0.05}_{-0.06}\times$10$^{12}$ & 1.7 & 2.0 & \ldots \\
& HNC & 800 & 9$^{+1}_{-1}$ & 2.84$^{+0.45}_{-0.45}\times$10$^{11}$ & 3.34$^{+0.45}_{-0.46}\times$10$^{11}$ & 0.3 & 0.3 & \ldots \\
V4046~Sgr & HCN & 180 & 22$^{+7}_{-4}$ & 4.65$^{+0.73}_{-0.54}\times$10$^{12}$ & 4.28$^{+0.08}_{-0.08}\times$10$^{13}$ & \ldots & 8.6 & 8.0 \\
& HNC & 180 & [22] & 4.60$^{+1.64}_{-1.62}\times$10$^{12}$\tablenotemark{a} & \ldots & \ldots & \ldots & \ldots  \\
\enddata
\tablecomments{The HNC column densities in GM~Aur and LkCa~15 were calculated by adopting their respective HCN rotational temperatures (brackets). The HCN maximum radius is obtained from spatially-resolved observations \citep{Bergner19, Guzman21, Phuong21}. HCN and HNC are assumed to be co-spatial. The HCN and HNC J=1--0 fluxes were taken from \citet{Phuong21}.}
\tablenotetext{a}{The HNC column density in the V4046~Sgr disk is the mean of the independently-derived HNC J=3--2 and J=4--3 column densities (see Section~\ref{sec:HNC_in_V4046Sgr}).}
\end{deluxetable*}

For HNC, on the other hand, multi-line detections only permitted T$_{\rm{rot}}$ determinations for the J1604 and GG~Tau disks. The derived HNC T$_{\rm{rot}}$ values are similar to those of HCN for GG~Tau (9~K), while the T$_{\rm{rot}}$ of HNC in J1604 (13~K) is less than half that of HCN. This indicates a potential disk-to-disk diversity in the location of the HNC reservoir. For GM~Aur and LkCa~15, we instead adopted the T$_{\rm{rot}}$ of HCN to derive the HNC column density. We also followed the same procedure for V4046~Sgr, despite having two HNC lines detected (see Section \ref{sec:HNC_in_V4046Sgr} for more details). In general, the HNC N$_{\rm{T}}$ values are on the order of 10$^{12}$~cm$^{-2}$, with the exception of GG~Tau, which like HCN, shows a reduction in N$_{\rm{T}}$ by a factor of 10 compared the rest of the sample. For those disks with multi-line HNC detections, the inferred optical depths are modest (${\approx}$0.3-0.6), which is consistent with the lower inferred column densities relative to HCN. For LkCa~15, GM~Aur, and V4046~Sgr, where we could not infer $\tau$, we adopted the N$_{\rm{T}}^{\rm{thin}}$ value in our subsequent analysis. This does not significantly influence our conclusions, as the typical $\tau$-correction to HNC in the other sources only results in a ${\approx}$10\% boost in N$_{\rm{T}}$, which is well within our uncertainties.

\subsubsection{HNC Excitation in V4046~Sgr} \label{sec:HNC_in_V4046Sgr}

As noted in Section~\ref{sec:integrated_fluxes}, the HNC J=4--3 flux in the V4046~Sgr disk is more than twice that of the J=3--2 transition. As a result, a rotational diagram fit yields a column density of ${\approx}4\times10^{12}$~cm$^{-2}$ and a relatively high rotational temperature of ${\approx}70\pm20$~K, with both transitions being optically-thin. This elevated T$_{\rm{rot}}$ could indicate unusually warm HNC-emitting gas, i.e., distinct emitting vertical layers, or non-LTE conditions. However, with only two lines available, we cannot robustly distinguish between these possibilities. Moreover, the inferred HNC rotational temperature is substantially higher than that derived for HCN and appears inconsistent with the well-characterized temperature structure of the V4046~Sgr disk \citep[e.g.,][]{Rosenfeld12, Law22, Galloway25}. We therefore adopt the HCN rotational temperature, which is well-constrained, and fit the HNC J=3--2 and J=4--3 transitions independently. The resulting column densities for each line are consistent to each other, and throughout this work we adopt their mean value of $4.60^{+1.64}_{-1.62}\times10^{12}$~cm$^{-2}$.

\section{Discussion} \label{sec:discussion}

\subsection{Comparison to Full Disk Models and Observations} \label{sec:vs_full_disk_models}

To gain insight into HNC chemistry in disks, we first compare our observations to the existing model predictions of \citet{Long21}. Specifically, \citet{Long21} used the 2D thermochemical code DALI \citep{Bruderer12, Bruderer13} and an updated nitrogen chemistry network \citep{Visser18} to explore how the HNC and HCN line emission and abundances depend on stellar and disk parameters. These models are constructed for full disks, i.e., no central cavities. We begin by considering full-disk models, not because they represent the physical structure of our transition disk sample, but as an initial diagnostic test. If full disk models are able to explain our observations, we can infer that the presence or absence of a dust cavity (and consequent alternations in disk properties) is likely not a dominant factor in setting HNC chemistry.

\subsubsection{Full Disk Model Predictions}

Based on the models of \citet{Long21}, HNC and HCN are expected to be abundant in the warm surface layers and, at large radii, in the cold midplane, leading to a ring-shaped emission morphology. The HNC-to-HCN line ratio is predicted to increase from ${<}$0.1 in the inner (${\lesssim}$100~au) disk, where HNC is preferentially destroyed in warm gas, to ${\approx}$0.8 at larger radii. Higher J lines of HNC are brighter in models with increased flaring angles and higher levels of UV radiation, while the abundance and spatial distribution of both HNC and HCN are sensitive to the degree of carbon and oxygen depletion.
 
\subsubsection{Unusually Bright HNC in Transition Disks} \label{sec:bright_HNC_in_trans_disk}

\begin{figure*}
    \centering
    \includegraphics[width=\linewidth]{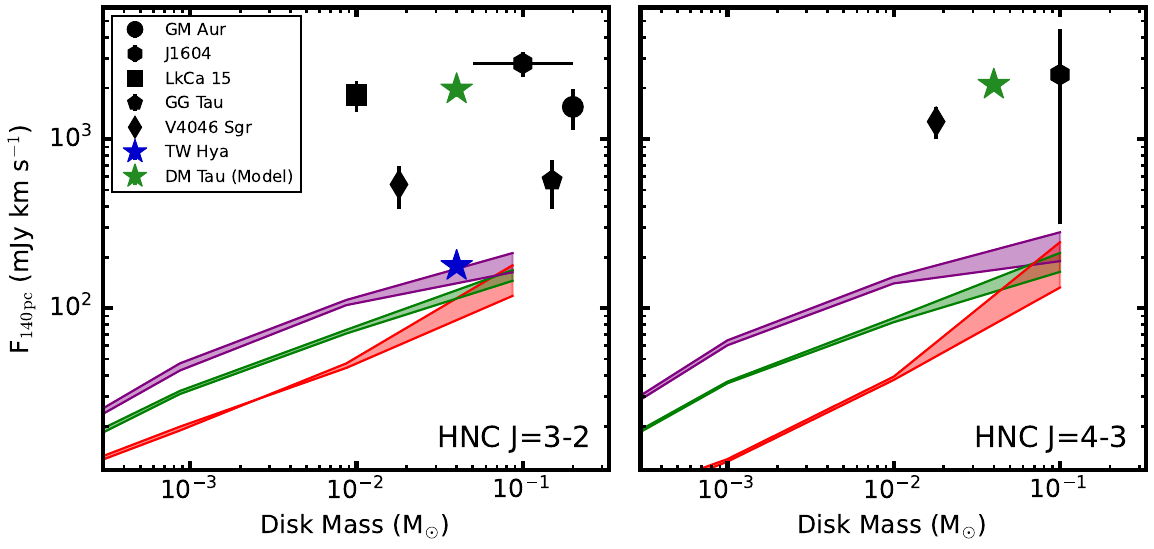}
    \caption{Disk-integrated HNC J=3--2 (\textit{left}) and J=4--3 (\textit{right}) fluxes measured in our sample (black points) and full disk models of \citet{Long21} (shaded lines) versus disk mass. Line fluxes for the TW~Hya disk are taken from \citet{Graninger15}. The DM~Tau fluxes are extracted from the model of \citet{Long24}. All line fluxes are normalized to a distance of 140~pc. Each color shows a different power-law index for scale height, and the upper and lower lines indicate several disk flaring indices. Disk mass estimates are from GM~Aur \citep{Schwarz21}, J1604 \citep{Yoshida25}, LkCa~15 \citep{Sturm23}, GG~Tau \citep{Guilloteau99, Andrews14}, V4046~Sgr \citep{Ruiz19}, TW~Hya \citep{Bergin13}, and DM~Tau \citep{Andrews11}.}
    \label{fig:long_model_comparison}
\end{figure*}

Figure \ref{fig:long_model_comparison} shows the disk-integrated HNC J=3--2 and J=4--3 fluxes in our transition disk sample and the full disk model predictions from \citet{Long21}, as a function of disk mass. The observationally-measured HNC fluxes exceed model predictions in all cases by at least a factor of a few and up to more than an order of magnitude. Although a few of the disk masses in our sample (GM~Aur, GG~Tau, J1604) are greater than the highest mass (0.1~M$_{\odot}$) considered in \citet{Long21}, reasonable extrapolations of the existing full-disk models cannot reproduce the high HNC fluxes.

We also included the HNC J=3--2 flux of the TW~Hya disk from \citet{Graninger15} in this comparison. TW~Hya has an HNC flux well-reproduced by full disk models, even though TW~Hya hosts a transition disk with a compact mm dust cavity with a radius of ${\approx}$1~au \citep[e.g.,][]{Calvet02, Andrews16}. Despite this, its dust distribution is distinct from the disks in our sample, which have central cavities of 10s to 100s of au. Thus, it is unsurprising that the outer disk chemistry of TW~Hya, as traced by our HNC observations, appears similar to that of full disks. 

An earlier indication that transition disks exhibit bright HNC emission was noted by \citet{Long21}, who detected HNC J=1--0 in the J1608 transition disk. This disk hosts a large dust cavity with a radius of 75~au \citep{vanderMarel18} and has an HNC flux that is one order of magnitude greater than full disk models. Moreover, HNC emission in J1608 was marginally spatially-resolved and showed a centrally-peaked profile within the inner dust cavity, which contrasts with full disk model predictions of an outer, ringed-distribution. \citet{Long21} ascribed these discrepancies to the different UV environments expected in transition disks. Given that disks in our sample have cavity sizes comparable to J1608 and show similar HNC flux excesses, our observations provide further evidence that the transition disk structure enhances HNC chemistry.

\begin{figure*}
    \centering
    \includegraphics[width=0.9\linewidth]{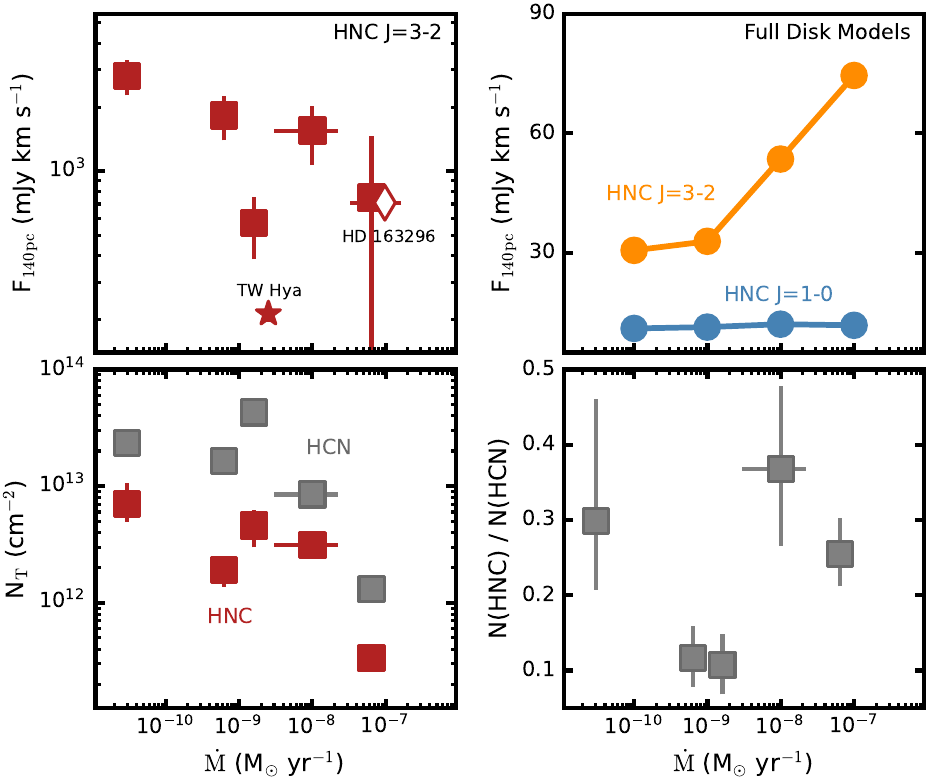}
    \caption{HNC and HCN properties as a function of stellar mass accretion rate. \textit{Top row:} Disk-integrated HNC J=3--2 fluxes (\textit{left}) and full disk model predictions (\textit{right}) of \citet{Long21} for the J=3--2 (dark orange) and J=1--0 (light blue) lines. We take HNC line fluxes for the TW~Hya and HD~163296 disks from \citet{Graninger15}. \textit{Bottom row:} HCN and HNC column densities (\textit{left}) and the HNC-to-HCN column density ratio (\textit{right}).}
    \label{fig:mass_accretion_vs_flux_HNC}
\end{figure*}

\subsubsection{HNC and Stellar Mass Accretion Rate} \label{sec:HNC_vs_Stellar_Mdot}

Chemical models show that UV radiation strongly influences cyanide chemistry in disks \citep[e.g.,][]{Aikawa99, Fogel11, Chapillon12, Cleeves13, Visser18}. Specifically, H$_2$ is vibrationally excited (to H$_2^*$) via FUV pumping \citep{Tielens85}, which then reacts with atomic N to form NH, which then reacts with C$^+$ to produce CN$^+$, and then, in turn, HCNH$^+$, initializing one of the main formation pathways of HCN and HNC as described in Appendix C of \citet{Visser18}. To assess how the observed chemistry depends on the UV field strength, we compared our measured HNC and HCN properties with stellar mass accretion rates ($\dot{M}$), as a proxy for excess UV flux. Our sample includes host stars with $\dot{M}$ from ${\gtrsim}$10$^{-11}$ to 10$^{-7}$~M$_{\odot}$~yr$^{-1}$, which correspond to order-of-magnitude variations in FUV flux \citep[e.g., see Figure 1 in][]{Visser18}.

The top left panel of Figure \ref{fig:mass_accretion_vs_flux_HNC} shows distance-normalized HNC J=3--2 fluxes from our SMA sample versus $\dot{M}$. We also included fluxes from the literature for HD~163296 and TW~Hya \citep{Graninger15}. The HNC J=3--2 fluxes decrease with larger $\dot{M}$, which is the opposite of the models of \citet{Long21} shown in the upper right panel of Figure \ref{fig:mass_accretion_vs_flux_HNC}. These models predict monotonically increasing HNC J=3--2 fluxes, with a steep increase at $\dot{M} > 10^{-9}$~M$_{\odot}$~yr$^{-1}$, which is attributed to UV-driven HNC production in the disk surface layers. Both HNC and HCN show decreasing column densities with $\dot{M}$, as shown in the bottom left panel of Figure \ref{fig:mass_accretion_vs_flux_HNC}. Thus, at least in a disk-averaged sense, the photodissociation of HCN and HNC must be more significant than any corresponding increase in H$_2^*$ production. The decline in both N(HCN) and N(HNC) is similar in relative magnitude, which also indicates that there is not significant conversion of HNC into HCN, e.g., as suggested in other ISM settings \citep{Bublitz19}. This yields consistent HNC-to-HCN column density ratios\footnote{\citet{Aguado17} suggest that HNC may have a larger photodissociation cross section and be photodissociated faster then HCN by factors of ${\approx}$2-10 depending on the radiation field strength. However, we do not find any evidence for selective HNC photodissociation in our data, as both HCN and HNC column densities decline at nearly the same rate across three orders of magnitude in $\dot{M}$.} of ${\approx}$0.1-0.4, which span no more than a factor of a few. Thus, the most likely explanation for the decreasing HNC fluxes and column densities is photodissociation from intrinsically stronger UV fields (i.e., from higher accretion rates) combined with increased UV penetration into the outer disk due to the large central dust cavities. Similar effects have been seen in spatially-resolved disk observations, namely CN-to-HCN column density ratios that increase from unity to ${\sim}$100 across inner to outer disk radii and local peaks in the CN-to-HCN column density ratio at the location of mm dust gaps \citep{Bergner21}. Both of these trends were interpreted as tracing the increased UV penetration in lower-density and less-UV-shielded regions \citep{Bergner21}. Overall, we note that the sample sizes considered here are small, so these trends, while intriguing, remain tentative and will require further confirmation in additional HNC observations of larger disk samples.

\begin{figure}
    \centering
    \includegraphics[width=\linewidth]{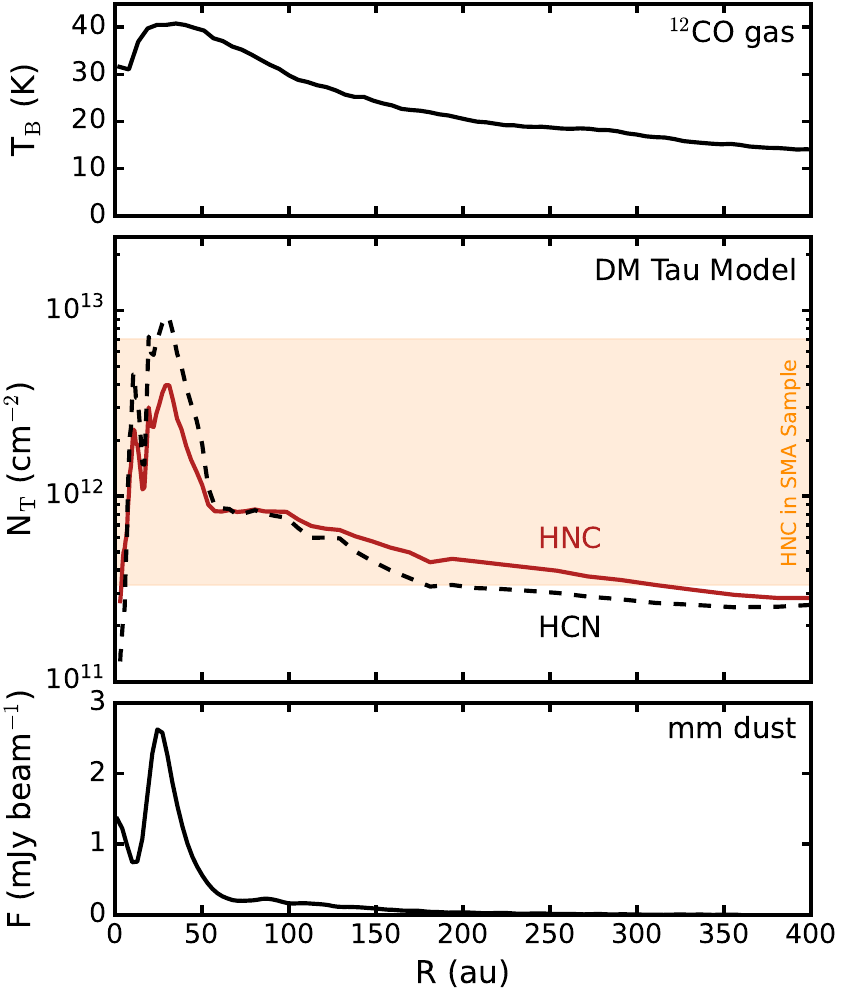}
    \caption{HNC (solid) and HCN (dashed) column densities (\textit{middle}) from the DM~Tau model of \citet{Long24}. The range of derived HNC column densities in our disk sample is shaded in orange. The $^{12}$CO gas temperature (\textit{top}) and mm dust (\textit{bottom}) radial profiles are from \citet{Curone25} and \citet{Galloway25}, respectively.}
    \label{fig:DM_Tau_model_Details}
\end{figure}

\subsection{Origins of HNC Chemistry in Transition Disks} \label{sec:origins_HNC_discussion}

We found compelling evidence that HNC chemistry is distinct in transition disks hosting large dust cavities. However, to move beyond disk-averaged inferences and investigate how local conditions influence HNC chemistry, we require chemical models, where we can track how HNC abundances, reaction rates, and gas conditions vary across the central cavity and over the entire disk extent.

\begin{figure*}
    \centering
    \includegraphics[width=0.49\linewidth]{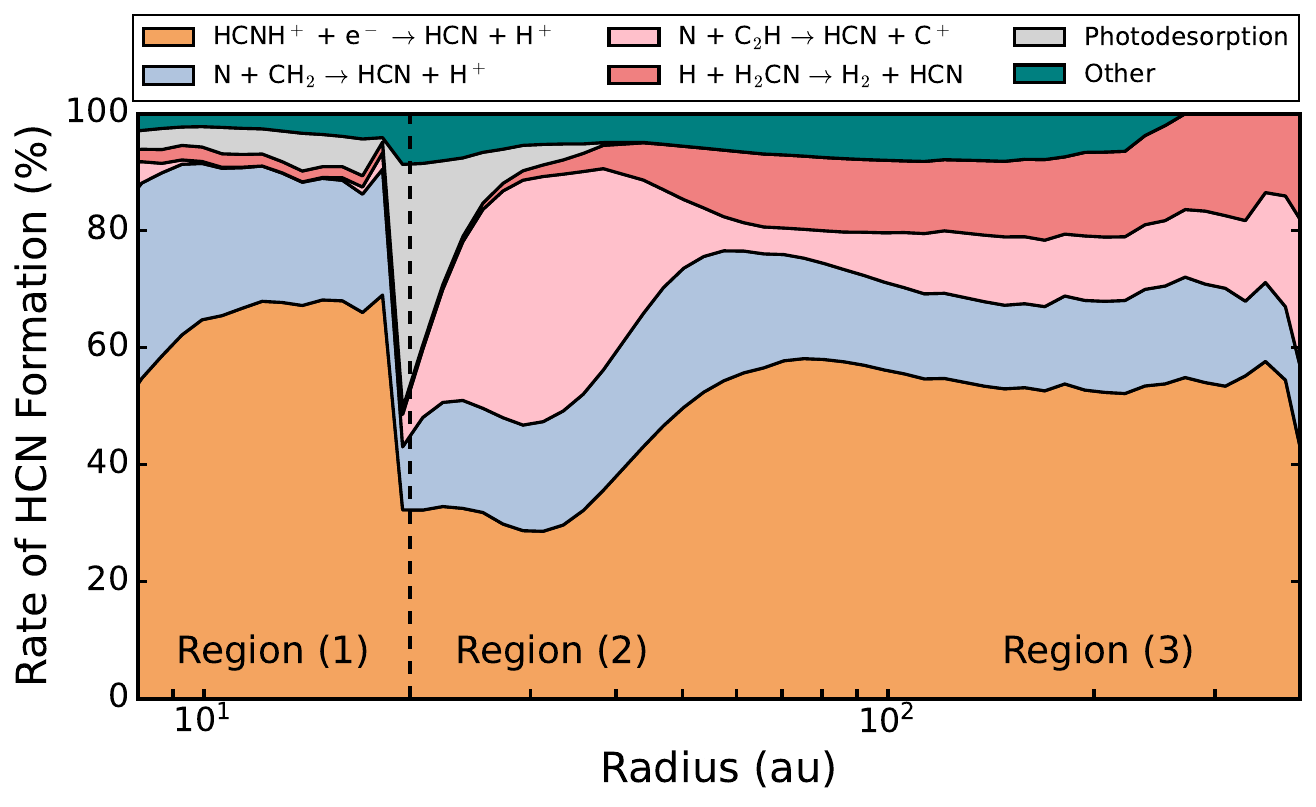}
    \includegraphics[width=0.49\linewidth]{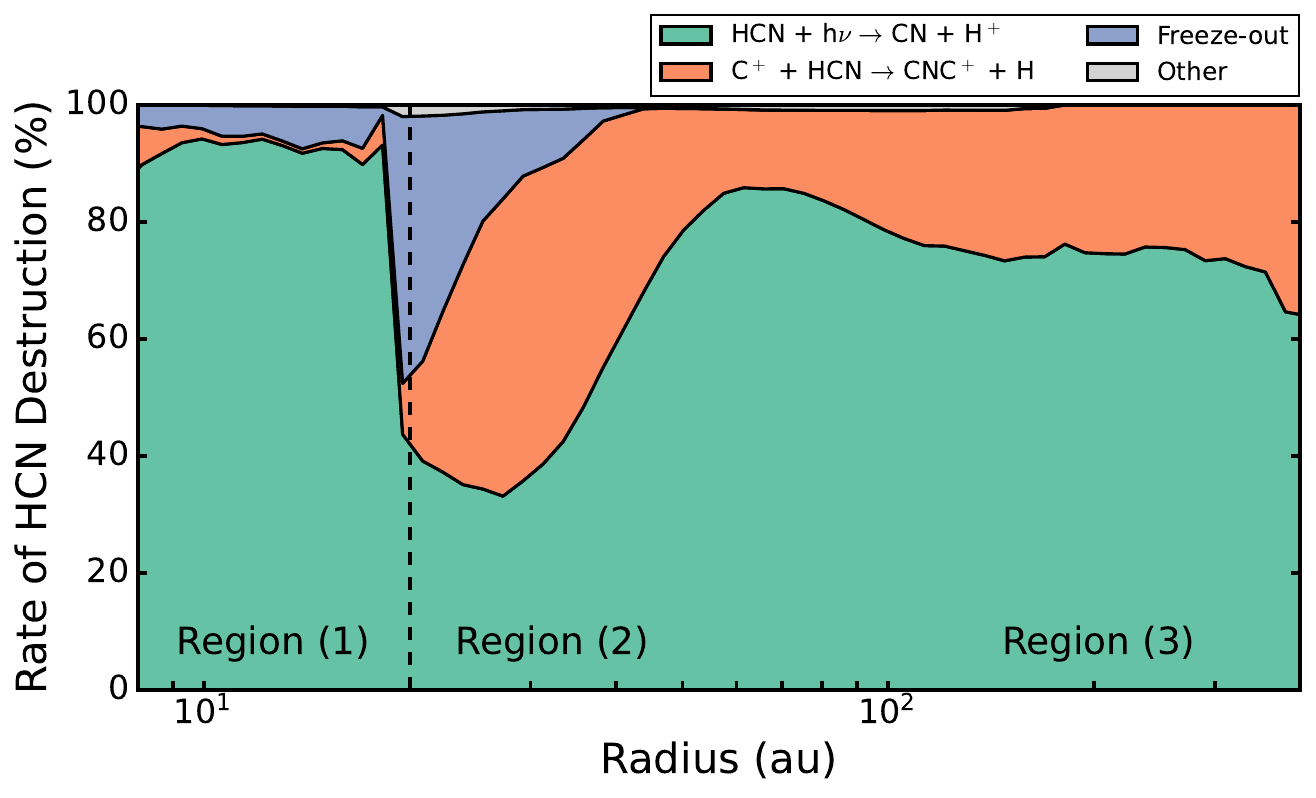}    \includegraphics[width=0.49\linewidth]{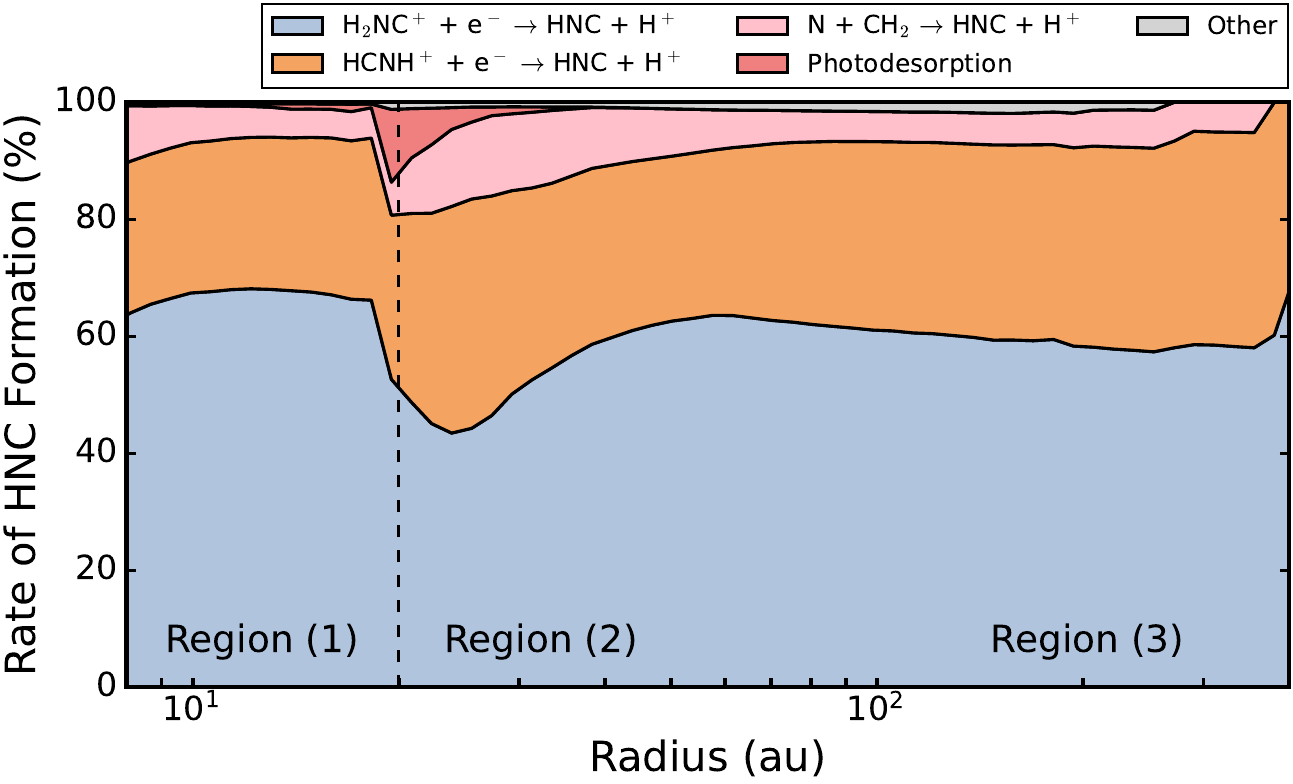}
    \includegraphics[width=0.49\linewidth]{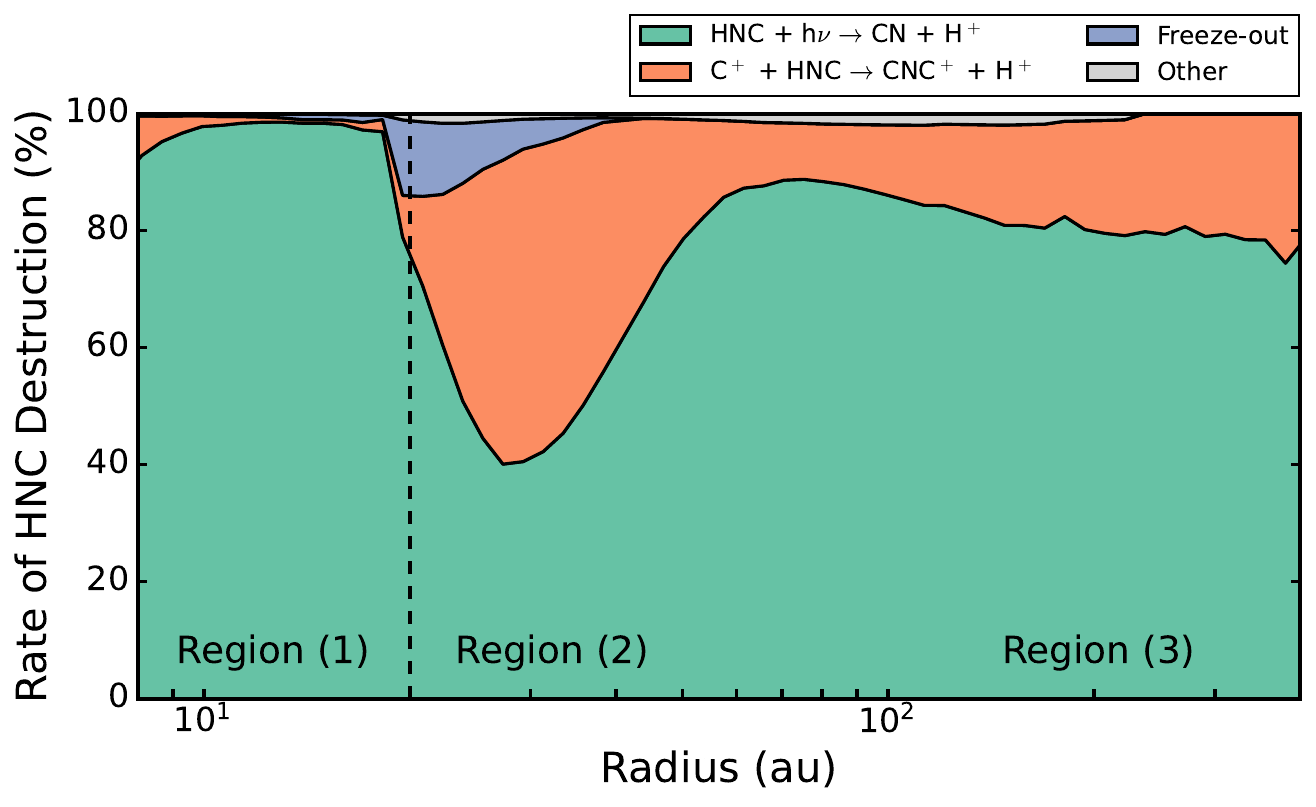}
    \caption{Fractional contribution of the most important reactions to the formation (\textit{left}) and destruction (\textit{right}) of HCN (\textit{top}) and HNC (\textit{bottom}) in the DM~Tau disk model. Colors correspond to individual reactions and the following regions of interest are annotated, i.e., region (1) is the dust cavity; region (2) is the mm dust disk; and region (3) is the outer disk.}
    \label{fig:reacation_rates}
\end{figure*}

\subsubsection{The DM~Tau Transition Disk Model} \label{sec:DMTau_model_setup}

To do so, we make use of an existing chemical model of the DM~Tau transition disk from \citet{Long24}. This model has a continuous gas distribution based on $^{12}$CO observations from \citet{Flaherty20} and dust depletion in the region $R < R_{\rm{cav}}=19$~au, as constrained from continuum visibilities and SED fitting by \citet{Andrews11}. Input UV and X-ray fluxes were obtained by scaling observed stellar UV/X-ray spectra for TW~Hya \citep{kastner99, herczeg04} to DM Tau's brighter FUV luminosity by a factor of 2.16 and an X-ray luminosity of 2.5$\times$10$^{29}$~erg~s$^{-1}$. Their model suite was tuned to explore the ionization conditions of the DM~Tau system based on ALMA observations of HCO$^+$, H$^{13}$CO$^+$, and N$_2$H$^+$ and was run for 1~Myr to reach a pseudo-steady state. We adopted the best-fit model, which has a moderately reduced CR ionization rate ($\zeta_{\rm{CR}}$ $\sim$ 10$^{-18}$~s$^{-1}$) and hard X-ray spectrum representing stellar flaring conditions. 

The DM~Tau system is an ideal proxy for exploring HNC chemistry, as it shares numerous characteristics with our transition disk sample. DM~Tau is a young (${\sim}$3-7~Myr), nearby (143~pc) T~Tauri star with a similar stellar mass (${\approx}$0.5~M$_{\odot}$) and spectral type (M1) that hosts a large mm dust cavity and an extended gas disk \citep{Simon00, Dartois03, Kudo18, Pegues20, Hashimoto21, Francis22, Gaia23}. The DM~Tau disk has also been extensively studied with a well-constrained thermal and dust structure, as shown in Figure \ref{fig:DM_Tau_model_Details}.

\subsubsection{Predicted HNC and HCN Chemistry}\label{sec:DMTau_HNCandHCN}

Following the approach outlined in \citet{Long24}, we computed the disk-integrated HNC line fluxes based on the modeled abundances using the \texttt{LIME} non-LTE radiative transfer code \citep{Brinch10} and collisional rates from the LAMDA database \citep{Faure07, Dumouchel10, Vera17}. The model fluxes of HNC J=3--2 (${\approx}$1.97~Jy~km~s$^{-1}$) and J=4--3 (${\approx}$2.11~Jy~km~s$^{-1}$) are consistent with our transition disk sample and also exceed full disk model predictions, as shown in Figure \ref{fig:long_model_comparison}.

Figure \ref{fig:DM_Tau_model_Details} shows HNC and HCN column density radial profiles extracted from the DM~Tau model. For comparison, we also included the known gas temperature, as inferred from optically-thick $^{12}$CO line emission \citep{Galloway25}, and mm dust continuum profiles \citep{Curone25}, respectively. The modeled HNC column densities are in good agreement with the range of values measured in our sample. Both HNC and HCN show similar spatial distributions, peaking near the inner dust ring followed by a steady decline at larger radii. Interior to, and just beyond, this dust ring (i.e., within ${\approx}$60-70~au), HCN has a larger column density than HNC, but for all radii beyond this, the relationship reverses with N(HNC) $>$ N(HCN). As a result, the HNC-to-HCN column density ratio has a minimum beyond the cavity at ${\approx}$20-30~au and a maximum in the outer disk at a few 100~au. This is broadly consistent with models that predict HNC to be abundant in the cold outer disk but is preferentially destroyed in the hot inner disk gas \citep[e.g.,][]{Visser18, Long21}.

We emphasize that the best-fit model from \citet{Long24} was not tuned to cyanide chemistry, and it is thus striking that the predicted HNC line fluxes and column densities are in broad agreement with our transition disk sample.

\subsubsection{What Drives HNC Chemistry in Transition Disks?}

To determine which reactions set HCN and HNC chemistry across the DM~Tau disk, we compute the fractional contribution of each reaction in the model at each radial bin summed over the disk vertical extent in that bin. This approach is motivated by prior models, which indicate that cyanide abundances are dominated by only a handful of formation and destruction reactions \citep[e.g.,][]{Semenov11, Visser18}. We also expect HNC and HCN to form in a thin layer (z/r$\sim$0.1-0.2) above the midplane \citep[e.g.,][]{Agundez18}, and thus we do not expect these reactions to vary significantly across the disk height. To aid in interpretation, we further divided reactions into formation and destruction pathways. Figure \ref{fig:reacation_rates} shows a summary of the most significant reactions in the DM~Tau disk model as a function of radius. At most radii, especially in the outer disk, three reactions combined account for nearly 80\% of the total formation or destruction rates, with the exception of HCN formation, which is modestly more complex with several additional (less dominant) routes operating. 

The relative importance of these reactions have a radial dependence, which can be broadly divided into three disk regions: (1) within the central cavity (${<}$20~au); (2) beyond the cavity but within the disk region with abundant mm dust (20-60~au); and (3) the outer disk with limited or no mm dust (${>}$60~au). Each region is labeled in Figure \ref{fig:reacation_rates}. Below, we discuss the formation and destruction pathways in detail for both species:

\indent \textit{Formation}: The dissociative recombination of HCNH$^+$ is the dominant formation pathway of both HNC and HCN, as predicted by numerous previous models \citep[e.g,.][]{Visser18, Long21}. For HNC, there is also a significant contribution from the dissociative recombination of the other stable isomer of protonated hydrogen cyanide, i.e., H$_2$NC$^+$. While the relative contributions of the two protonated hydrogen cyanide isomers will be set by their respective, but uncertain, branching ratios, both depend in similar ways on irradiation-driven pathways \citep{Loison14}. HCN and HNC also have production channels via atomic nitrogen reacting with small hydrocarbons (CH$_2$, C$_2$H). The additional HCN pathway of H $+$ H$_2$CN is also related to hydrocarbon chemistry as H$_2$CN is efficiently produced by the N $+$ CH$_3$ reaction \citep[e.g.,][]{Marston89, Hebrard12}. The relatively larger (although still ${<}$10\%) component of HCN formation from unlabeled reactions in Figure \ref{fig:reacation_rates} is made up of numerous, roughly co-equal reactions involving atomic N and other less abundant hydrocarbons (e.g., C$_2$H$_2^+$, CH$_3$).

Dissociative recombination reactions are dominant in regions 1 and 3, due to the absence or reduction in dust density and increasing UV penetration depths. This is because there are many routes to form either HCNH$^+$ or  H$_2$NC$^+$, all of which depend on irradiation, e.g., sublimating NH$_3$ ice, proton transfer reactions from HCO$^+$, H$_3^+$ and other ions, as well as N$_2$ $+$ He$^+$ \citep{Visser18}. The presence of significant dust in region (2) results in photodesorption becoming efficient and leads to a modest decline in the relative contribution of the aforementioned gas-phase routes.

\indent \textit{Destruction}: Photodissociation and ion-driven processes (i.e., via C$^+$) dominate both HCN and HNC destruction, consistent with our findings in Section \ref{sec:HNC_vs_Stellar_Mdot} and prior models \citep[e.g.,][]{Loison14}. There is also a minor contribution from freeze-out onto dust in the midplane. The highest contributions from photodissociation are, in order of relative significance, in regions (1) and (3), which have reduced dust shielding. Correspondingly, grain-mediated loss channels only contribute in region (2). The C$^+ +$ HNC and C$^+ +$ HCN reactions are most important beyond the cavity, with maximum contributions in region (2), as increased dust shielding here lessens the relative importance of photodissociation. Both HNC and HCN appear similar in their destruction mechanisms with HNC being modestly more sensitive (${\lesssim}$10\%) to photodissociation at a given radius.

Overall, we find that the production of HNC is tightly linked to the radiation environment, potentially even more so than HCN. While the temperature-dependent isomerization reactions between HNC and HCN (i.e., HNC + H $\to$ HCN + H) are present in the DM~Tau model, they are not significant contributors to the formation or destruction of either species. If true in broader samples, this suggests that the HNC-to-HCN ratio may not be a straightforward tracer of disk gas temperature (see Section \ref{sec:disk_prop_ratio_discu}). Our models also show that contributions from the dissociative recombination of both stable isomers of protonated hydrogen cyanide (HCNH$^+$, H$_2$NC$^+$) are important in reproducing HNC chemistry and should be included in future modeling efforts.

\begin{figure*}
    \centering
    \includegraphics[width=0.85\linewidth]{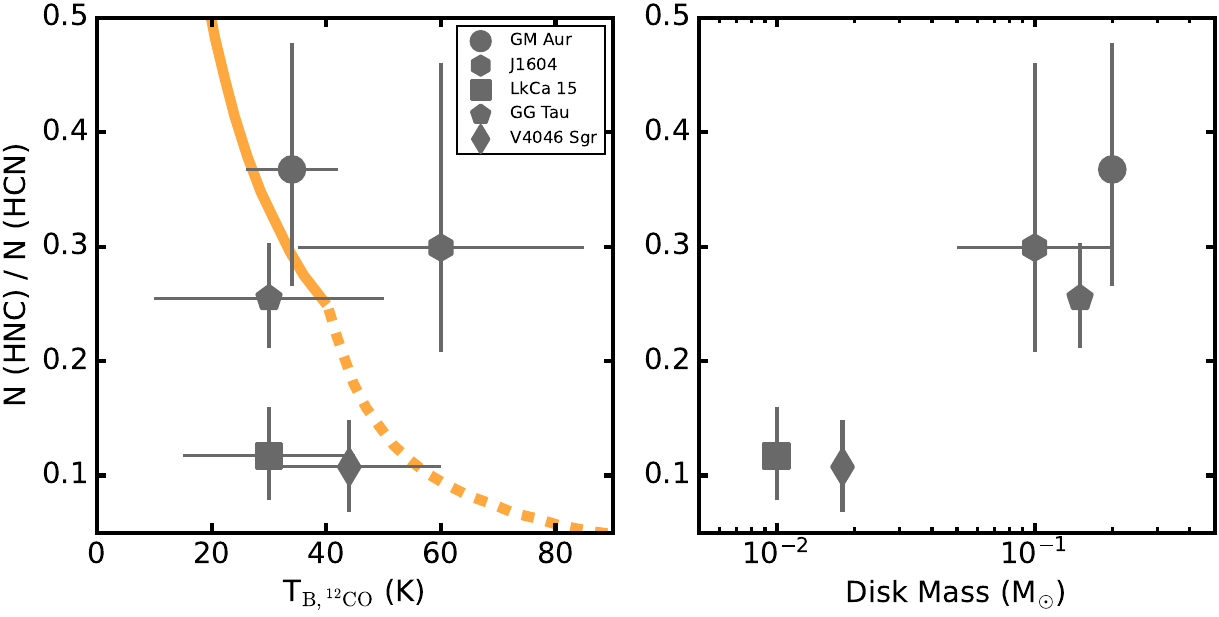}
    \caption{HNC-to-HCN column density ratio versus $^{12}$CO brightness temperature (\textit{left}) and disk mass (\textit{right}). Temperature error bars show the T$_{\rm{B}}$ range over the radial extent in which the HNC-to-HCN ratio was measured and were extracted from literature T$_{\rm{B}}$ profiles \citep{Law21_thermal, Wolfer23, Galloway25}. The orange line shows the line ratio-gas temperature correlation inferred in \citet{Hacar20}, where solid and dashed lines indicate the low- and high-temperature fits, respectively. Disk mass references are the same as those listed in Figure \ref{fig:long_model_comparison}.}
    \label{fig:temp_disk_mass_corr}
\end{figure*}

If we consider the results of this chemical modeling and the trends observed in Section \ref{sec:vs_full_disk_models}, we find that photodissociation is significant in destroying both HNC and HCN and that systems with higher UV fluxes and accretion rates result in diminished cyanide abundances. However, for a fixed UV-field strength, the disk structure appears more important, i.e., more efficient HNC production in transition disks than full disks due to increased UV penetration depths. While there may appear to be tension between these explanations, i.e., UV-driven production versus dissociation, the former conclusion is based on comparing full and transition disks with similar properties (e.g., mass, accretion rates), while the latter is inferred from a disk sample spanning several orders of magnitude in stellar radiation properties.

\subsection{The HNC-to-HCN Ratio as a Tracer of Bulk Disk Properties} \label{sec:disk_prop_ratio_discu}

While the HNC-to-HCN ratio is a useful tracer of gas properties in the ISM -- most notably as a chemical thermometer \citep[e.g.,][]{Hacar20} -- its utility is less clear in planet-forming disks, given their complex thermal and radiation structures and the presence of elemental and compositional gradients. Here, we investigate how the disk-averaged HNC-to-HCN column density ratio measured in our sample relates to the two key properties of gas temperature and disk mass.

The left panel of Figure \ref{fig:temp_disk_mass_corr} shows the HNC-to-HCN column density ratio as a function of gas temperature, as measured from optically-thick $^{12}$CO line emission \citep[e.g.,][]{Weaver18, Law21_thermal}. We find no clear trends, with the caveat that our disk sample only includes T~Tauri systems with similar stellar masses. Thus, our sample spans a limited range of temperatures (T$_{\rm{B, ^{12}CO}}{\approx}$20-40~K) with the exception of the modestly-warmer (${\approx}$60~K) J1604 disk. Notably, it is J1604 that deviates the most from the correlation identified by \citet{Hacar20}, as shown in Figure \ref{fig:temp_disk_mass_corr}, which suggests that disks are perhaps unlikely to follow this ISM-calibrated relation. However, HNC observations in substantially cooler and warmer disk systems, i.e., around very-low mass stars and Herbigs, respectively, are needed to fully explore the HNC-to-HCN column density ratio as a disk gas temperature.

The right panel of Figure \ref{fig:temp_disk_mass_corr} shows the same comparison, but with disk gas mass. Here, we instead identify a positive trend between the HNC-to-HCN column density ratio and disk mass, which is consistent with the expectation that HNC is more abundant in cooler gas, as more massive, and thus larger, disks have larger reservoirs of cool gas. This is further supported by positive trends between the size of the mm dust disk with both the HNC-to-HCN flux (Figure \ref{fig:flux_plots}) and column density (Appendix \ref{sec:Ncol_correlation_plots}) ratios. This explanation may, at first, appear at odds with the lack of a firm temperature trend. However, we caution that gas temperatures inferred using $^{12}$CO, while a useful proxy, are not without caveats. The $^{12}$CO emission only probes the uppermost molecular layers \citep[e.g.,][]{Law21_thermal, Law22}, which is likely distinct from the more shielded HNC emission closer to the midplane. Moreover, models \citep{Long21} predict multiple HNC reservoirs (i.e., midplane and warm surface layers), and our excitation analysis (Section \ref{sec:rot_diags}) suggests that the HNC emitting region can vary on a disk-to-disk basis (even with respect to HCN). Nonetheless, we also checked for potential trends by instead using the HCN rotational temperatures (from Table \ref{tab:rotate_data}) but found none.

We also note that this trend is consistent with early model predictions that indicated HNC is particularly sensitive to the total column density of the disk \citep{Aikawa99}. Given the difficulty in determining disk gas masses \citep[e.g.,][]{Miotello23}, the potential presence of a positive HNC-to-HCN trend means that HNC observations, if calibrated across a wider disk sample, may offer a powerful opportunity to constrain one of the most fundamental but elusive disk properties.

\section{Conclusions} \label{sec:conclusions}

We presented an analysis of SMA observations of the J=3--2 and J=4--3 lines of HNC and HCN in a sample of five transition disks around the T~Tauri stars GM~Aur, J1604, LkCa~15, GG~Tau, and V4046~Sgr. We conclude the following:

\begin{enumerate}
    \item We detected at least one line of both HCN and HNC in each source, including the first disk detection of HNC J=4--3. We nearly doubled the number of HNC disk detections reported in the literature to date, and our high detection rates suggest an abundant reservoir of HNC gas in transition disks.
    \item We measured disk-integrated HNC-to-HCN flux ratios, which ranged from ${\approx}$0.1-0.7 with a median ratio of 0.37. Three of the disks (GM~Aur, J1604, LkCa~15) show ratios higher than seen in prior observations and predicted from full disk models. 
    \item Using a rotational diagram analysis, we derived column densities, and when possible, rotational temperatures for both molecules. HCN and HNC column densities are on the order of 10$^{13}$~cm$^{-2}$ and 10$^{12}$~cm$^{-2}$, respectively, with the exception of the GG~Tau disk, which shows an one-order-of-magnitude reduction in both species. The inferred HNC-to-HCN column density ratios are ${\approx}$0.1-0.4.
    \item For two disks, we derived both HCN and HNC rotational temperatures. For one disk (J1604), the rotational temperature of HCN is twice that of HNC, whereas they are equal in the other disk (GG~Tau). This suggests that there may be disk-to-disk variation in the location of the HNC reservoir.
    \item The measured HNC fluxes in our transition disk sample exceed full disk model predictions in all cases, ranging from a factor of a few to more than an order of magnitude. This indicates that the structure of transition disks is responsible for altering the HNC chemistry and highlights the important role cavities play in setting disk conditions, even at radii larger than the cavity itself.
    \item To explore the origins of HNC, we make use of an existing chemical model of the similar transition disk around DM~Tau. The production of HNC is tightly linked to the radiation environment, even more so than HCN. Our models show that contributions from the dissociative recombination of both stable isomers of protonated hydrogen cyanide (i.e., HCNH$^+$, H$_2$NC$^+$), which have numerous UV-driven formation pathways, are important in accurately reproducing observed HNC chemistry in transition disks.
    \item The disk-integrated HNC-to-HCN column density ratio positively correlates with the disk gas mass. This is likely due to HNC being more abundant in the larger reservoirs of cooler gas available in more massive disks.    
\end{enumerate}

\begin{acknowledgments}

The authors thank the anonymous referee for valuable comments that improved the content and presentation of this work.

The authors wish to recognize and acknowledge the very significant cultural role and reverence that the summit of Maunakea has always had within the indigenous Hawaiian community. We are most fortunate to have had the opportunity to conduct observations from this mountain.

K.G. acknowledges support from a Virginia Initiative on Cosmic Origins (VICO) summer undergraduate fellowship. Support for C.J.L. was provided by NASA through the NASA Hubble Fellowship grant No. HST-HF2-51535.001-A awarded by the Space Telescope Science Institute, which is operated by the Association of Universities for Research in Astronomy, Inc., for NASA, under contract NAS5-26555. L.I.C. also acknowledges support from the Research Corporation for Science Advancement Cottrell Scholarship Award 28249, the David and Lucille Packard Foundation, and NSF AAG Award 2205698. R.L.G. acknowledges funding from the French Agence Nationale de la Recherche (ANR) through the project MAPSAJE (ANR-24-CE31-2126-01). V.V.G. gratefully acknowledges support from FONDECYT Regular 1221352, ANID CATA-BASAL project FB210003 and ANID -- Millennium Science Initiative Program -- Center Code NCN2024\_001. 
\end{acknowledgments}

\clearpage
\appendix

\section{Teardrop Plots} \label{sec:teardrops}

\begin{figure*}[h]
    \includegraphics[width=1.0\linewidth]{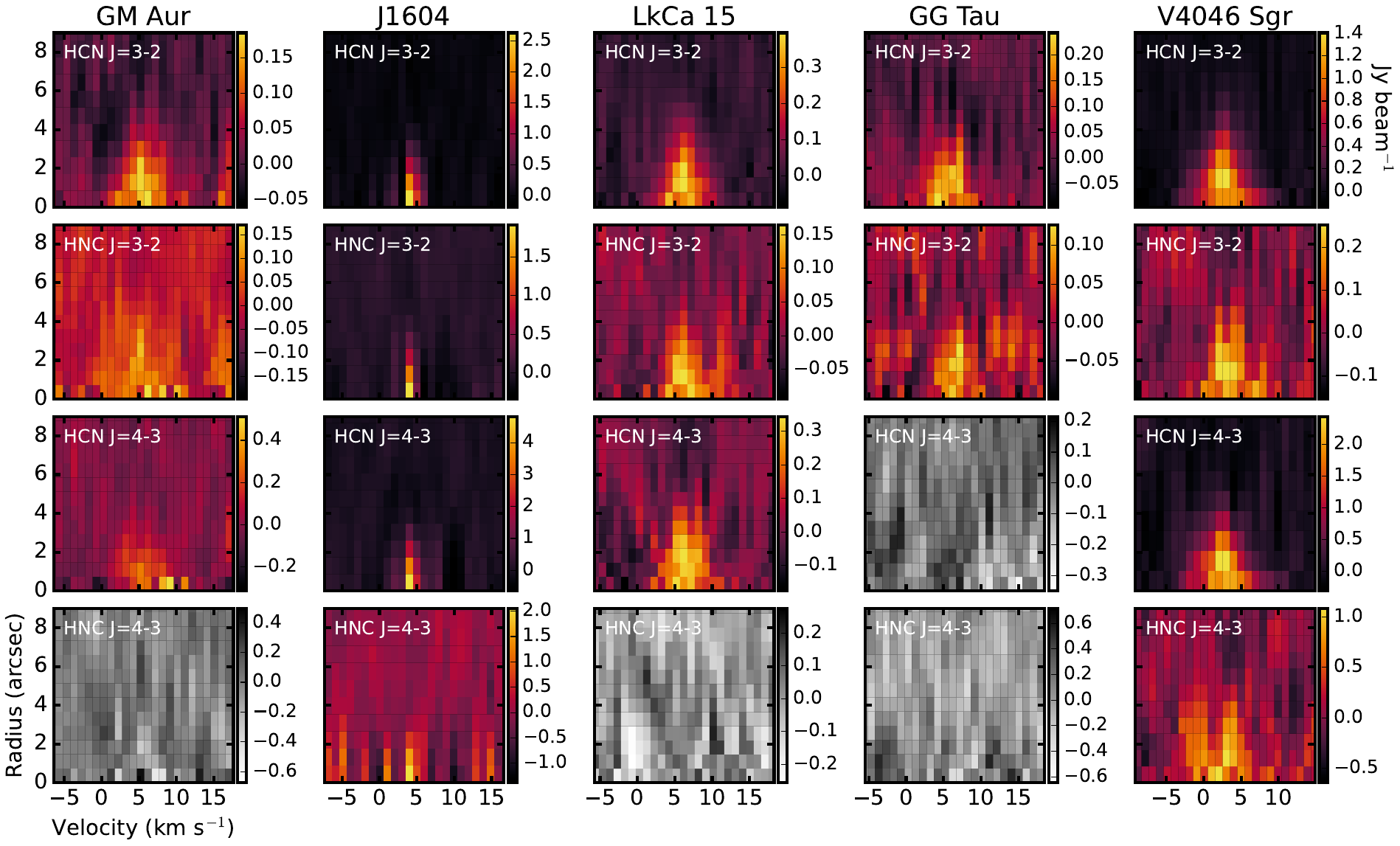}
    \caption{Gallery of teardrop plots of the HCN and HNC J=3--2 and J=4--3 lines (\textit{rows}) for each disk (\textit{columns}) in our sample. All image cubes are shown at 1~km~s$^{-1}$ resolution. Gray panels indicate non-detections.}
    \label{fig:teardrop_gallery}
\end{figure*}

To enhance our SNR and confirm the reported line detections, we leveraged the Keplerian rotation and known disk geometry (Table \ref{tab:source_prop}) to shift-and-stack the emission signature in each line and disk \citep[e.g.,][]{Teague16, Yen16}. Figure \ref{fig:teardrop_gallery} shows the resulting `teardrop' plots. All detected lines exhibit clear signatures of Keplerian-rotating gas. In particular, the coherence of the signal for HNC J=3--2 in the GG~Tau disk and HNC J=4--3 in the J1604 disk provides further evidence of the robustness of these tentative detections discussed in the main text.

\clearpage
\section{Correlation Plots} \label{sec:Ncol_correlation_plots}

\begin{figure*}[h]
    \includegraphics[width=\linewidth]{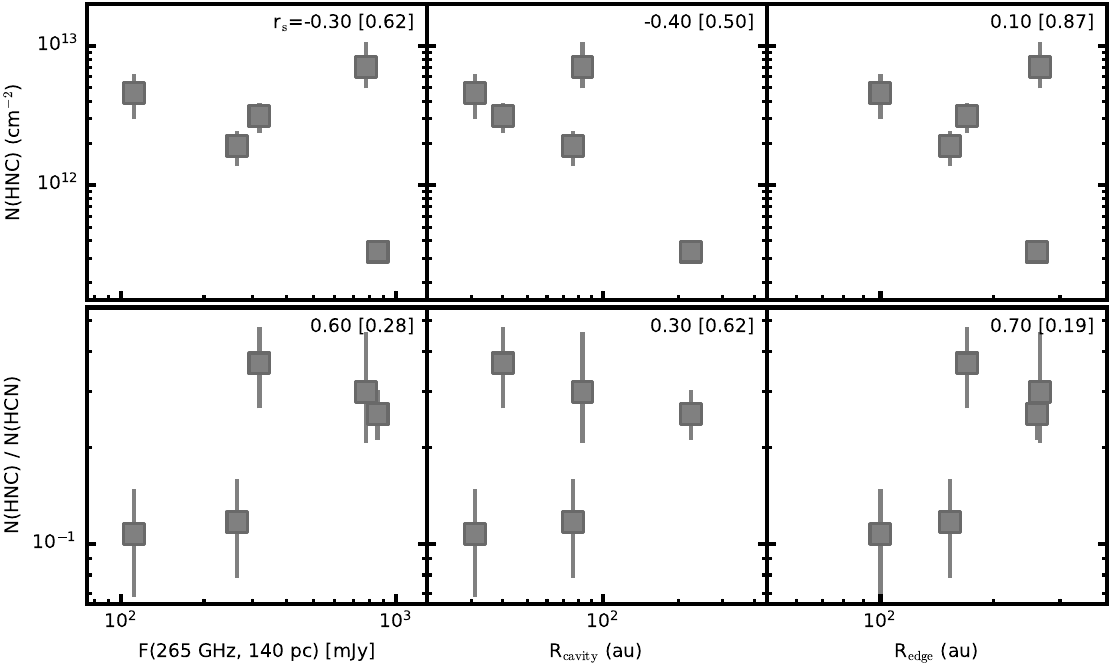}
    \caption{Disk-integrated HNC column density (\textit{top)} and HNC-to-HCN column density ratios (\textit{bottom)}. Spearman correlation coefficients and associated p-values are shown in the upper right corner of each panel.}
    \label{fig:ncol_corr_plots_gallery}
\end{figure*}

Figure \ref{fig:ncol_corr_plots_gallery} shows the full gallery of correlation plots for the HNC column density and HNC-to-HCN column density ratios versus various disk parameters. As in Section \ref{sec:line_flux_ratios}, the Spearman correlation coefficients were computed for each set of parameters.

\software{\texttt{bettermoments} \citep{teague_bettermoments_2018}, CASA \citep{McMullin_etal_2007, CASATeam20}, \texttt{emcee} \citep{Foreman13}, \texttt{linmix} \citep{Kelly07}, \texttt{LIME} \citep{Brinch10}, \texttt{pyuvdata} \citep{Hazelton17}, \texttt{gofish} \citep{Teague19JOSS}, Matplotlib \citep{Hunter07}, NumPy \citep{vanderWalt_etal_2011}, SciPy \citep{Virtanen20}}

\facilities{SMA}

\bibliography{HNC_SMA}{}

@ARTICLE{herczeg04,
       author = {{Herczeg}, Gregory J. and {Wood}, Brian E. and {Linsky}, Jeffrey L. and {Valenti}, Jeff A. and {Johns-Krull}, Christopher M.},
        title = "{The Far-Ultraviolet Spectra of TW Hydrae. II. Models of H$_{2}$ Fluorescence in a Disk}",
      journal = {\apj},
         year = 2004,
        month = may,
       volume = {607},
       number = {1},
        pages = {369-383},
          doi = {10.1086/383340},
archivePrefix = {arXiv},
       eprint = {astro-ph/0402238},
 primaryClass = {astro-ph},
       adsurl = {https://ui.adsabs.harvard.edu/abs/2004ApJ...607..369H}
}

@ARTICLE{Stuber25,
       author = {{Stuber}, Sophia K. and {Schinnerer}, Eva and {Usero}, Antonio and {Bigiel}, Frank and {den Brok}, Jakob and {Pety}, Jerome and {Neumann}, Lukas and {Jim{\'e}nez-Donaire}, Mar{\'\i}a J. and {Sun}, Jiayi and {Querejeta}, Miguel and {Barnes}, Ashley. T. and {Be{\v{s}}li{\'c}}, Ivana and {Cao}, Yixian and {Dale}, Daniel A. and {Eibensteiner}, Cosima and {Gleis}, Damian and {Glover}, Simon C.~O. and {Grasha}, Kathryn and {Klessen}, Ralf S. and {Liu}, Daizhong and {Meidt}, Sharon and {Pan}, Hsi-An and {Saito}, Toshiki and {Thorp}, Mallory and {Williams}, Thomas G.},
        title = "{The SWAN view of dense gas in the Whirlpool: A cloud-scale comparison of N$_{2}$H$^{+}$, HCO$^{+}$, HNC, and HCN emission in M51}",
      journal = {\aap},
         year = 2025,
        month = oct,
       volume = {702},
          eid = {A66},
        pages = {A66},
          doi = {10.1051/0004-6361/202554473},
archivePrefix = {arXiv},
       eprint = {2507.19439},
 primaryClass = {astro-ph.GA},
       adsurl = {https://ui.adsabs.harvard.edu/abs/2025A&A...702A..66S}
}

@ARTICLE{kastner99,
       author = {{Kastner}, Joel H. and {Huenemoerder}, David P. and {Schulz}, Norbert S. and {Weintraub}, David A.},
        title = "{X-Ray Spectroscopy of the Nearby, Classical T Tauri Star TW Hydrae}",
      journal = {\apj},
         year = 1999,
        month = nov,
       volume = {525},
       number = {2},
        pages = {837-844},
          doi = {10.1086/307946},
archivePrefix = {arXiv},
       eprint = {astro-ph/9905126},
 primaryClass = {astro-ph},
       adsurl = {https://ui.adsabs.harvard.edu/abs/1999ApJ...525..837K}
}

@ARTICLE{Teague25,
       author = {{Teague}, Richard and {Benisty}, Myriam and {Facchini}, Stefano and {Fukagawa}, Misato and {Pinte}, Christophe and {Andrews}, Sean M. and {Bae}, Jaehan and {Barraza-Alfaro}, Marcelo and {Cataldi}, Gianni and {Cuello}, Nicol{\'a}s and {Curone}, Pietro and {Czekala}, Ian and {Fasano}, Daniele and {Flock}, Mario and {Galloway-Sprietsma}, Maria and {Garg}, Himanshi and {Hall}, Cassandra and {Hammond}, Iain and {Hilder}, Thomas and {Huang}, Jane and {Ilee}, John D. and {Izquierdo}, Andr{\'e}s F. and {Kanagawa}, Kazuhiro and {Lesur}, Geoffroy and {Lodato}, Giuseppe and {Longarini}, Cristiano and {Loomis}, Ryan A. and {Masset}, Fr{\'e}d{\'e}ric and {Menard}, Francois and {Orihara}, Ryuta and {Price}, Daniel J. and {Rosotti}, Giovanni and {Stadler}, Jochen and {Testi}, Leonardo and {Yen}, Hsi-Wei and {Wafflard-Fernandez}, Gaylor and {Wilner}, David J. and {Winter}, Andrew J. and {W{\"o}lfer}, Lisa and {Yoshida}, Tomohiro C. and {Zawadzki}, Brianna},
        title = "{exoALMA. I. Science Goals, Project Design, and Data Products}",
      journal = {\apjl},
         year = 2025,
        month = may,
       volume = {984},
       number = {1},
          eid = {L6},
        pages = {L6},
          doi = {10.3847/2041-8213/adc43b},
archivePrefix = {arXiv},
       eprint = {2504.18688},
 primaryClass = {astro-ph.EP},
       adsurl = {https://ui.adsabs.harvard.edu/abs/2025ApJ...984L...6T}
}

@ARTICLE{Carpenter14,
       author = {{Carpenter}, John M. and {Ricci}, Luca and {Isella}, Andrea},
        title = "{An ALMA Continuum Survey of Circumstellar Disks in the Upper Scorpius OB Association}",
      journal = {\apj},
         year = 2014,
        month = may,
       volume = {787},
       number = {1},
          eid = {42},
        pages = {42},
          doi = {10.1088/0004-637X/787/1/42},
archivePrefix = {arXiv},
       eprint = {1404.0387},
 primaryClass = {astro-ph.SR},
       adsurl = {https://ui.adsabs.harvard.edu/abs/2014ApJ...787...42C}
}

@ARTICLE{Dong17,
       author = {{Dong}, Ruobing and {van der Marel}, Nienke and {Hashimoto}, Jun and {Chiang}, Eugene and {Akiyama}, Eiji and {Liu}, Hauyu Baobab and {Muto}, Takayuki and {Knapp}, Gillian R. and {Tsukagoshi}, Takashi and {Brown}, Joanna and {Bruderer}, Simon and {Koyamatsu}, Shin and {Kudo}, Tomoyuki and {Ohashi}, Nagayoshi and {Rich}, Evan and {Satoshi}, Mayama and {Takami}, Michihiro and {Wisniewski}, John and {Yang}, Yi and {Zhu}, Zhaohuan and {Tamura}, Motohide},
        title = "{The Sizes and Depletions of the Dust and Gas Cavities in the Transitional Disk J160421.7-213028}",
      journal = {\apj},
         year = 2017,
        month = feb,
       volume = {836},
       number = {2},
          eid = {201},
        pages = {201},
          doi = {10.3847/1538-4357/aa5abf},
archivePrefix = {arXiv},
       eprint = {1701.05189},
 primaryClass = {astro-ph.SR},
       adsurl = {https://ui.adsabs.harvard.edu/abs/2017ApJ...836..201D}
}

@ARTICLE{Donati19,
       author = {{Donati}, J. -F. and {Bouvier}, J. and {Alencar}, S.~H. and {Hill}, C. and {Carmona}, A. and {Folsom}, C.~P. and {M{\'e}nard}, F. and {Gregory}, S.~G. and {Hussain}, G.~A. and {Grankin}, K. and {Moutou}, C. and {Malo}, L. and {Takami}, M. and {Herczeg}, G.~J. and {MaTYSSE Collaboration}},
        title = "{The magnetic propeller accretion regime of LkCa 15}",
      journal = {\mnras},
         year = 2019,
        month = feb,
       volume = {483},
       number = {1},
        pages = {L1-L5},
          doi = {10.1093/mnrasl/sly207},
archivePrefix = {arXiv},
       eprint = {1811.04810},
 primaryClass = {astro-ph.SR},
       adsurl = {https://ui.adsabs.harvard.edu/abs/2019MNRAS.483L...1D}
}

@article{loomis_distribution_2018,
	title = {The {Distribution} and {Excitation} of {CH} ₃ {CN} in a {Solar} {Nebula} {Analog}},
	volume = {859},
	issn = {0004-637X, 1538-4357},
	url = {https://iopscience.iop.org/article/10.3847/1538-4357/aac169},
	doi = {10.3847/1538-4357/aac169},
	number = {2},
	urldate = {2024-06-21},
	journal = {The Astrophysical Journal},
	author = {Loomis, Ryan A. and Cleeves, L. Ilsedore and Öberg, Karin I. and Aikawa, Yuri and Bergner, Jennifer and Furuya, Kenji and Guzman, V. V. and Walsh, Catherine},
	month = jun,
	year = {2018},
	pages = {131},
}

@article{goldsmith_population_1999,
	title = {Population {Diagram} {Analysis} of {Molecular} {Line} {Emission}},
	volume = {517},
	issn = {0004-637X, 1538-4357},
	url = {https://iopscience.iop.org/article/10.1086/307195},
	doi = {10.1086/307195},
	language = {en},
	number = {1},
	urldate = {2024-07-09},
	journal = {The Astrophysical Journal},
	author = {Goldsmith, Paul F. and Langer, William D.},
	month = may,
	year = {1999},
	pages = {209--225},
}

@ARTICLE{Donati11,
       author = {{Donati}, J. -F. and {Gregory}, S.~G. and {Montmerle}, T. and {Maggio}, A. and {Argiroffi}, C. and {Sacco}, G. and {Hussain}, G. and {Kastner}, J. and {Alencar}, S.~H.~P. and {Audard}, M. and {Bouvier}, J. and {Damiani}, F. and {G{\"u}del}, M. and {Huenemoerder}, D. and {Wade}, G.~A.},
        title = "{The close classical T Tauri binary V4046 Sgr: complex magnetic fields and distributed mass accretion}",
      journal = {\mnras},
         year = 2011,
        month = nov,
       volume = {417},
       number = {3},
        pages = {1747-1759},
          doi = {10.1111/j.1365-2966.2011.19366.x},
archivePrefix = {arXiv},
       eprint = {1109.2447},
 primaryClass = {astro-ph.SR},
       adsurl = {https://ui.adsabs.harvard.edu/abs/2011MNRAS.417.1747D}
}

@ARTICLE{Wolfer23,
       author = {{W{\"o}lfer}, L. and {Facchini}, S. and {van der Marel}, N. and {van Dishoeck}, E.~F. and {Benisty}, M. and {Bohn}, A.~J. and {Francis}, L. and {Izquierdo}, A.~F. and {Teague}, R.~D.},
        title = "{Kinematics and brightness temperatures of transition discs. A survey of gas substructures as seen with ALMA}",
      journal = {\aap},
         year = 2023,
        month = feb,
       volume = {670},
          eid = {A154},
        pages = {A154},
          doi = {10.1051/0004-6361/202243601},
archivePrefix = {arXiv},
       eprint = {2208.09494},
 primaryClass = {astro-ph.EP},
       adsurl = {https://ui.adsabs.harvard.edu/abs/2023A&A...670A.154W}
}

@ARTICLE{Galloway25,
       author = {{Galloway-Sprietsma}, Maria and {Bae}, Jaehan and {Izquierdo}, Andr{\'e}s F. and {Stadler}, Jochen and {Longarini}, Cristiano and {Teague}, Richard and {Andrews}, Sean M. and {Winter}, Andrew J. and {Benisty}, Myriam and {Facchini}, Stefano and {Rosotti}, Giovanni and {Zawadzki}, Brianna and {Pinte}, Christophe and {Fasano}, Daniele and {Barraza-Alfaro}, Marcelo and {Cataldi}, Gianni and {Cuello}, Nicol{\'a}s and {Curone}, Pietro and {Czekala}, Ian and {Flock}, Mario and {Fukagawa}, Misato and {Gardner}, Charles H. and {Garg}, Himanshi and {Hall}, Cassandra and {Huang}, Jane and {Ilee}, John D. and {Kanagawa}, Kazuhiro and {Lesur}, Geoffroy and {Lodato}, Giuseppe and {Loomis}, Ryan A. and {Menard}, Francois and {Orihara}, Ryuta and {Price}, Daniel J. and {Wafflard-Fernandez}, Gaylor and {Wilner}, David J. and {W{\"o}lfer}, Lisa and {Yen}, Hsi-Wei and {Yoshida}, Tomohiro C.},
        title = "{exoALMA. V. Gaseous Emission Surfaces and Temperature Structures}",
      journal = {\apjl},
         year = 2025,
        month = may,
       volume = {984},
       number = {1},
          eid = {L10},
        pages = {L10},
          doi = {10.3847/2041-8213/adc437},
archivePrefix = {arXiv},
       eprint = {2504.19902},
 primaryClass = {astro-ph.EP},
       adsurl = {https://ui.adsabs.harvard.edu/abs/2025ApJ...984L..10G}
}

@ARTICLE{Law22,
       author = {{Law}, Charles J. and {Crystian}, Sage and {Teague}, Richard and {{\"O}berg}, Karin I. and {Rich}, Evan A. and {Andrews}, Sean M. and {Bae}, Jaehan and {Flaherty}, Kevin and {Guzm{\'a}n}, Viviana V. and {Huang}, Jane and {Ilee}, John D. and {Kastner}, Joel H. and {Loomis}, Ryan A. and {Long}, Feng and {P{\'e}rez}, Laura M. and {P{\'e}rez}, Sebasti{\'a}n and {Qi}, Chunhua and {Rosotti}, Giovanni P. and {Ru{\'\i}z-Rodr{\'\i}guez}, Dary and {Tsukagoshi}, Takashi and {Wilner}, David J.},
        title = "{CO Line Emission Surfaces and Vertical Structure in Midinclination Protoplanetary Disks}",
      journal = {\apj},
         year = 2022,
        month = jun,
       volume = {932},
       number = {2},
          eid = {114},
        pages = {114},
          doi = {10.3847/1538-4357/ac6c02},
archivePrefix = {arXiv},
       eprint = {2205.01776},
 primaryClass = {astro-ph.EP},
       adsurl = {https://ui.adsabs.harvard.edu/abs/2022ApJ...932..114L}
}

@ARTICLE{Virtanen20,
       author = {{Virtanen}, Pauli and {Gommers}, Ralf and {Oliphant}, Travis E. and {Haberland}, Matt and {Reddy}, Tyler and {Cournapeau}, David and {Burovski}, Evgeni and {Peterson}, Pearu and {Weckesser}, Warren and {Bright}, Jonathan and {van der Walt}, St{\'e}fan J. and {Brett}, Matthew and {Wilson}, Joshua and {Millman}, K. Jarrod and {Mayorov}, Nikolay and {Nelson}, Andrew R.~J. and {Jones}, Eric and {Kern}, Robert and {Larson}, Eric and {Carey}, C.~J. and {Polat}, {\.I}lhan and {Feng}, Yu and {Moore}, Eric W. and {VanderPlas}, Jake and {Laxalde}, Denis and {Perktold}, Josef and {Cimrman}, Robert and {Henriksen}, Ian and {Quintero}, E.~A. and {Harris}, Charles R. and {Archibald}, Anne M. and {Ribeiro}, Ant{\^o}nio H. and {Pedregosa}, Fabian and {van Mulbregt}, Paul and {SciPy 1. 0 Contributors}},
        title = "{SciPy 1.0: fundamental algorithms for scientific computing in Python}",
      journal = {Nature Methods},
         year = 2020,
        month = feb,
       volume = {17},
        pages = {261-272},
          doi = {10.1038/s41592-019-0686-2},
archivePrefix = {arXiv},
       eprint = {1907.10121},
 primaryClass = {cs.MS},
       adsurl = {https://ui.adsabs.harvard.edu/abs/2020NatMe..17..261V}
}

@ARTICLE{Fairlamb15,
       author = {{Fairlamb}, J.~R. and {Oudmaijer}, R.~D. and {Mendigut{\'\i}a}, I. and {Ilee}, J.~D. and {van den Ancker}, M.~E.},
        title = "{A spectroscopic survey of Herbig Ae/Be stars with X-shooter - I. Stellar parameters and accretion rates}",
      journal = {\mnras},
         year = 2015,
        month = oct,
       volume = {453},
       number = {1},
        pages = {976-1001},
          doi = {10.1093/mnras/stv1576},
archivePrefix = {arXiv},
       eprint = {1507.05967},
 primaryClass = {astro-ph.SR},
       adsurl = {https://ui.adsabs.harvard.edu/abs/2015MNRAS.453..976F}
}

@ARTICLE{Wichittanakom20,
       author = {{Wichittanakom}, C. and {Oudmaijer}, R.~D. and {Fairlamb}, J.~R. and {Mendigut{\'\i}a}, I. and {Vioque}, M. and {Ababakr}, K.~M.},
        title = "{The accretion rates and mechanisms of Herbig Ae/Be stars}",
      journal = {\mnras},
         year = 2020,
        month = mar,
       volume = {493},
       number = {1},
        pages = {234-249},
          doi = {10.1093/mnras/staa169},
archivePrefix = {arXiv},
       eprint = {2001.05971},
 primaryClass = {astro-ph.SR},
       adsurl = {https://ui.adsabs.harvard.edu/abs/2020MNRAS.493..234W}
}

@ARTICLE{Andrews18,
       author = {{Andrews}, Sean M. and {Huang}, Jane and {P{\'e}rez}, Laura M. and {Isella}, Andrea and {Dullemond}, Cornelis P. and {Kurtovic}, Nicol{\'a}s T. and {Guzm{\'a}n}, Viviana V. and {Carpenter}, John M. and {Wilner}, David J. and {Zhang}, Shangjia and {Zhu}, Zhaohuan and {Birnstiel}, Tilman and {Bai}, Xue-Ning and {Benisty}, Myriam and {Hughes}, A. Meredith and {{\"O}berg}, Karin I. and {Ricci}, Luca},
        title = "{The Disk Substructures at High Angular Resolution Project (DSHARP). I. Motivation, Sample, Calibration, and Overview}",
      journal = {\apjl},
         year = 2018,
        month = dec,
       volume = {869},
       number = {2},
          eid = {L41},
        pages = {L41},
          doi = {10.3847/2041-8213/aaf741},
archivePrefix = {arXiv},
       eprint = {1812.04040},
 primaryClass = {astro-ph.SR},
       adsurl = {https://ui.adsabs.harvard.edu/abs/2018ApJ...869L..41A}
}

@ARTICLE{Calvet02,
       author = {{Calvet}, Nuria and {D'Alessio}, Paola and {Hartmann}, Lee and {Wilner}, David and {Walsh}, Andrew and {Sitko}, Michael},
        title = "{Evidence for a Developing Gap in a 10 Myr Old Protoplanetary Disk}",
      journal = {\apj},
         year = 2002,
        month = apr,
       volume = {568},
       number = {2},
        pages = {1008-1016},
          doi = {10.1086/339061},
archivePrefix = {arXiv},
       eprint = {astro-ph/0201425},
 primaryClass = {astro-ph},
       adsurl = {https://ui.adsabs.harvard.edu/abs/2002ApJ...568.1008C}
}

@ARTICLE{Andrews16,
       author = {{Andrews}, Sean M. and {Wilner}, David J. and {Zhu}, Zhaohuan and {Birnstiel}, Tilman and {Carpenter}, John M. and {P{\'e}rez}, Laura M. and {Bai}, Xue-Ning and {{\"O}berg}, Karin I. and {Hughes}, A. Meredith and {Isella}, Andrea and {Ricci}, Luca},
        title = "{Ringed Substructure and a Gap at 1 au in the Nearest Protoplanetary Disk}",
      journal = {\apjl},
         year = 2016,
        month = apr,
       volume = {820},
       number = {2},
          eid = {L40},
        pages = {L40},
          doi = {10.3847/2041-8205/820/2/L40},
archivePrefix = {arXiv},
       eprint = {1603.09352},
 primaryClass = {astro-ph.EP},
       adsurl = {https://ui.adsabs.harvard.edu/abs/2016ApJ...820L..40A}
}

@ARTICLE{Kastner15,
       author = {{Kastner}, Joel H. and {Qi}, Chunhua and {Gorti}, Uma and {Hily-Blant}, Pierre and {Oberg}, Karin and {Forveille}, Thierry and {Andrews}, Sean and {Wilner}, David},
        title = "{A Ring of C$_{2}$H in the Molecular Disk Orbiting TW Hya}",
      journal = {\apj},
         year = 2015,
        month = jun,
       volume = {806},
       number = {1},
          eid = {75},
        pages = {75},
          doi = {10.1088/0004-637X/806/1/75},
archivePrefix = {arXiv},
       eprint = {1504.05980},
 primaryClass = {astro-ph.SR},
       adsurl = {https://ui.adsabs.harvard.edu/abs/2015ApJ...806...75K}
}

@ARTICLE{Herczeg23,
       author = {{Herczeg}, Gregory J. and {Chen}, Yuguang and {Donati}, Jean-Francois and {Dupree}, Andrea K. and {Walter}, Frederick M. and {Hillenbrand}, Lynne A. and {Johns-Krull}, Christopher M. and {Manara}, Carlo F. and {G{\"u}nther}, Hans Moritz and {Fang}, Min and {Schneider}, P. Christian and {Valenti}, Jeff A. and {Alencar}, Silvia H.~P. and {Venuti}, Laura and {Alcal{\'a}}, Juan Manuel and {Frasca}, Antonio and {Arulanantham}, Nicole and {Linsky}, Jeffrey L. and {Bouvier}, Jerome and {Brickhouse}, Nancy S. and {Calvet}, Nuria and {Espaillat}, Catherine C. and {Campbell-White}, Justyn and {Carpenter}, John M. and {Chang}, Seok-Jun and {Cruz}, Kelle L. and {Dahm}, S.~E. and {Eisl{\"o}ffel}, Jochen and {Edwards}, Suzan and {Fischer}, William J. and {Guo}, Zhen and {Henning}, Thomas and {Ji}, Tao and {Jose}, Jessy and {Kastner}, Joel H. and {Launhardt}, Ralf and {Principe}, David A. and {Robinson}, Connor E. and {Serna}, Javier and {Siwak}, Michal and {Sterzik}, Michael F. and {Takasao}, Shinsuke},
        title = "{Twenty-five Years of Accretion onto the Classical T Tauri Star TW Hya}",
      journal = {\apj},
         year = 2023,
        month = oct,
       volume = {956},
       number = {2},
          eid = {102},
        pages = {102},
          doi = {10.3847/1538-4357/acf468},
archivePrefix = {arXiv},
       eprint = {2308.14590},
 primaryClass = {astro-ph.SR},
       adsurl = {https://ui.adsabs.harvard.edu/abs/2023ApJ...956..102H}
}

@ARTICLE{Andrews20,
       author = {{Andrews}, Sean M.},
        title = "{Observations of Protoplanetary Disk Structures}",
      journal = {\araa},
         year = 2020,
        month = aug,
       volume = {58},
        pages = {483-528},
          doi = {10.1146/annurev-astro-031220-010302},
archivePrefix = {arXiv},
       eprint = {2001.05007},
 primaryClass = {astro-ph.EP},
       adsurl = {https://ui.adsabs.harvard.edu/abs/2020ARA&A..58..483A}
}

@ARTICLE{Kastner18,
       author = {{Kastner}, Joel H. and {Qi}, C. and {Dickson-Vandervelde}, D.~A. and {Hily-Blant}, P. and {Forveille}, T. and {Andrews}, S. and {Gorti}, U. and {{\"O}berg}, K. and {Wilner}, D.},
        title = "{A Subarcsecond ALMA Molecular Line Imaging Survey of the Circumbinary, Protoplanetary Disk Orbiting V4046 Sgr}",
      journal = {\apj},
         year = 2018,
        month = aug,
       volume = {863},
       number = {1},
          eid = {106},
        pages = {106},
          doi = {10.3847/1538-4357/aacff7},
archivePrefix = {arXiv},
       eprint = {1806.10553},
 primaryClass = {astro-ph.SR},
       adsurl = {https://ui.adsabs.harvard.edu/abs/2018ApJ...863..106K}
}

@ARTICLE{Law21_MAPSIII,
       author = {{Law}, Charles J. and {Loomis}, Ryan A. and {Teague}, Richard and {{\"O}berg}, Karin I. and {Czekala}, Ian and {Andrews}, Sean M. and {Huang}, Jane and {Aikawa}, Yuri and {Alarc{\'o}n}, Felipe and {Bae}, Jaehan and {Bergin}, Edwin A. and {Bergner}, Jennifer B. and {Boehler}, Yann and {Booth}, Alice S. and {Bosman}, Arthur D. and {Calahan}, Jenny K. and {Cataldi}, Gianni and {Cleeves}, L. Ilsedore and {Furuya}, Kenji and {Guzm{\'a}n}, Viviana V. and {Ilee}, John D. and {Le Gal}, Romane and {Liu}, Yao and {Long}, Feng and {M{\'e}nard}, Fran{\c{c}}ois and {Nomura}, Hideko and {Qi}, Chunhua and {Schwarz}, Kamber R. and {Sierra}, Anibal and {Tsukagoshi}, Takashi and {Yamato}, Yoshihide and {van't Hoff}, Merel L.~R. and {Walsh}, Catherine and {Wilner}, David J. and {Zhang}, Ke},
        title = "{Molecules with ALMA at Planet-forming Scales (MAPS). III. Characteristics of Radial Chemical Substructures}",
      journal = {\apjs},
         year = 2021,
        month = nov,
       volume = {257},
       number = {1},
          eid = {3},
        pages = {3},
          doi = {10.3847/1538-4365/ac1434},
archivePrefix = {arXiv},
       eprint = {2109.06210},
 primaryClass = {astro-ph.EP},
       adsurl = {https://ui.adsabs.harvard.edu/abs/2021ApJS..257....3L}
}

@ARTICLE{Simon00,
       author = {{Simon}, M. and {Dutrey}, A. and {Guilloteau}, S.},
        title = "{Dynamical Masses of T Tauri Stars and Calibration of Pre-Main-Sequence Evolution}",
      journal = {\apj},
         year = 2000,
        month = dec,
       volume = {545},
       number = {2},
        pages = {1034-1043},
          doi = {10.1086/317838},
archivePrefix = {arXiv},
       eprint = {astro-ph/0008370},
 primaryClass = {astro-ph},
       adsurl = {https://ui.adsabs.harvard.edu/abs/2000ApJ...545.1034S}
}

@ARTICLE{Dartois03,
       author = {{Dartois}, E. and {Dutrey}, A. and {Guilloteau}, S.},
        title = "{Structure of the DM Tau Outer Disk: Probing the vertical kinetic temperature gradient}",
      journal = {\aap},
         year = 2003,
        month = feb,
       volume = {399},
        pages = {773-787},
          doi = {10.1051/0004-6361:20021638},
       adsurl = {https://ui.adsabs.harvard.edu/abs/2003A&A...399..773D}
}

@ARTICLE{Pegues20,
       author = {{Pegues}, Jamila and {{\"O}berg}, Karin I. and {Bergner}, Jennifer B. and {Loomis}, Ryan A. and {Qi}, Chunhua and {Le Gal}, Romane and {Cleeves}, L. Ilsedore and {Guzm{\'a}n}, Viviana V. and {Huang}, Jane and {J{\o}rgensen}, Jes K. and {Andrews}, Sean M. and {Blake}, Geoffrey A. and {Carpenter}, John M. and {Schwarz}, Kamber R. and {Williams}, Jonathan P. and {Wilner}, David J.},
        title = "{An ALMA Survey of H$_{2}$CO in Protoplanetary Disks}",
      journal = {\apj},
         year = 2020,
        month = feb,
       volume = {890},
       number = {2},
          eid = {142},
        pages = {142},
          doi = {10.3847/1538-4357/ab64d9},
archivePrefix = {arXiv},
       eprint = {2002.12525},
 primaryClass = {astro-ph.SR},
       adsurl = {https://ui.adsabs.harvard.edu/abs/2020ApJ...890..142P}
}

@ARTICLE{Vera17,
       author = {{Hern{\'a}ndez Vera}, M. and {Lique}, F. and {Dumouchel}, F. and {Hily-Blant}, P. and {Faure}, A.},
        title = "{The rotational excitation of the HCN and HNC molecules by H$_{2}$ revisited}",
      journal = {\mnras},
         year = 2017,
        month = jun,
       volume = {468},
       number = {1},
        pages = {1084-1091},
          doi = {10.1093/mnras/stx422},
       adsurl = {https://ui.adsabs.harvard.edu/abs/2017MNRAS.468.1084H}
}

@ARTICLE{Faure07,
       author = {{Faure}, Alexandre and {Varambhia}, Hemal N. and {Stoecklin}, Thierry and {Tennyson}, Jonathan},
        title = "{Electron-impact rotational and hyperfine excitation of HCN, HNC, DCN and DNC}",
      journal = {\mnras},
         year = 2007,
        month = dec,
       volume = {382},
       number = {2},
        pages = {840-848},
          doi = {10.1111/j.1365-2966.2007.12416.x},
archivePrefix = {arXiv},
       eprint = {0709.1904},
 primaryClass = {astro-ph},
       adsurl = {https://ui.adsabs.harvard.edu/abs/2007MNRAS.382..840F}
}

@ARTICLE{Dumouchel10,
       author = {{Dumouchel}, F. and {Faure}, A. and {Lique}, F.},
        title = "{The rotational excitation of HCN and HNC by He: temperature dependence of the collisional rate coefficients}",
      journal = {\mnras},
         year = 2010,
        month = aug,
       volume = {406},
       number = {4},
        pages = {2488-2492},
          doi = {10.1111/j.1365-2966.2010.16826.x},
       adsurl = {https://ui.adsabs.harvard.edu/abs/2010MNRAS.406.2488D}
}

@ARTICLE{Brinch10,
       author = {{Brinch}, C. and {Hogerheijde}, M.~R.},
        title = "{LIME - a flexible, non-LTE line excitation and radiation transfer method for millimeter and far-infrared wavelengths}",
      journal = {\aap},
         year = 2010,
        month = nov,
       volume = {523},
          eid = {A25},
        pages = {A25},
          doi = {10.1051/0004-6361/201015333},
archivePrefix = {arXiv},
       eprint = {1008.1492},
 primaryClass = {astro-ph.SR},
       adsurl = {https://ui.adsabs.harvard.edu/abs/2010A&A...523A..25B}
}

@ARTICLE{Graninger14,
       author = {{Graninger}, Dawn M. and {Herbst}, Eric and {{\"O}berg}, Karin I. and {Vasyunin}, Anton I.},
        title = "{The HNC/HCN Ratio in Star-forming Regions}",
      journal = {\apj},
         year = 2014,
        month = may,
       volume = {787},
       number = {1},
          eid = {74},
        pages = {74},
          doi = {10.1088/0004-637X/787/1/74},
archivePrefix = {arXiv},
       eprint = {1404.5338},
 primaryClass = {astro-ph.GA},
       adsurl = {https://ui.adsabs.harvard.edu/abs/2014ApJ...787...74G}
}

@ARTICLE{Mendes12,
       author = {{Mendes}, Mario B. and {Buhr}, Henrik and {Berg}, Max H. and {Froese}, Michael and {Grieser}, Manfred and {Heber}, Oded and {Jordon-Thaden}, Brandon and {Krantz}, Claude and {Novotn{\'y}}, Old{\v{r}}ich and {Novotny}, Steffen and {Orlov}, Dmitry A. and {Petrignani}, Annemieke and {Rappaport}, Michael L. and {Repnow}, Roland and {Schwalm}, Dirk and {Shornikov}, Andrey and {St{\"u}tzel}, Julia and {Zajfman}, Daniel and {Wolf}, Andreas},
        title = "{Cold Electron Reactions Producing the Energetic Isomer of Hydrogen Cyanide in Interstellar Clouds}",
      journal = {\apjl},
         year = 2012,
        month = feb,
       volume = {746},
       number = {1},
          eid = {L8},
        pages = {L8},
          doi = {10.1088/2041-8205/746/1/L8},
       adsurl = {https://ui.adsabs.harvard.edu/abs/2012ApJ...746L...8M}
}

@ARTICLE{Schilke92,
       author = {{Schilke}, P. and {Walmsley}, C.~M. and {Pineau Des Forets}, G. and {Roueff}, E. and {Flower}, D.~R. and {Guilloteau}, S.},
        title = "{A study of HCN, HNC and their isotopometers in OMC-1. I. Abundances and chemistry.}",
      journal = {\aap},
         year = 1992,
        month = mar,
       volume = {256},
        pages = {595-612},
       adsurl = {https://ui.adsabs.harvard.edu/abs/1992A&A...256..595S}
}

@ARTICLE{Hebrard12,
       author = {{H{\'e}brard}, E. and {Dobrijevic}, M. and {Loison}, J.~C. and {Bergeat}, A. and {Hickson}, K.~M.},
        title = "{Neutral production of hydrogen isocyanide (HNC) and hydrogen cyanide (HCN) in Titan's upper atmosphere}",
      journal = {\aap},
         year = 2012,
        month = may,
       volume = {541},
          eid = {A21},
        pages = {A21},
          doi = {10.1051/0004-6361/201218837},
       adsurl = {https://ui.adsabs.harvard.edu/abs/2012A&A...541A..21H}
}

@ARTICLE{Preibisch05,
       author = {{Preibisch}, Thomas and {Kim}, Yong-Cheol and {Favata}, Fabio and {Feigelson}, Eric D. and {Flaccomio}, Ettore and {Getman}, Konstantin and {Micela}, Giusi and {Sciortino}, Salvatore and {Stassun}, Keivan and {Stelzer}, Beate and {Zinnecker}, Hans},
        title = "{The Origin of T Tauri X-Ray Emission: New Insights from the Chandra Orion Ultradeep Project}",
      journal = {\apjs},
         year = 2005,
        month = oct,
       volume = {160},
       number = {2},
        pages = {401-422},
          doi = {10.1086/432891},
archivePrefix = {arXiv},
       eprint = {astro-ph/0506526},
 primaryClass = {astro-ph},
       adsurl = {https://ui.adsabs.harvard.edu/abs/2005ApJS..160..401P}
}

@ARTICLE{Tielens85,
       author = {{Tielens}, A.~G.~G.~M. and {Hollenbach}, D.},
        title = "{Photodissociation regions. I. Basic model.}",
      journal = {\apj},
         year = 1985,
        month = apr,
       volume = {291},
        pages = {722-746},
          doi = {10.1086/163111},
       adsurl = {https://ui.adsabs.harvard.edu/abs/1985ApJ...291..722T}
}

@ARTICLE{Aikawa99,
       author = {{Aikawa}, Y. and {Herbst}, E.},
        title = "{Molecular evolution in protoplanetary disks. Two-dimensional distributions and column densities of gaseous molecules}",
      journal = {\aap},
         year = 1999,
        month = nov,
       volume = {351},
        pages = {233-246},
       adsurl = {https://ui.adsabs.harvard.edu/abs/1999A&A...351..233A}
}

@ARTICLE{Fogel11,
       author = {{Fogel}, Jeffrey K.~J. and {Bethell}, Thomas J. and {Bergin}, Edwin A. and {Calvet}, Nuria and {Semenov}, Dmitry},
        title = "{Chemistry of a Protoplanetary Disk with Grain Settling and Ly{\ensuremath{\alpha}} Radiation}",
      journal = {\apj},
         year = 2011,
        month = jan,
       volume = {726},
       number = {1},
          eid = {29},
        pages = {29},
          doi = {10.1088/0004-637X/726/1/29},
archivePrefix = {arXiv},
       eprint = {1011.0446},
 primaryClass = {astro-ph.SR},
       adsurl = {https://ui.adsabs.harvard.edu/abs/2011ApJ...726...29F}
}

@ARTICLE{Cleeves13,
       author = {{Cleeves}, L. Ilsedore and {Adams}, Fred C. and {Bergin}, Edwin A.},
        title = "{Exclusion of Cosmic Rays in Protoplanetary Disks: Stellar and Magnetic Effects}",
      journal = {\apj},
         year = 2013,
        month = jul,
       volume = {772},
       number = {1},
          eid = {5},
        pages = {5},
          doi = {10.1088/0004-637X/772/1/5},
archivePrefix = {arXiv},
       eprint = {1306.0902},
 primaryClass = {astro-ph.SR},
       adsurl = {https://ui.adsabs.harvard.edu/abs/2013ApJ...772....5C}
}

@ARTICLE{Aguado17,
       author = {{Aguado}, Alfredo and {Roncero}, Octavio and {Zanchet}, Alexandre and {Ag{\'u}ndez}, Marcelino and {Cernicharo}, Jos{\'e}},
        title = "{The Photodissociation of HCN and HNC: Effects on the HNC/HCN Abundance Ratio in the Interstellar Medium}",
      journal = {\apj},
         year = 2017,
        month = mar,
       volume = {838},
       number = {1},
          eid = {33},
        pages = {33},
          doi = {10.3847/1538-4357/aa63ee},
       adsurl = {https://ui.adsabs.harvard.edu/abs/2017ApJ...838...33A}
}

@ARTICLE{Lee24,
       author = {{Lee}, Jeong-Eun and {Kim}, Chul-Hwan and {Lee}, Seokho and {Lee}, Seonjae and {Baek}, Giseon and {Yun}, Hyeong-Sik and {Aikawa}, Yuri and {Johnstone}, Doug and {Herczeg}, Gregory J. and {Cieza}, Lucas},
        title = "{ALMA Spectral Survey of an Eruptive Young Star, V883 Ori (ASSAY). I. What Triggered the Current Episode of Eruption?}",
      journal = {\apj},
         year = 2024,
        month = may,
       volume = {966},
       number = {1},
          eid = {119},
        pages = {119},
          doi = {10.3847/1538-4357/ad3106},
archivePrefix = {arXiv},
       eprint = {2403.03436},
 primaryClass = {astro-ph.SR},
       adsurl = {https://ui.adsabs.harvard.edu/abs/2024ApJ...966..119L}
}

@ARTICLE{Yoshida22,
       author = {{Yoshida}, Tomohiro C. and {Nomura}, Hideko and {Tsukagoshi}, Takashi and {Furuya}, Kenji and {Ueda}, Takahiro},
        title = "{Discovery of Line Pressure Broadening and Direct Constraint on Gas Surface Density in a Protoplanetary Disk}",
      journal = {\apjl},
         year = 2022,
        month = sep,
       volume = {937},
       number = {1},
          eid = {L14},
        pages = {L14},
          doi = {10.3847/2041-8213/ac903a},
archivePrefix = {arXiv},
       eprint = {2209.03367},
 primaryClass = {astro-ph.EP},
       adsurl = {https://ui.adsabs.harvard.edu/abs/2022ApJ...937L..14Y}
}

@ARTICLE{Kelly07,
       author = {{Kelly}, Brandon C.},
        title = "{Some Aspects of Measurement Error in Linear Regression of Astronomical Data}",
      journal = {\apj},
         year = 2007,
        month = aug,
       volume = {665},
       number = {2},
        pages = {1489-1506},
          doi = {10.1086/519947},
archivePrefix = {arXiv},
       eprint = {0705.2774},
 primaryClass = {astro-ph},
       adsurl = {https://ui.adsabs.harvard.edu/abs/2007ApJ...665.1489K}
}

@ARTICLE{Kastner14,
       author = {{Kastner}, Joel H. and {Hily-Blant}, Pierre and {Rodriguez}, David R. and {Punzi}, Kristina and {Forveille}, Thierry},
        title = "{Unbiased Millimeter-wave Line Surveys of TW Hya and V4046 Sgr: The Enhanced C$_{2}$H and CN Abundances of Evolved Protoplanetary Disks}",
      journal = {\apj},
         year = 2014,
        month = sep,
       volume = {793},
       number = {1},
          eid = {55},
        pages = {55},
          doi = {10.1088/0004-637X/793/1/55},
archivePrefix = {arXiv},
       eprint = {1408.5918},
 primaryClass = {astro-ph.SR},
       adsurl = {https://ui.adsabs.harvard.edu/abs/2014ApJ...793...55K}
}

@ARTICLE{Guzman15,
       author = {{Guzm{\'a}n}, V.~V. and {{\"O}berg}, K.~I. and {Loomis}, R. and {Qi}, C.},
        title = "{Cyanide Photochemistry and Nitrogen Fractionation in the MWC 480 Disk}",
      journal = {\apj},
         year = 2015,
        month = nov,
       volume = {814},
       number = {1},
          eid = {53},
        pages = {53},
          doi = {10.1088/0004-637X/814/1/53},
archivePrefix = {arXiv},
       eprint = {1511.03313},
 primaryClass = {astro-ph.SR},
       adsurl = {https://ui.adsabs.harvard.edu/abs/2015ApJ...814...53G}
}

@ARTICLE{Guilloteau16,
       author = {{Guilloteau}, S. and {Reboussin}, L. and {Dutrey}, A. and {Chapillon}, E. and {Wakelam}, V. and {Pi{\'e}tu}, V. and {Di Folco}, E. and {Semenov}, D. and {Henning}, Th.},
        title = "{Chemistry in disks. X. The molecular content of protoplanetary disks in Taurus}",
      journal = {\aap},
         year = 2016,
        month = aug,
       volume = {592},
          eid = {A124},
        pages = {A124},
          doi = {10.1051/0004-6361/201527088},
archivePrefix = {arXiv},
       eprint = {1604.05028},
 primaryClass = {astro-ph.EP},
       adsurl = {https://ui.adsabs.harvard.edu/abs/2016A&A...592A.124G}
}

@ARTICLE{Oberg10,
       author = {{{\"O}berg}, Karin I. and {Qi}, Chunhua and {Fogel}, Jeffrey K.~J. and {Bergin}, Edwin A. and {Andrews}, Sean M. and {Espaillat}, Catherine and {van Kempen}, Tim A. and {Wilner}, David J. and {Pascucci}, Ilaria},
        title = "{The Disk Imaging Survey of Chemistry with SMA. I. Taurus Protoplanetary Disk Data}",
      journal = {\apj},
         year = 2010,
        month = sep,
       volume = {720},
       number = {1},
        pages = {480-493},
          doi = {10.1088/0004-637X/720/1/480},
archivePrefix = {arXiv},
       eprint = {1007.1476},
 primaryClass = {astro-ph.GA},
       adsurl = {https://ui.adsabs.harvard.edu/abs/2010ApJ...720..480O}
}

@ARTICLE{Terwisga19,
       author = {{van Terwisga}, S.~E. and {van Dishoeck}, E.~F. and {Cazzoletti}, P. and {Facchini}, S. and {Trapman}, L. and {Williams}, J.~P. and {Manara}, C.~F. and {Miotello}, A. and {van der Marel}, N. and {Ansdell}, M. and {Hogerheijde}, M.~R. and {Tazzari}, M. and {Testi}, L.},
        title = "{The ALMA Lupus protoplanetary disk survey: evidence for compact gas disks and molecular rings from CN}",
      journal = {\aap},
         year = 2019,
        month = mar,
       volume = {623},
          eid = {A150},
        pages = {A150},
          doi = {10.1051/0004-6361/201834257},
archivePrefix = {arXiv},
       eprint = {1811.03071},
 primaryClass = {astro-ph.SR},
       adsurl = {https://ui.adsabs.harvard.edu/abs/2019A&A...623A.150V}
}

@ARTICLE{Thi04,
       author = {{Thi}, W. -F. and {van Zadelhoff}, G. -J. and {van Dishoeck}, E.~F.},
        title = "{Organic molecules in protoplanetary disks around T Tauri and Herbig Ae stars}",
      journal = {\aap},
         year = 2004,
        month = oct,
       volume = {425},
        pages = {955-972},
          doi = {10.1051/0004-6361:200400026},
archivePrefix = {arXiv},
       eprint = {astro-ph/0406577},
 primaryClass = {astro-ph},
       adsurl = {https://ui.adsabs.harvard.edu/abs/2004A&A...425..955T}
}

@ARTICLE{Chapillon12,
       author = {{Chapillon}, E. and {Guilloteau}, S. and {Dutrey}, A. and {Pi{\'e}tu}, V. and {Gu{\'e}lin}, M.},
        title = "{Chemistry in disks. VI. CN and HCN in protoplanetary disks}",
      journal = {\aap},
         year = 2012,
        month = jan,
       volume = {537},
          eid = {A60},
        pages = {A60},
          doi = {10.1051/0004-6361/201116762},
archivePrefix = {arXiv},
       eprint = {1109.5595},
 primaryClass = {astro-ph.GA},
       adsurl = {https://ui.adsabs.harvard.edu/abs/2012A&A...537A..60C}
}

@ARTICLE{vanderMarel18,
       author = {{van der Marel}, Nienke and {Williams}, Jonathan P. and {Ansdell}, M. and {Manara}, Carlo F. and {Miotello}, Anna and {Tazzari}, Marco and {Testi}, Leonardo and {Hogerheijde}, Michiel and {Bruderer}, Simon and {van Terwisga}, Sierk E. and {van Dishoeck}, Ewine F.},
        title = "{New Insights into the Nature of Transition Disks from a Complete Disk Survey of the Lupus Star-forming Region}",
      journal = {\apj},
         year = 2018,
        month = feb,
       volume = {854},
       number = {2},
          eid = {177},
        pages = {177},
          doi = {10.3847/1538-4357/aaaa6b},
archivePrefix = {arXiv},
       eprint = {1801.06154},
 primaryClass = {astro-ph.EP},
       adsurl = {https://ui.adsabs.harvard.edu/abs/2018ApJ...854..177V}
}

@ARTICLE{Visser18,
       author = {{Visser}, Ruud and {Bruderer}, Simon and {Cazzoletti}, Paolo and {Facchini}, Stefano and {Heays}, Alan N. and {van Dishoeck}, Ewine F.},
        title = "{Nitrogen isotope fractionation in protoplanetary disks}",
      journal = {\aap},
         year = 2018,
        month = jul,
       volume = {615},
          eid = {A75},
        pages = {A75},
          doi = {10.1051/0004-6361/201731898},
archivePrefix = {arXiv},
       eprint = {1802.02841},
 primaryClass = {astro-ph.SR},
       adsurl = {https://ui.adsabs.harvard.edu/abs/2018A&A...615A..75V}
}

@ARTICLE{SantaMaria23,
       author = {{Santa-Maria}, M.~G. and {Goicoechea}, J.~R. and {Pety}, J. and {Gerin}, M. and {Orkisz}, J.~H. and {Le Petit}, F. and {Einig}, L. and {Palud}, P. and {de Souza Magalhaes}, V. and {Be{\v{s}}li{\'c}}, I. and {Segal}, L. and {Bardeau}, S. and {Bron}, E. and {Chainais}, P. and {Chanussot}, J. and {Gratier}, P. and {Guzm{\'a}n}, V.~V. and {Hughes}, A. and {Languignon}, D. and {Levrier}, F. and {Lis}, D.~C. and {Liszt}, H.~S. and {Le Bourlot}, J. and {Oya}, Y. and {{\"O}berg}, K. and {Peretto}, N. and {Roueff}, E. and {Roueff}, A. and {Sievers}, A. and {Thouvenin}, P. -A. and {Yamamoto}, S.},
        title = "{HCN emission from translucent gas and UV-illuminated cloud edges revealed by wide-field IRAM 30 m maps of the Orion B GMC. Revisiting its role as a tracer of the dense gas reservoir for star formation}",
      journal = {\aap},
         year = 2023,
        month = nov,
       volume = {679},
          eid = {A4},
        pages = {A4},
          doi = {10.1051/0004-6361/202346598},
archivePrefix = {arXiv},
       eprint = {2309.03186},
 primaryClass = {astro-ph.GA},
       adsurl = {https://ui.adsabs.harvard.edu/abs/2023A&A...679A...4S}
}

@ARTICLE{Bergin16,
       author = {{Bergin}, Edwin A. and {Du}, Fujun and {Cleeves}, L. Ilsedore and {Blake}, G.~A. and {Schwarz}, K. and {Visser}, R. and {Zhang}, K.},
        title = "{Hydrocarbon Emission Rings in Protoplanetary Disks Induced by Dust Evolution}",
      journal = {\apj},
         year = 2016,
        month = nov,
       volume = {831},
       number = {1},
          eid = {101},
        pages = {101},
          doi = {10.3847/0004-637X/831/1/101},
archivePrefix = {arXiv},
       eprint = {1609.06337},
 primaryClass = {astro-ph.EP},
       adsurl = {https://ui.adsabs.harvard.edu/abs/2016ApJ...831..101B}
}

@ARTICLE{Oberg15,
       author = {{{\"O}berg}, Karin I. and {Guzm{\'a}n}, Viviana V. and {Furuya}, Kenji and {Qi}, Chunhua and {Aikawa}, Yuri and {Andrews}, Sean M. and {Loomis}, Ryan and {Wilner}, David J.},
        title = "{The comet-like composition of a protoplanetary disk as revealed by complex cyanides}",
      journal = {\nat},
         year = 2015,
        month = apr,
       volume = {520},
       number = {7546},
        pages = {198-201},
          doi = {10.1038/nature14276},
archivePrefix = {arXiv},
       eprint = {1505.06347},
 primaryClass = {astro-ph.GA},
       adsurl = {https://ui.adsabs.harvard.edu/abs/2015Natur.520..198O}
}

@ARTICLE{Calahan23,
       author = {{Calahan}, Jenny K. and {Bergin}, Edwin A. and {Bosman}, Arthur D. and {Rich}, Evan A. and {Andrews}, Sean M. and {Bergner}, Jennifer B. and {Cleeves}, L. Ilsedore and {Guzm{\'a}n}, Viviana V. and {Huang}, Jane and {Ilee}, John D. and {Law}, Charles J. and {Le Gal}, Romane and {{\"O}berg}, Karin I. and {Teague}, Richard and {Walsh}, Catherine and {Wilner}, David J. and {Zhang}, Ke},
        title = "{UV-driven chemistry as a signpost of late-stage planet formation}",
      journal = {Nature Astronomy},
         year = 2023,
        month = jan,
       volume = {7},
        pages = {49-56},
          doi = {10.1038/s41550-022-01831-8},
archivePrefix = {arXiv},
       eprint = {2212.05539},
 primaryClass = {astro-ph.EP},
       adsurl = {https://ui.adsabs.harvard.edu/abs/2023NatAs...7...49C}
}

@ARTICLE{Marston89,
       author = {{Marston}, G. and {Nesbitt}, F.~L. and {Stief}, L.~J.},
        title = "{Branching ratios in the N + CH3 reaction - Formation of the methylene amidogen (H2CN) radical}",
      journal = {\jcp},
         year = 1989,
        month = sep,
       volume = {91},
        pages = {3483-3491},
          doi = {10.1063/1.456878},
       adsurl = {https://ui.adsabs.harvard.edu/abs/1989JChPh..91.3483M}
}

@ARTICLE{Ilee21,
       author = {{Ilee}, John D. and {Walsh}, Catherine and {Booth}, Alice S. and {Aikawa}, Yuri and {Andrews}, Sean M. and {Bae}, Jaehan and {Bergin}, Edwin A. and {Bergner}, Jennifer B. and {Bosman}, Arthur D. and {Cataldi}, Gianni and {Cleeves}, L. Ilsedore and {Czekala}, Ian and {Guzm{\'a}n}, Viviana V. and {Huang}, Jane and {Law}, Charles J. and {Le Gal}, Romane and {Loomis}, Ryan A. and {M{\'e}nard}, Fran{\c{c}}ois and {Nomura}, Hideko and {{\"O}berg}, Karin I. and {Qi}, Chunhua and {Schwarz}, Kamber R. and {Teague}, Richard and {Tsukagoshi}, Takashi and {Wilner}, David J. and {Yamato}, Yoshihide and {Zhang}, Ke},
        title = "{Molecules with ALMA at Planet-forming Scales (MAPS). IX. Distribution and Properties of the Large Organic Molecules HC$_{3}$N, CH$_{3}$CN, and c-C$_{3}$H$_{2}$}",
      journal = {\apjs},
         year = 2021,
        month = nov,
       volume = {257},
       number = {1},
          eid = {9},
        pages = {9},
          doi = {10.3847/1538-4365/ac1441},
archivePrefix = {arXiv},
       eprint = {2109.06319},
 primaryClass = {astro-ph.EP},
       adsurl = {https://ui.adsabs.harvard.edu/abs/2021ApJS..257....9I}
}

@ARTICLE{Aikawa21,
       author = {{Aikawa}, Yuri and {Cataldi}, Gianni and {Yamato}, Yoshihide and {Zhang}, Ke and {Booth}, Alice S. and {Furuya}, Kenji and {Andrews}, Sean M. and {Bae}, Jaehan and {Bergin}, Edwin A. and {Bergner}, Jennifer B. and {Bosman}, Arthur D. and {Cleeves}, L. Ilsedore and {Czekala}, Ian and {Guzm{\'a}n}, Viviana V. and {Huang}, Jane and {Ilee}, John D. and {Law}, Charles J. and {Le Gal}, Romane and {Loomis}, Ryan A. and {M{\'e}nard}, Fran{\c{c}}ois and {Nomura}, Hideko and {{\"O}berg}, Karin I. and {Qi}, Chunhua and {Schwarz}, Kamber R. and {Teague}, Richard and {Tsukagoshi}, Takashi and {Walsh}, Catherine and {Wilner}, David J.},
        title = "{Molecules with ALMA at Planet-forming Scales (MAPS). XIII. HCO$^{+}$ and Disk Ionization Structure}",
      journal = {\apjs},
         year = 2021,
        month = nov,
       volume = {257},
       number = {1},
          eid = {13},
        pages = {13},
          doi = {10.3847/1538-4365/ac143c},
archivePrefix = {arXiv},
       eprint = {2109.06419},
 primaryClass = {astro-ph.SR},
       adsurl = {https://ui.adsabs.harvard.edu/abs/2021ApJS..257...13A}
}

@ARTICLE{Martinez22,
       author = {{Martinez-Brunner}, Rafael and {Casassus}, Simon and {P{\'e}rez}, Sebasti{\'a}n and {Hales}, Antonio and {Weber}, Philipp and {C{\'a}rcamo}, Miguel and {Arce-Tord}, Carla and {Cieza}, Lucas and {Garufi}, Antonio and {Marino}, Sebasti{\'a}n and {Zurlo}, Alice},
        title = "{High-resolution ALMA observations of V4046 Sgr: a circumbinary disc with a thin ring}",
      journal = {\mnras},
         year = 2022,
        month = feb,
       volume = {510},
       number = {1},
        pages = {1248-1257},
          doi = {10.1093/mnras/stab3440},
archivePrefix = {arXiv},
       eprint = {2111.12668},
 primaryClass = {astro-ph.EP},
       adsurl = {https://ui.adsabs.harvard.edu/abs/2022MNRAS.510.1248M}
}

@ARTICLE{Bergner19,
       author = {{Bergner}, Jennifer B. and {{\"O}berg}, Karin I. and {Bergin}, Edwin A. and {Loomis}, Ryan A. and {Pegues}, Jamila and {Qi}, Chunhua},
        title = "{A Survey of C$_{2}$H, HCN, and C$^{18}$O in Protoplanetary Disks}",
      journal = {\apj},
         year = 2019,
        month = may,
       volume = {876},
       number = {1},
          eid = {25},
        pages = {25},
          doi = {10.3847/1538-4357/ab141e},
archivePrefix = {arXiv},
       eprint = {1904.09315},
 primaryClass = {astro-ph.EP},
       adsurl = {https://ui.adsabs.harvard.edu/abs/2019ApJ...876...25B}
}

@ARTICLE{Bergner21,
       author = {{Bergner}, Jennifer B. and {{\"O}berg}, Karin I. and {Guzm{\'a}n}, Viviana V. and {Law}, Charles J. and {Loomis}, Ryan A. and {Cataldi}, Gianni and {Bosman}, Arthur D. and {Aikawa}, Yuri and {Andrews}, Sean M. and {Bergin}, Edwin A. and {Booth}, Alice S. and {Cleeves}, L. Ilsedore and {Czekala}, Ian and {Huang}, Jane and {Ilee}, John D. and {Le Gal}, Romane and {Long}, Feng and {Nomura}, Hideko and {M{\'e}nard}, Fran{\c{c}}ois and {Qi}, Chunhua and {Schwarz}, Kamber R. and {Teague}, Richard and {Tsukagoshi}, Takashi and {Walsh}, Catherine and {Wilner}, David J. and {Yamato}, Yoshihide},
        title = "{Molecules with ALMA at Planet-forming Scales (MAPS). XI. CN and HCN as Tracers of Photochemistry in Disks}",
      journal = {\apjs},
         year = 2021,
        month = nov,
       volume = {257},
       number = {1},
          eid = {11},
        pages = {11},
          doi = {10.3847/1538-4365/ac143a},
archivePrefix = {arXiv},
       eprint = {2109.06694},
 primaryClass = {astro-ph.SR},
       adsurl = {https://ui.adsabs.harvard.edu/abs/2021ApJS..257...11B}
}

@ARTICLE{Riaz18,
       author = {{Riaz}, B. and {Thi}, W. -F. and {Caselli}, P.},
        title = "{Chemical tracers in proto-brown dwarfs: CN, HCN, and HNC observations}",
      journal = {\mnras},
         year = 2018,
        month = dec,
       volume = {481},
       number = {4},
        pages = {4662-4679},
          doi = {10.1093/mnras/sty2583},
archivePrefix = {arXiv},
       eprint = {1809.10164},
 primaryClass = {astro-ph.SR},
       adsurl = {https://ui.adsabs.harvard.edu/abs/2018MNRAS.481.4662R}
}

@ARTICLE{Paneque24,
       author = {{Paneque-Carre{\~n}o}, T. and {Izquierdo}, A.~F. and {Teague}, R. and {Miotello}, A. and {Bergin}, E.~A. and {Loomis}, R. and {van Dishoeck}, E.~F.},
        title = "{High turbulence in the IM Lup protoplanetary disk. Direct observational constraints from CN and C$_{2}$H emission}",
      journal = {\aap},
         year = 2024,
        month = apr,
       volume = {684},
          eid = {A174},
        pages = {A174},
          doi = {10.1051/0004-6361/202347757},
archivePrefix = {arXiv},
       eprint = {2312.04618},
 primaryClass = {astro-ph.EP},
       adsurl = {https://ui.adsabs.harvard.edu/abs/2024A&A...684A.174P}
}

@ARTICLE{Oberg21,
       author = {{{\"O}berg}, Karin I. and {Guzm{\'a}n}, Viviana V. and {Walsh}, Catherine and {Aikawa}, Yuri and {Bergin}, Edwin A. and {Law}, Charles J. and {Loomis}, Ryan A. and {Alarc{\'o}n}, Felipe and {Andrews}, Sean M. and {Bae}, Jaehan and {Bergner}, Jennifer B. and {Boehler}, Yann and {Booth}, Alice S. and {Bosman}, Arthur D. and {Calahan}, Jenny K. and {Cataldi}, Gianni and {Cleeves}, L. Ilsedore and {Czekala}, Ian and {Furuya}, Kenji and {Huang}, Jane and {Ilee}, John D. and {Kurtovic}, Nicolas T. and {Le Gal}, Romane and {Liu}, Yao and {Long}, Feng and {M{\'e}nard}, Fran{\c{c}}ois and {Nomura}, Hideko and {P{\'e}rez}, Laura M. and {Qi}, Chunhua and {Schwarz}, Kamber R. and {Sierra}, Anibal and {Teague}, Richard and {Tsukagoshi}, Takashi and {Yamato}, Yoshihide and {van't Hoff}, Merel L.~R. and {Waggoner}, Abygail R. and {Wilner}, David J. and {Zhang}, Ke},
        title = "{Molecules with ALMA at Planet-forming Scales (MAPS). I. Program Overview and Highlights}",
      journal = {\apjs},
         year = 2021,
        month = nov,
       volume = {257},
       number = {1},
          eid = {1},
        pages = {1},
          doi = {10.3847/1538-4365/ac1432},
archivePrefix = {arXiv},
       eprint = {2109.06268},
 primaryClass = {astro-ph.EP},
       adsurl = {https://ui.adsabs.harvard.edu/abs/2021ApJS..257....1O}
}

@ARTICLE{Graninger15,
       author = {{Graninger}, Dawn and {{\"O}berg}, Karin I. and {Qi}, Chunhua and {Kastner}, Joel},
        title = "{HNC in Protoplanetary Disks}",
      journal = {\apjl},
         year = 2015,
        month = jul,
       volume = {807},
       number = {1},
          eid = {L15},
        pages = {L15},
          doi = {10.1088/2041-8205/807/1/L15},
archivePrefix = {arXiv},
       eprint = {1506.03820},
 primaryClass = {astro-ph.SR},
       adsurl = {https://ui.adsabs.harvard.edu/abs/2015ApJ...807L..15G}
}

@ARTICLE{Rosenfeld12,
       author = {{Rosenfeld}, Katherine A. and {Andrews}, Sean M. and {Wilner}, David J. and {Stempels}, H.~C.},
        title = "{A Disk-based Dynamical Mass Estimate for the Young Binary V4046 Sgr}",
      journal = {\apj},
         year = 2012,
        month = nov,
       volume = {759},
       number = {2},
          eid = {119},
        pages = {119},
          doi = {10.1088/0004-637X/759/2/119},
archivePrefix = {arXiv},
       eprint = {1209.4407},
 primaryClass = {astro-ph.SR},
       adsurl = {https://ui.adsabs.harvard.edu/abs/2012ApJ...759..119R}
}

@ARTICLE{Long24,
       author = {{Long}, Deryl E. and {Cleeves}, L. Ilsedore and {Adams}, Fred C. and {Andrews}, Sean and {Bergin}, Edwin A. and {Guzm{\'a}n}, Viviana V. and {Huang}, Jane and {Hughes}, A. Meredith and {Qi}, Chunhua and {Schwarz}, Kamber and {Simon}, Jacob B. and {Wilner}, David},
        title = "{Exploring the Complex Ionization Environment of the Turbulent DM Tau Disk}",
      journal = {\apj},
         year = 2024,
        month = sep,
       volume = {972},
       number = {1},
          eid = {88},
        pages = {88},
          doi = {10.3847/1538-4357/ad5c67},
archivePrefix = {arXiv},
       eprint = {2406.18657},
 primaryClass = {astro-ph.SR},
       adsurl = {https://ui.adsabs.harvard.edu/abs/2024ApJ...972...88L}
}

@ARTICLE{Facchini20,
       author = {{Facchini}, S. and {Benisty}, M. and {Bae}, J. and {Loomis}, R. and {Perez}, L. and {Ansdell}, M. and {Mayama}, S. and {Pinilla}, P. and {Teague}, R. and {Isella}, A. and {Mann}, A.},
        title = "{Annular substructures in the transition disks around LkCa 15 and J1610}",
      journal = {\aap},
         year = 2020,
        month = jul,
       volume = {639},
          eid = {A121},
        pages = {A121},
          doi = {10.1051/0004-6361/202038027},
archivePrefix = {arXiv},
       eprint = {2005.02712},
 primaryClass = {astro-ph.EP},
       adsurl = {https://ui.adsabs.harvard.edu/abs/2020A&A...639A.121F}
}

@ARTICLE{White99,
       author = {{White}, Russel J. and {Ghez}, A.~M. and {Reid}, I. Neill and {Schultz}, Greg},
        title = "{A Test of Pre-Main-Sequence Evolutionary Models across the Stellar/Substellar Boundary Based on Spectra of the Young Quadruple GG Tauri}",
      journal = {\apj},
         year = 1999,
        month = aug,
       volume = {520},
       number = {2},
        pages = {811-821},
          doi = {10.1086/307494},
archivePrefix = {arXiv},
       eprint = {astro-ph/9902318},
 primaryClass = {astro-ph},
       adsurl = {https://ui.adsabs.harvard.edu/abs/1999ApJ...520..811W}
}

@ARTICLE{Long22,
       author = {{Long}, Feng and {Andrews}, Sean M. and {Zhang}, Shangjia and {Qi}, Chunhua and {Benisty}, Myriam and {Facchini}, Stefano and {Isella}, Andrea and {Wilner}, David J. and {Bae}, Jaehan and {Huang}, Jane and {Loomis}, Ryan A. and {{\"O}berg}, Karin I. and {Zhu}, Zhaohuan},
        title = "{ALMA Detection of Dust Trapping around Lagrangian Points in the LkCa 15 Disk}",
      journal = {\apjl},
         year = 2022,
        month = sep,
       volume = {937},
       number = {1},
          eid = {L1},
        pages = {L1},
          doi = {10.3847/2041-8213/ac8b10},
archivePrefix = {arXiv},
       eprint = {2209.05535},
 primaryClass = {astro-ph.EP},
       adsurl = {https://ui.adsabs.harvard.edu/abs/2022ApJ...937L...1L}
}

@ARTICLE{Long21,
       author = {{Long}, Feng and {Bosman}, Arthur D. and {Cazzoletti}, Paolo and {van Dishoeck}, Ewine F. and {{\"O}berg}, Karin I. and {Facchini}, Stefano and {Tazzari}, Marco and {Guzm{\'a}n}, Viviana V. and {Testi}, Leonardo},
        title = "{Exploring HNC and HCN line emission as probes of the protoplanetary disk temperature}",
      journal = {\aap},
         year = 2021,
        month = mar,
       volume = {647},
          eid = {A118},
        pages = {A118},
          doi = {10.1051/0004-6361/202039336},
archivePrefix = {arXiv},
       eprint = {2102.06338},
 primaryClass = {astro-ph.EP},
       adsurl = {https://ui.adsabs.harvard.edu/abs/2021A&A...647A.118L}
}

@ARTICLE{Loison14,
       author = {{Loison}, Jean-Christophe and {Wakelam}, Valentine and {Hickson}, Kevin M.},
        title = "{The interstellar gas-phase chemistry of HCN and HNC}",
      journal = {\mnras},
         year = 2014,
        month = sep,
       volume = {443},
       number = {1},
        pages = {398-410},
          doi = {10.1093/mnras/stu1089},
archivePrefix = {arXiv},
       eprint = {1406.1696},
 primaryClass = {astro-ph.GA},
       adsurl = {https://ui.adsabs.harvard.edu/abs/2014MNRAS.443..398L}
}

@ARTICLE{Tasa25,
       author = {{Tasa-Chaveli}, A. and {Fuente}, A. and {Esplugues}, G. and {Navarro-Almaida}, D. and {Majumdar}, L. and {Rayalacheruvu}, P. and {Rivi{\`e}re-Marichalar}, P. and {Rodr{\'\i}guez-Baras}, M.},
        title = "{Gas phase Elemental abundances in Molecular cloudS (GEMS): XI. The evolution of HCN, HNC, and N$_{2}$H$^{+}$ isotopic ratios in starless cores}",
      journal = {\aap},
         year = 2025,
        month = aug,
       volume = {700},
          eid = {A226},
        pages = {A226},
          doi = {10.1051/0004-6361/202554121},
archivePrefix = {arXiv},
       eprint = {2507.10380},
 primaryClass = {astro-ph.GA},
       adsurl = {https://ui.adsabs.harvard.edu/abs/2025A&A...700A.226T}
}

@ARTICLE{Francis22,
       author = {{Francis}, Logan and {Marel}, Nienke van der and {Johnstone}, Doug and {Akiyama}, Eiji and {Bruderer}, Simon and {Dong}, Ruobing and {Hashimoto}, Jun and {Liu}, Hauyu Baobab and {Muto}, Takayuki and {Yang}, Yi},
        title = "{Gap Opening and Inner Disk Structure in the Strongly Accreting Transition Disk of DM Tau}",
      journal = {\aj},
         year = 2022,
        month = sep,
       volume = {164},
       number = {3},
          eid = {105},
        pages = {105},
          doi = {10.3847/1538-3881/ac7ffb},
archivePrefix = {arXiv},
       eprint = {2208.01598},
 primaryClass = {astro-ph.EP},
       adsurl = {https://ui.adsabs.harvard.edu/abs/2022AJ....164..105F}
}

@ARTICLE{Semaniak01,
       author = {{Semaniak}, J. and {Minaev}, B.~F. and {Derkatch}, A.~M. and {Hellberg}, F. and {Neau}, A. and {Ros{\'e}n}, S. and {Thomas}, R. and {Larsson}, M. and {Danared}, H. and {Pa{\'a}l}, A. and {af Ugglas}, M.},
        title = "{Dissociative Recombination of HCNH$^{+}$: Absolute Cross-Sections and Branching Ratios}",
      journal = {\apjs},
         year = 2001,
        month = aug,
       volume = {135},
       number = {2},
        pages = {275-283},
          doi = {10.1086/321797},
       adsurl = {https://ui.adsabs.harvard.edu/abs/2001ApJS..135..275S}
}

@ARTICLE{Agundez18,
       author = {{Ag{\'u}ndez}, Marcelino and {Roueff}, Evelyne and {Le Petit}, Franck and {Le Bourlot}, Jacques},
        title = "{The chemistry of disks around T Tauri and Herbig Ae/Be stars}",
      journal = {\aap},
         year = 2018,
        month = aug,
       volume = {616},
          eid = {A19},
        pages = {A19},
          doi = {10.1051/0004-6361/201732518},
archivePrefix = {arXiv},
       eprint = {1803.09450},
 primaryClass = {astro-ph.GA},
       adsurl = {https://ui.adsabs.harvard.edu/abs/2018A&A...616A..19A}
}

@ARTICLE{Bruderer12,
       author = {{Bruderer}, S. and {van Dishoeck}, E.~F. and {Doty}, S.~D. and {Herczeg}, G.~J.},
        title = "{The warm gas atmosphere of the HD 100546 disk seen by Herschel. Evidence of a gas-rich, carbon-poor atmosphere?}",
      journal = {\aap},
         year = 2012,
        month = may,
       volume = {541},
          eid = {A91},
        pages = {A91},
          doi = {10.1051/0004-6361/201118218},
archivePrefix = {arXiv},
       eprint = {1201.4860},
 primaryClass = {astro-ph.SR},
       adsurl = {https://ui.adsabs.harvard.edu/abs/2012A&A...541A..91B}
}

@ARTICLE{Bruderer13,
       author = {{Bruderer}, Simon},
        title = "{Survival of molecular gas in cavities of transition disks. I. CO}",
      journal = {\aap},
         year = 2013,
        month = nov,
       volume = {559},
          eid = {A46},
        pages = {A46},
          doi = {10.1051/0004-6361/201321171},
archivePrefix = {arXiv},
       eprint = {1308.2966},
 primaryClass = {astro-ph.SR},
       adsurl = {https://ui.adsabs.harvard.edu/abs/2013A&A...559A..46B}
}

@misc{teague_bettermoments_2018,
	title = {Bettermoments: {A} {Robust} {Method} {To} {Measure} {Line} {Centroids}},
	publisher = {Zenodo},
	author = {Teague, Richard and Foreman-Mackey, Daniel},
	month = sep,
	year = {2018},
	doi = {10.5281/zenodo.1419754},
	note = {Version Number: v1.0},
}

@ARTICLE{Flaherty20,
       author = {{Flaherty}, Kevin and {Hughes}, A. Meredith and {Simon}, Jacob B. and {Qi}, Chunhua and {Bai}, Xue-Ning and {Bulatek}, Alyssa and {Andrews}, Sean M. and {Wilner}, David J. and {K{\'o}sp{\'a}l}, {\'A}gnes},
        title = "{Measuring Turbulent Motion in Planet-forming Disks with ALMA: A Detection around DM Tau and Nondetections around MWC 480 and V4046 Sgr}",
      journal = {\apj},
         year = 2020,
        month = jun,
       volume = {895},
       number = {2},
          eid = {109},
        pages = {109},
          doi = {10.3847/1538-4357/ab8cc5},
archivePrefix = {arXiv},
       eprint = {2004.12176},
 primaryClass = {astro-ph.SR},
       adsurl = {https://ui.adsabs.harvard.edu/abs/2020ApJ...895..109F}
}

@ARTICLE{Keppler20,
       author = {{Keppler}, M. and {Penzlin}, A. and {Benisty}, M. and {van Boekel}, R. and {Henning}, T. and {van Holstein}, R.~G. and {Kley}, W. and {Garufi}, A. and {Ginski}, C. and {Brandner}, W. and {Bertrang}, G.~H. -M. and {Boccaletti}, A. and {de Boer}, J. and {Bonavita}, M. and {Brown Sevilla}, S. and {Chauvin}, G. and {Dominik}, C. and {Janson}, M. and {Langlois}, M. and {Lodato}, G. and {Maire}, A. -L. and {M{\'e}nard}, F. and {Pantin}, E. and {Pinte}, C. and {Stolker}, T. and {Szul{\'a}gyi}, J. and {Thebault}, P. and {Villenave}, M. and {Zurlo}, A. and {Rabou}, P. and {Feautrier}, P. and {Feldt}, M. and {Madec}, F. and {Wildi}, F.},
        title = "{Gap, shadows, spirals, and streamers: SPHERE observations of binary-disk interactions in GG Tauri A}",
      journal = {\aap},
         year = 2020,
        month = jul,
       volume = {639},
          eid = {A62},
        pages = {A62},
          doi = {10.1051/0004-6361/202038032},
archivePrefix = {arXiv},
       eprint = {2005.09037},
 primaryClass = {astro-ph.SR},
       adsurl = {https://ui.adsabs.harvard.edu/abs/2020A&A...639A..62K}
}

@ARTICLE{Weaver18,
       author = {{Weaver}, Erik and {Isella}, Andrea and {Boehler}, Yann},
        title = "{Empirical Temperature Measurement in Protoplanetary Disks}",
      journal = {\apj},
         year = 2018,
        month = feb,
       volume = {853},
       number = {2},
          eid = {113},
        pages = {113},
          doi = {10.3847/1538-4357/aaa481},
archivePrefix = {arXiv},
       eprint = {1801.03478},
 primaryClass = {astro-ph.EP},
       adsurl = {https://ui.adsabs.harvard.edu/abs/2018ApJ...853..113W}
}

@ARTICLE{Kudo18,
       author = {{Kudo}, Tomoyuki and {Hashimoto}, Jun and {Muto}, Takayuki and {Liu}, Hauyu Baobab and {Dong}, Ruobing and {Hasegawa}, Yasuhiro and {Tsukagoshi}, Takashi and {Konishi}, Mihoko},
        title = "{A Spatially Resolved au-scale Inner Disk around DM Tau}",
      journal = {\apjl},
         year = 2018,
        month = nov,
       volume = {868},
       number = {1},
          eid = {L5},
        pages = {L5},
          doi = {10.3847/2041-8213/aaeb1c},
archivePrefix = {arXiv},
       eprint = {1810.13148},
 primaryClass = {astro-ph.EP},
       adsurl = {https://ui.adsabs.harvard.edu/abs/2018ApJ...868L...5K}
}

@ARTICLE{Gaia23,
       author = {{Gaia Collaboration} and {Vallenari}, A. and {Brown}, A.~G.~A. and {Prusti}, T. and {de Bruijne}, J.~H.~J. and {Arenou}, F. and {Babusiaux}, C. and {Biermann}, M. and {Creevey}, O.~L. and {Ducourant}, C. and {Evans}, D.~W. and {Eyer}, L. and {Guerra}, R. and {Hutton}, A. and {Jordi}, C. and {Klioner}, S.~A. and {Lammers}, U.~L. and {Lindegren}, L. and {Luri}, X. and {Mignard}, F. and {Panem}, C. and {Pourbaix}, D. and {Randich}, S. and {Sartoretti}, P. and {Soubiran}, C. and {Tanga}, P. and {Walton}, N.~A. and {Bailer-Jones}, C.~A.~L. and {Bastian}, U. and {Drimmel}, R. and {Jansen}, F. and {Katz}, D. and {Lattanzi}, M.~G. and {van Leeuwen}, F. and {Bakker}, J. and {Cacciari}, C. and {Casta{\~n}eda}, J. and {De Angeli}, F. and {Fabricius}, C. and {Fouesneau}, M. and {Fr{\'e}mat}, Y. and {Galluccio}, L. and {Guerrier}, A. and {Heiter}, U. and {Masana}, E. and {Messineo}, R. and {Mowlavi}, N. and {Nicolas}, C. and {Nienartowicz}, K. and {Pailler}, F. and {Panuzzo}, P. and {Riclet}, F. and {Roux}, W. and {Seabroke}, G.~M. and {Sordo}, R. and {Th{\'e}venin}, F. and {Gracia-Abril}, G. and {Portell}, J. and {Teyssier}, D. and {Altmann}, M. and {Andrae}, R. and {Audard}, M. and {Bellas-Velidis}, I. and {Benson}, K. and {Berthier}, J. and {Blomme}, R. and {Burgess}, P.~W. and {Busonero}, D. and {Busso}, G. and {C{\'a}novas}, H. and {Carry}, B. and {Cellino}, A. and {Cheek}, N. and {Clementini}, G. and {Damerdji}, Y. and {Davidson}, M. and {de Teodoro}, P. and {Nu{\~n}ez Campos}, M. and {Delchambre}, L. and {Dell'Oro}, A. and {Esquej}, P. and {Fern{\'a}ndez-Hern{\'a}ndez}, J. and {Fraile}, E. and {Garabato}, D. and {Garc{\'\i}a-Lario}, P. and {Gosset}, E. and {Haigron}, R. and {Halbwachs}, J. -L. and {Hambly}, N.~C. and {Harrison}, D.~L. and {Hern{\'a}ndez}, J. and {Hestroffer}, D. and {Hodgkin}, S.~T. and {Holl}, B. and {Jan{\ss}en}, K. and {Jevardat de Fombelle}, G. and {Jordan}, S. and {Krone-Martins}, A. and {Lanzafame}, A.~C. and {L{\"o}ffler}, W. and {Marchal}, O. and {Marrese}, P.~M. and {Moitinho}, A. and {Muinonen}, K. and {Osborne}, P. and {Pancino}, E. and {Pauwels}, T. and {Recio-Blanco}, A. and {Reyl{\'e}}, C. and {Riello}, M. and {Rimoldini}, L. and {Roegiers}, T. and {Rybizki}, J. and {Sarro}, L.~M. and {Siopis}, C. and {Smith}, M. and {Sozzetti}, A. and {Utrilla}, E. and {van Leeuwen}, M. and {Abbas}, U. and {{\'A}brah{\'a}m}, P. and {Abreu Aramburu}, A. and {Aerts}, C. and {Aguado}, J.~J. and {Ajaj}, M. and {Aldea-Montero}, F. and {Altavilla}, G. and {{\'A}lvarez}, M.~A. and {Alves}, J. and {Anders}, F. and {Anderson}, R.~I. and {Anglada Varela}, E. and {Antoja}, T. and {Baines}, D. and {Baker}, S.~G. and {Balaguer-N{\'u}{\~n}ez}, L. and {Balbinot}, E. and {Balog}, Z. and {Barache}, C. and {Barbato}, D. and {Barros}, M. and {Barstow}, M.~A. and {Bartolom{\'e}}, S. and {Bassilana}, J. -L. and {Bauchet}, N. and {Becciani}, U. and {Bellazzini}, M. and {Berihuete}, A. and {Bernet}, M. and {Bertone}, S. and {Bianchi}, L. and {Binnenfeld}, A. and {Blanco-Cuaresma}, S. and {Blazere}, A. and {Boch}, T. and {Bombrun}, A. and {Bossini}, D. and {Bouquillon}, S. and {Bragaglia}, A. and {Bramante}, L. and {Breedt}, E. and {Bressan}, A. and {Brouillet}, N. and {Brugaletta}, E. and {Bucciarelli}, B. and {Burlacu}, A. and {Butkevich}, A.~G. and {Buzzi}, R. and {Caffau}, E. and {Cancelliere}, R. and {Cantat-Gaudin}, T. and {Carballo}, R. and {Carlucci}, T. and {Carnerero}, M.~I. and {Carrasco}, J.~M. and {Casamiquela}, L. and {Castellani}, M. and {Castro-Ginard}, A. and {Chaoul}, L. and {Charlot}, P. and {Chemin}, L. and {Chiaramida}, V. and {Chiavassa}, A. and {Chornay}, N. and {Comoretto}, G. and {Contursi}, G. and {Cooper}, W.~J. and {Cornez}, T. and {Cowell}, S. and {Crifo}, F. and {Cropper}, M. and {Crosta}, M. and {Crowley}, C. and {Dafonte}, C. and {Dapergolas}, A. and {David}, M. and {David}, P. and {de Laverny}, P. and {De Luise}, F. and {De March}, R. and {De Ridder}, J. and {de Souza}, R. and {de Torres}, A. and {del Peloso}, E.~F. and {del Pozo}, E. and {Delbo}, M. and {Delgado}, A. and {Delisle}, J. -B. and {Demouchy}, C. and {Dharmawardena}, T.~E. and {Di Matteo}, P. and {Diakite}, S. and {Diener}, C. and {Distefano}, E. and {Dolding}, C. and {Edvardsson}, B. and {Enke}, H. and {Fabre}, C. and {Fabrizio}, M. and {Faigler}, S. and {Fedorets}, G. and {Fernique}, P. and {Fienga}, A. and {Figueras}, F. and {Fournier}, Y. and {Fouron}, C. and {Fragkoudi}, F. and {Gai}, M. and {Garcia-Gutierrez}, A. and {Garcia-Reinaldos}, M. and {Garc{\'\i}a-Torres}, M. and {Garofalo}, A. and {Gavel}, A. and {Gavras}, P. and {Gerlach}, E. and {Geyer}, R. and {Giacobbe}, P. and {Gilmore}, G. and {Girona}, S. and {Giuffrida}, G. and {Gomel}, R. and {Gomez}, A. and {Gonz{\'a}lez-N{\'u}{\~n}ez}, J. and {Gonz{\'a}lez-Santamar{\'\i}a}, I. and {Gonz{\'a}lez-Vidal}, J.~J. and {Granvik}, M. and {Guillout}, P. and {Guiraud}, J. and {Guti{\'e}rrez-S{\'a}nchez}, R. and {Guy}, L.~P. and {Hatzidimitriou}, D. and {Hauser}, M. and {Haywood}, M. and {Helmer}, A. and {Helmi}, A. and {Sarmiento}, M.~H. and {Hidalgo}, S.~L. and {Hilger}, T. and {H{\l}adczuk}, N. and {Hobbs}, D. and {Holland}, G. and {Huckle}, H.~E. and {Jardine}, K. and {Jasniewicz}, G. and {Jean-Antoine Piccolo}, A. and {Jim{\'e}nez-Arranz}, {\'O}. and {Jorissen}, A. and {Juaristi Campillo}, J. and {Julbe}, F. and {Karbevska}, L. and {Kervella}, P. and {Khanna}, S. and {Kontizas}, M. and {Kordopatis}, G. and {Korn}, A.~J. and {K{\'o}sp{\'a}l}, {\'A}. and {Kostrzewa-Rutkowska}, Z. and {Kruszy{\'n}ska}, K. and {Kun}, M. and {Laizeau}, P. and {Lambert}, S. and {Lanza}, A.~F. and {Lasne}, Y. and {Le Campion}, J. -F. and {Lebreton}, Y. and {Lebzelter}, T. and {Leccia}, S. and {Leclerc}, N. and {Lecoeur-Taibi}, I. and {Liao}, S. and {Licata}, E.~L. and {Lindstr{\o}m}, H.~E.~P. and {Lister}, T.~A. and {Livanou}, E. and {Lobel}, A. and {Lorca}, A. and {Loup}, C. and {Madrero Pardo}, P. and {Magdaleno Romeo}, A. and {Managau}, S. and {Mann}, R.~G. and {Manteiga}, M. and {Marchant}, J.~M. and {Marconi}, M. and {Marcos}, J. and {Marcos Santos}, M.~M.~S. and {Mar{\'\i}n Pina}, D. and {Marinoni}, S. and {Marocco}, F. and {Marshall}, D.~J. and {Martin Polo}, L. and {Mart{\'\i}n-Fleitas}, J.~M. and {Marton}, G. and {Mary}, N. and {Masip}, A. and {Massari}, D. and {Mastrobuono-Battisti}, A. and {Mazeh}, T. and {McMillan}, P.~J. and {Messina}, S. and {Michalik}, D. and {Millar}, N.~R. and {Mints}, A. and {Molina}, D. and {Molinaro}, R. and {Moln{\'a}r}, L. and {Monari}, G. and {Mongui{\'o}}, M. and {Montegriffo}, P. and {Montero}, A. and {Mor}, R. and {Mora}, A. and {Morbidelli}, R. and {Morel}, T. and {Morris}, D. and {Muraveva}, T. and {Murphy}, C.~P. and {Musella}, I. and {Nagy}, Z. and {Noval}, L. and {Oca{\~n}a}, F. and {Ogden}, A. and {Ordenovic}, C. and {Osinde}, J.~O. and {Pagani}, C. and {Pagano}, I. and {Palaversa}, L. and {Palicio}, P.~A. and {Pallas-Quintela}, L. and {Panahi}, A. and {Payne-Wardenaar}, S. and {Pe{\~n}alosa Esteller}, X. and {Penttil{\"a}}, A. and {Pichon}, B. and {Piersimoni}, A.~M. and {Pineau}, F. -X. and {Plachy}, E. and {Plum}, G. and {Poggio}, E. and {Pr{\v{s}}a}, A. and {Pulone}, L. and {Racero}, E. and {Ragaini}, S. and {Rainer}, M. and {Raiteri}, C.~M. and {Rambaux}, N. and {Ramos}, P. and {Ramos-Lerate}, M. and {Re Fiorentin}, P. and {Regibo}, S. and {Richards}, P.~J. and {Rios Diaz}, C. and {Ripepi}, V. and {Riva}, A. and {Rix}, H. -W. and {Rixon}, G. and {Robichon}, N. and {Robin}, A.~C. and {Robin}, C. and {Roelens}, M. and {Rogues}, H.~R.~O. and {Rohrbasser}, L. and {Romero-G{\'o}mez}, M. and {Rowell}, N. and {Royer}, F. and {Ruz Mieres}, D. and {Rybicki}, K.~A. and {Sadowski}, G. and {S{\'a}ez N{\'u}{\~n}ez}, A. and {Sagrist{\`a} Sell{\'e}s}, A. and {Sahlmann}, J. and {Salguero}, E. and {Samaras}, N. and {Sanchez Gimenez}, V. and {Sanna}, N. and {Santove{\~n}a}, R. and {Sarasso}, M. and {Schultheis}, M. and {Sciacca}, E. and {Segol}, M. and {Segovia}, J.~C. and {S{\'e}gransan}, D. and {Semeux}, D. and {Shahaf}, S. and {Siddiqui}, H.~I. and {Siebert}, A. and {Siltala}, L. and {Silvelo}, A. and {Slezak}, E. and {Slezak}, I. and {Smart}, R.~L. and {Snaith}, O.~N. and {Solano}, E. and {Solitro}, F. and {Souami}, D. and {Souchay}, J. and {Spagna}, A. and {Spina}, L. and {Spoto}, F. and {Steele}, I.~A. and {Steidelm{\"u}ller}, H. and {Stephenson}, C.~A. and {S{\"u}veges}, M. and {Surdej}, J. and {Szabados}, L. and {Szegedi-Elek}, E. and {Taris}, F. and {Taylor}, M.~B. and {Teixeira}, R. and {Tolomei}, L. and {Tonello}, N. and {Torra}, F. and {Torra}, J. and {Torralba Elipe}, G. and {Trabucchi}, M. and {Tsounis}, A.~T. and {Turon}, C. and {Ulla}, A. and {Unger}, N. and {Vaillant}, M.~V. and {van Dillen}, E. and {van Reeven}, W. and {Vanel}, O. and {Vecchiato}, A. and {Viala}, Y. and {Vicente}, D. and {Voutsinas}, S. and {Weiler}, M. and {Wevers}, T. and {Wyrzykowski}, {\L}. and {Yoldas}, A. and {Yvard}, P. and {Zhao}, H. and {Zorec}, J. and {Zucker}, S. and {Zwitter}, T.},
        title = "{Gaia Data Release 3. Summary of the content and survey properties}",
      journal = {\aap},
         year = 2023,
        month = jun,
       volume = {674},
          eid = {A1},
        pages = {A1},
          doi = {10.1051/0004-6361/202243940},
archivePrefix = {arXiv},
       eprint = {2208.00211},
 primaryClass = {astro-ph.GA},
       adsurl = {https://ui.adsabs.harvard.edu/abs/2023A&A...674A...1G}
}

@ARTICLE{Manara20,
       author = {{Manara}, C.~F. and {Natta}, A. and {Rosotti}, G.~P. and {Alcal{\'a}}, J.~M. and {Nisini}, B. and {Lodato}, G. and {Testi}, L. and {Pascucci}, I. and {Hillenbrand}, L. and {Carpenter}, J. and {Scholz}, A. and {Fedele}, D. and {Frasca}, A. and {Mulders}, G. and {Rigliaco}, E. and {Scardoni}, C. and {Zari}, E.},
        title = "{X-shooter survey of disk accretion in Upper Scorpius. I. Very high accretion rates at age > 5 Myr}",
      journal = {\aap},
         year = 2020,
        month = jul,
       volume = {639},
          eid = {A58},
        pages = {A58},
          doi = {10.1051/0004-6361/202037949},
archivePrefix = {arXiv},
       eprint = {2004.14232},
 primaryClass = {astro-ph.SR},
       adsurl = {https://ui.adsabs.harvard.edu/abs/2020A&A...639A..58M}
}

@ARTICLE{Wendeborn24,
       author = {{Wendeborn}, John and {Espaillat}, Catherine C. and {Lopez}, Sophia and {Thanathibodee}, Thanawuth and {Robinson}, Connor E. and {Pittman}, Caeley V. and {Calvet}, Nuria and {Flors}, Nicole and {Walter}, Fredrick M. and {K{\'o}sp{\'a}l}, {\'A}gnes and {Grankin}, Konstantin N. and {Mendigut{\'\i}a}, Ignacio and {G{\"u}nther}, Hans Moritz and {Eisl{\"o}ffel}, Jochen and {Guo}, Zhen and {France}, Kevin and {Fiorellino}, Eleonora and {Fischer}, William J. and {{\'A}brah{\'a}m}, P{\'e}ter and {Herczeg}, Gregory J.},
        title = "{A Multiwavelength, Multiepoch Monitoring Campaign of Accretion Variability in T Tauri Stars from the ODYSSEUS Survey. I. HST Far-UV and Near-UV Spectra}",
      journal = {\apj},
         year = 2024,
        month = aug,
       volume = {970},
       number = {2},
          eid = {118},
        pages = {118},
          doi = {10.3847/1538-4357/ad4a62},
archivePrefix = {arXiv},
       eprint = {2405.21038},
 primaryClass = {astro-ph.SR},
       adsurl = {https://ui.adsabs.harvard.edu/abs/2024ApJ...970..118W}
}

@ARTICLE{Espaillat19,
       author = {{Espaillat}, C.~C. and {Mac{\'\i}as}, E. and {Hern{\'a}ndez}, J. and {Robinson}, C.},
        title = "{Revealing the Star-Disk-Jet Connection in GM Aur Using Multiwavelength Variability}",
      journal = {\apjl},
         year = 2019,
        month = jun,
       volume = {877},
       number = {2},
          eid = {L34},
        pages = {L34},
          doi = {10.3847/2041-8213/ab2193},
archivePrefix = {arXiv},
       eprint = {1905.12789},
 primaryClass = {astro-ph.SR},
       adsurl = {https://ui.adsabs.harvard.edu/abs/2019ApJ...877L..34E}
}
\bibliographystyle{aasjournal}

\end{document}